\documentclass[twocolumn,aps,prb,nofootinbib]{revtex4-1}
\usepackage{graphicx}
\usepackage{times}
\usepackage{bm}
\usepackage[linktocpage=true]{hyperref}
\usepackage{pgfplots}
\usepackage{amsfonts}
\usepackage{amsmath}
\usepackage{braket}
\usepackage{relsize}
\usetikzlibrary{arrows}
\usepackage{enumerate}
\usetikzlibrary[patterns] 
\usetikzlibrary[snakes] 
\usetikzlibrary[shapes] 
\usepackage{pgfplots}
\usepackage{newfloat}
\DeclareFloatingEnvironment[name={Supplementary Figure}]{suppfigure}
\graphicspath{ {Figures/} }

\begin{document}

\title{Control and detection of Majorana bound states in quantum dot arrays}

\author{John P. T. Stenger\textsuperscript{1,2}}
\author{Benjamin D. Woods\textsuperscript{1}}
\author{Sergey M. Frolov\textsuperscript{2}}
\author{Tudor D. Stanescu\textsuperscript{1}}
\affiliation{\textsuperscript{1}Department of Physics and Astronomy, West Virginia University, Morgantown, WV 26506}
\affiliation{\textsuperscript{2}Department of Physics and Astronomy, University of Pittsburgh, Pittsburgh, PA 15260}

\begin{abstract}
We study the low-energy physics of a one-dimensional array of superconducting quantum dots realized by proximity coupling a semiconductor nanowire to multiple superconducting islands separated by narrow uncovered regions. The effective electrostatic potential inside the quantum dots and the uncovered regions can be controlled using potential gates. By performing detailed numerical calculations based on effective tight-binding models, we find that multiple low-energy sub-gap states consisting of partially overlapping Majorana bound states emerge generically in the vicinity of the uncovered regions. Explicit differential conductance calculations show that a robust zero-bias conductance peak is not inconsistent with the presence of such states localized throughout the system, hence the observation of such a peak does not demonstrate the realization of well-separated Majorana zero modes. However, we find that creating effective potential wells in the uncovered regions traps pairs of nearby partially overlapping Majorana bound states, which become less separated and acquire a finite gap that protects the pair of Majorana zero modes localized at the ends of the system. This behavior persists over a significant parameter range, suggesting that proximitized quantum dot arrays could provide a platform for highly controllable Majorana devices.
\end{abstract}

\maketitle

\section{Introduction}

The quest for the realization of topological superconductivity and Majorana zero modes (MZMs)\cite{Majorana1937,Read2000,Kitaev2001} in solid state   systems has gained significant momentum in recent years. Promising proposals for realizing these quantum states in topological insulator--superconductor  structures,\cite{Fu2008,Fu2009} atomic magnetic chains coupled to conventional superconductors,\cite{Vazifeh2013,NadjPerge2013,Pientka2013} or semiconductor nanowires with strong spin-orbit coupling and proximity induced superconductivity\cite{Sau2010a,Alicea2010,Oreg2010,Lutchyn2010,Sau2010} have stimulated lively experimental activity.\cite{Mourik2012,Deng2012,Das2012,Rokhinson2012,Churchill2013,Finck2013,NadjPerge2014,Wiedenmann2016} In particular,  improvements in materials science  and nanofabrication have led to significant progress in achieving experimental conditions consistent with the presence of non-Abelian MZMs in semiconductor-based  platforms.\cite{Chang2015,Albrecht2016,Deng2016,Chen2017,Zhang2017,Nichele2017,Zhang2018} 
Moreover, the development of high-quality two-dimensional semiconductor -superconductor structures\cite{Shabani2016,Kjaergaard2016,Suominen2017}
opens the possibility of using such systems as a platform for complex Majorana-based topological circuits.\cite{Alicea2011,Sau2011,Aasen2016,Karzig2017,Litinski2017}
While the main initial challenge -- the presence of  significant subgap conductance,\cite{Takei2013,Stanescu2014} which plagued the first generation of  experiments\cite{Mourik2012,Deng2012,Das2012,Churchill2013,Finck2013} -- has been overcome, there are still serious concerns related to the possible presence of (unwanted) quantum dots or non uniform parameters, which can lead to the formation of topologically-trivial low-energy states\cite{Kells2012,Chevallier2012,Prada2012,Roy2013,SanJose2013,Ojanen2013,Stanescu2014a,Lee2014,Cayao2015,Klinovaja2015,Fleckenstein2017,Liu2017a,Moore2018} that mimic the signatures of non-Abelian MZMs in local measurements at the end of the wire.  

Emergent MZMs -- also called zero-energy Majorana bound states (MBSs) -- are topologically-protected  and provide a natural basis
for fault-tolerant quantum computation.\cite{Nayak2008,DSarma2015,Stanescu2017} A key requirement for topological protection, which ensures the immunity of the Majorana-based qubit against local perturbations, is the non-locality (i.e., the spatial separation) of the MZMs. Specifically, in hybrid nanowire systems, the topological superconducting phase supports one pair of MZMs localized at the two ends of the wire.   
By contrast, the trivial low-energy states mimicking MZM signatures are partially separated Andreev bound states (ps-ABSs) consisting of a pair of component MBSs separated by a distance comparable to or larger than the characteristic Majorana  length-scale (but  less  than  the length of the wire).\cite{Stanescu2016,Moore2018}  It was recently argued that  the non-locality of MBSs in hybrid nanowires can be measured via the interaction
of the zero-energy state in the nanowire with a quantum-dot state at one end.\cite{Prada2017,Clarke2017,Deng2017,Schuray2017} This method can identify a ps-ABS, provided both constituent MBSs have a measurable coupling to the quantum dot. A natural question concerns the relevance a local measurement when the system supports multiple MBSs localized throughout the wire, for example when several impurities (or defects) effectively ``cut'' the wire into a chain of strongly coupled superconducting islands.   

\begin{figure}[t]
\begin{center}
\includegraphics[width=0.48\textwidth]{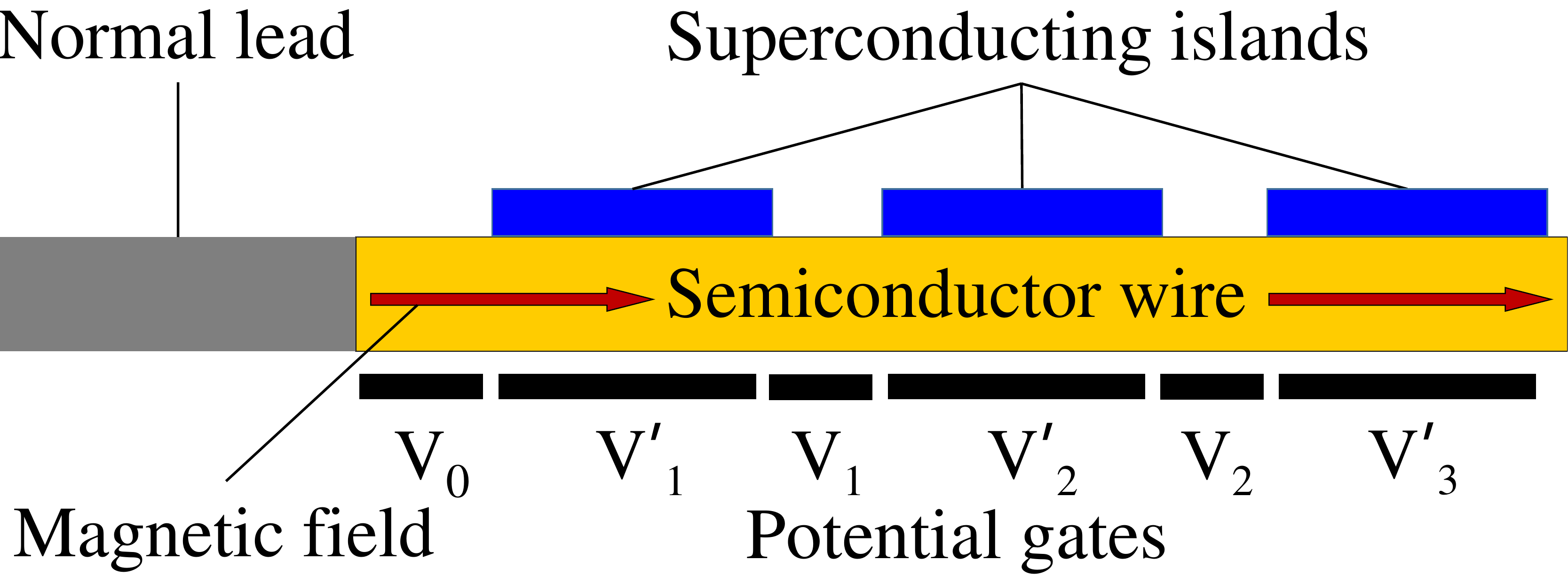}
\end{center}
\vspace{-1.5mm}

\caption{Schematic representation of the hybrid device. A SM wire with Rashba spin-orbit coupling is proximity-coupled to $N_{SC}$ superconducting islands (here $N_{SC}=3$) separated by small gaps, resulting in a 1D array of superconducting quantum dots. A magnetic field is applied along the wire. The electrostatic potential inside the proximitized and the uncovered regions is controlled by back gates ($V^\prime_i$ and $V_i$, respectively).  A normal metal lead is used to tunnel charge into the end of the chain through a tunnel barrier controlled by $V_0$.}
\label{Fig_01}
\end{figure}

A possible practical approach to the MZM non-locality challenge is to engineer hybrid  structures that are more controllable.  One such proposal involves 
a chain of gate-tunable quantum dots connected by s-wave superconductors.\cite{Sau2012a}  A similar type of structure, which is directly related to the Majorana nanowires used in current experiments, consists of chains of  proximitized nanowire segments (i.e. superconducting quantum dots or superconducting islands)  separated by narrow uncovered regions (see Fig. \ref{Fig_01}). The effective electrostatic potential inside both the quantum dots and the uncovered regions can be controlled using back gates. If the superconducting islands are weakly coupled (e.g., by creating large potential barriers in the uncovered regions), driving the system into a parameter regime corresponding to a topological superconducting phase (in long, homogeneous wires) will result in the emergence of low-energy ps-ABSs localized inside each island. The key question is whether or not these ps-ABSs can be ``merged'' into a pair of MZMs localized at the ends of the chain by controlling the barrier gates. 

In this paper we address the questions formulated above by systematically studying the low-energy physics of a chain of superconducting islands using an effective tight-binding model, which is solved numerically. The differential conductance for charge tunneling into the end of the wire from a normal lead\cite{Pientka2012,Prada2012,Lin2012,Rainis2013,Stanescu2014,Yan2014,Setiwan2015,Liu2017,Stenger2017} is calculated explicitly as a function of various control parameters. We find the observation of a robust zero-bias conductance peak does not guarantee the existence of a single pair of well separated MZMs. In general, the chain of coupled superconducting islands supports multiple MBSs localized throughout the system and having nearly zero energy within a significant range of Zeeman fields and back gate potentials. We also find that realizing a uniform effective potential, which is naively expected to lead to a topological state with a single pair of MZMs localized at the ends of the chain, by lowering the barrier potential between adjacent quantum dots requires fine tunning and may even be impossible if the barrier regions are too wide. Nonetheless, we identify  the optimal regime for realizing well separated MZMs in a quantum dot array as corresponding to (shallow) potential wells in the barrier regions between the dots. The potential wells act as ``traps'' for pairs of nearby MBSs, which overlap strongly and acquire an energy gap, while the unpaired, end-of-chain MBSs  develop into robust, topologically-protected MZMs. Our study shows that the details of the effective electrostatic potential are important for understanding the low-energy physics of a Majorana hybrid structure. Accurately capturing such details, particularly in the uncovered regions, may require solving a three-dimensional Schr\"{o}dinger-Poisson equation.\cite{Woods2018}  In addition, our results show that arrays of proximitized, gate-tunable quantum dots are versatile systems for realizing robust MZMs, but the unambiguous detection of such modes requires non-local probes, beyond end-of-chain charge tunneling.  

The remainder of the paper is organized as follows. In Sec. \ref{model} we describe the theoretical model used in our calculations. The tight-binding Hamiltonian for the normal-metal--semiconductor--superconductor structure is presented in Sec. \ref{TbHam}.  In Sec. \ref{Conductance} we briefly describe the method used for calculating the differential conductance, while in Sec. \ref{EffModel} we discuss the approach used to calculate the effective potential profile along the hybrid system.  
Our results are described in Sec. \ref{results}, starting with an overview of the general low-energy differential conductance features that characterize the device. In Sec. \ref{barrierPR} we establish the main regimes associated with the potential in the uncovered regions of the device based on a simplified effective model, while the details of the potential profile are considered explicitly in Sec.  \ref{effectivePMR}.  Our conclusions are provided in Sec. \ref{conclusion}.

\section{Theoretical model} \label{model}

In this section we introduce the theoretical model and briefly describe the methods used throughout this paper. In Sec. \ref{TbHam} we treat the nanowire within a simplified one dimensional (1D) chain model, but we incorporate the superconductor components explicitly using a self-energy formalism. Sec. \ref{Conductance} sketches the method used in the conductance calculations, while in Sec. \ref{EffModel} we derive a multi-orbital effective 1D model that incorporates the details of the electrostatic profile. 

\subsection{Tight-binding Hamiltonian} \label{TbHam}

The hybrid structure that we study consists of a semiconductor nanowire proximity coupled to superconductor islands separated by small uncovered regions (See Fig. \ref{Fig_01}). A magnetic field is applied along the wire, while multiple back gates enable the control of the electrostatic potential inside both the proximitized and the uncovered regions. A normal lead coupled to one of the ends is used for tunneling charge into the wire. The total Hamiltonian describing this hybrid system has the form

\begin{equation}
H = H_{NM}+ H_{SM} + \sum_j H_{SC,j} + T_{NM}+ \sum_j T_{SC,j} + H_{ext},                            \label{Htot}
\end{equation}
where the term $H_{NM}$ describes the normal metal (NM) lead, $H_{SM}$ represents the Hamiltonian of a semiconductor (SM) nanowire with Rashba-type spin-orbit coupling, $H_{SC,j}$ correspond to the superconductor (SC) islands, $T_{NM}$ and $T_{SC,j}$ describe the coupling of the SM nanowire to the NM lead and the SC islands, respectively, and $H_{ext}$ describes the external fields, including the magnetic field and gate-induced potentials.   
Explicitly, the Hamiltonian for the metallic lead can be written as 
\begin{equation}
H_{NM} = \sum_{i,\delta}t_{NM}{\rm a}^{\dagger}_{i}{\rm a}_{i+\delta}+ \mu_{NM}\sum_{i}{\rm a}^{\dagger}_{i}{\rm a}_{i},
\end{equation}
where $1\leq i \leq N_{NM}$ labels the position along the chain, $\delta=\pm 1$, $t_{NM}$ is the hopping matrix element, and $\mu_{NM}$ the chemical potential of the metallic lead. Using spinor notation, the electron creation operator on site $i$ is ${\rm a}_i^\dagger =({\rm a}_{i\uparrow}^\dagger  ~{\rm a}_{i\downarrow}^\dagger)$.
We neglect the effects of the external fields on the normal metal and the superconductor. The Hamiltonian that describes the semiconductor, including the applied fields, is
\begin{eqnarray}
H_{SM}&+&H_{ext} = \sum_{i, \delta}t_{SM}c^{\dagger}_{i}c_{i+\delta}+\sum_{i,j}\left(-\mu_{SM}+V_j(i)\right)c^{\dagger}_{i}c_{i} \nonumber \\
&+& i\frac{\alpha}{2}\sum_{i}\left(c^{\dagger}_{i+\delta}\hat{\sigma}_y c_i +h.c.\right) + E_Z\sum_{i}c^{\dagger}_{i}\hat{\sigma}_x c_{i}, \label{Ham_SM_EXT}
\end{eqnarray}
where $c_i^\dagger=(c_{i\uparrow}^\dagger  ~c_{i\downarrow}^\dagger)$ is the electron creation operator on the site $1 \leq i \leq N_{SM}$ of the semiconductor wire,  $t_{SM}$ is the nearest-neighbor hopping,  $\mu_{SM}$ is the chemical potential, and $\alpha$ is the Rashba spin-orbit coupling coefficient. The position-dependent potential $V_0(i)$ describes the charge tunneling barrier,  $V_{0}(i)=V_0\exp[-(i/\sigma)^2]$, where $\sigma$ is the width of the tunnel barrier, while $V_j(i)$, with $j\geq 1$, describes the effective potential  in the uncovered regions, 
\begin{equation}
V_{j}(i)=V_j\left[\frac{1}{1+\exp[i-M_{j+1}]}-\frac{1}{1+\exp[i-N_j]}\right],
\end{equation}
where $M_j$ and $N_j$ label the leftmost and rightmost sites of the j$^{\rm th}$ uncovered region, respectively. Finally, $E_Z$ represents the Zeeman splitting due to the applied magnetic field and the matrices $\hat{\sigma}_\mu$, with $\mu = x,y,z$,  represent Pauli matrices associated with the spin degree of freedom. For now, we assume vanishing effective potentials inside the proximitized regions, $V_j^\prime(i) = 0$. 
The superconducting islands are described at the mean field level by the Bogoliubov--de Gennes (BdG) Hamiltonian 
\begin{eqnarray}
H_{SC_j} &=& \sum_{i,\delta}t_{SC}a^{\dagger}_{i}a_{i+\delta} - \mu_{SC}\sum_{i}a^{\dagger}_{i}a_{i}  \nonumber \\
&+&  \Delta_{0}\sum_{i}(a_{i\uparrow}^{\dagger}a^{\dagger}_{i\downarrow}+a_{i\downarrow}a_{i\uparrow}),   
\end{eqnarray}
where $a_i^\dagger=(a_{i\uparrow}^\dagger  ~a_{i\downarrow}^\dagger)$ is the electron creation operator on the site  $i$ of the superconducting chains, $\delta$ is the nearest neighbor vector in the superconductor,
$t_{SC}$ is the nearest-neighbor hopping matrix element, $\mu_{SC}$ is the chemical potential, and $\Delta_0$ is the superconducting gap.
The coupling between the semiconductor wire and the metallic lead is described by the term
\begin{equation}
\label{msm}
T_{NM}= \tilde{t}_{NM-SM}\left(c^{\dagger}_{1}{\rm a}_{N_{NM}}+{\rm a}_{N_{NM}}^\dagger c_{1} \right), 
\end{equation}
where the nearest-neighbor hopping $\tilde{t}_{NM-SM}$ quantifies the coupling strength between the right end of the metallic lead and the left end of the nanowire.
Finally, the coupling between the semiconductor and the $j^{th}$ superconductor is described by the term
\begin{equation}
T_{SC,j}=\tilde{t}\sum_{i=M_j}^{N_{j}}\left(c_{i}^\dagger a_{i} + a_{i}^\dagger c_{i}\right), 
\label{msm}
\end{equation}
where $\tilde{t}$ is the hopping across the semiconductor-superconductor (SM-SC) interface, while $c_i$ designate the annihilation operator in the semiconductor and $a_i$ designates the annihilation operator on the surface of the superconductor.

In practice, rather than diagonalizing the full Hamiltonian, it is convenient to calculate the effective Green function for the semiconductor wire by integrating out the degrees of freedom of the superconductors.\cite{Stanescu2013} The proximity effect due to superconductor $j$ is captured by the self-energy term
\begin{equation}
\Sigma_{SC,j}(\omega)=\tilde{t}^2G_{SC,j}(\omega),  \label{Sig_sc}
\end{equation}
where $G_{SC,j}(\omega)$ is the Green function of the $j^{th}$ superconductor at the SM-SC interface. 
Assuming that the superconductors are truly bulk systems (i.e., wide enough and thick enough), the self-energy becomes local,\cite{Stanescu2013} $\Sigma_{SC,j}(\omega; i, i') = \delta_{i,i'}\Sigma_{SC,j}(\omega)$, with
\begin{equation}
\Sigma_{SC,j}(\omega)=-|\tilde{t}|^2\nu_{SC}\left(\frac{\omega\tau_0+\Delta_0\tau_x}{\sqrt{\Delta_0^2-\omega^2}}+\zeta\tau_z\right), \label{fullscse}
\end{equation}
where $\nu_{SC} = \sqrt{4t_{SC}\mu_{SC}-\mu_{SC}^2}/({2 t_{SC}^2})$ is the surface density of states of the bulk superconductor at the chemical potential and $\zeta= (2t_{SC}\mu_{SC}-4t_{SC}^2)/(\mu_{SC}^2-4t_{SC}\mu_{SC})$ is a proximity-induced shift of the SM chemical potential. The matrices $\hat{\tau}_\mu$, with $\mu = x,y,z$,  represent Pauli matrices associated with particle-hole degree of freedom.  For simplicity, we assume that the superconducting islands are identical, so that the self energy will be the same for each proximitized segment.  

\subsection{Conductance calculations} \label{Conductance}

The differential conductance is calculated using the  Blonder-Tinkham-Klawijk (BTK) formalism.\cite{Blonder1982}  In essence, this involves 
solving the BdG equation for the total Hamiltonian given by Eq. (\ref{Htot}) with appropriate boundary conditions.\cite{Lin2012}
\begin{eqnarray}
\label{shro}
&~&\sum_{i^\prime=0}^{N_{tot}}\sum_{\sigma^\prime}({\cal H}_{i\sigma,i^\prime\sigma^\prime}-\omega\delta_{i,i^\prime}\delta_{\sigma,\sigma^\prime})\Psi_{i^\prime\sigma^\prime}=0  \\
&~&{\rm for}~ ~ i=1,\dots,N_{tot},  ~~\sigma=\pm 1, \nonumber
\end{eqnarray}
where $N_{tot}=N_{NM}+N_{SC}$ and ${\cal H}$ is the first quantized representation of $H_{tot}$.  We apply plane wave boundary conditions to the wave vector $\Psi$ on the leftmost two sites of the normal lead, $i=0,1$.  Note that these boundary conditions are expressed in terms of the normal and anomalous reflection coefficients.\cite{Blonder1982,Lin2012} 
Solving Eq. \ref{shro} provides the values of these reflection coefficients, which in turn,  determine the differential conductance. Explicitly, we have    
\begin{equation}
\frac{dI}{dV}=\frac{e^2}{h}\!\sum_{\sigma,\sigma^\prime}\left(1\!-\!|[r_N(V)]_{\sigma,\sigma^\prime}|^2\!+\!|[r_A(V)]_{\sigma,\sigma^\prime}|^2\right), \label{dIdV}
\end{equation}
where $r_N$ is the matrix of normal reflection coefficients and $r_A$ is the matrix of the anomalous coefficients, with the indices $\sigma$ and $\sigma'$ running over the spin degrees of freedom. 

Instead of solving the BdG equation (\ref{shro}) for the full Hamiltonian, we can integrate out the superconducting degrees of freedom and calculate the reflection coefficients by solving  the equation\cite{Stenger2017}
\begin{equation}
\widetilde{\Psi}^{(\sigma)}=({\cal H}_{NM}+{\cal H}_{SM}+{\cal T}_{NM}+\sum_j{\Sigma_{SC,j}}+Q(\omega)-\omega)^{-1}J^{(\sigma)}(\omega), \label{GQ1}
\end{equation}  
where $\Sigma_{SC,j}$ is given by Eq. (\ref{fullscse}), ${\cal H}_{NM}$, ${\cal H}_{SM}$, ${\cal T}_{NM}$ are the first quantized representations of $H_{NM}$, $H_{SM}$, $T_{NM}$, the wave vector $\widetilde{\Psi}$ differs from $\Psi$ by phase factors multiplying the reflection coefficients,   $J^{(\sigma)}$ is a vector determined by the boundary conditions for the incoming current and $Q$ is a matrix determined by the boundary conditions for the reflected current. The details of the formalism can be found in Ref. \onlinecite{Stenger2017}.

\subsection{Effective potential calculations}\label{EffModel}

The theoretical model described above, which is ubiquitous in the literature on Majorana nanowires, is based on some rather arbitrary assumptions regarding the position dependence of the effective potential, i.e. the functions $V_j(i)$ and  $V^\prime_j(i)$. In reality, this potential depends on many system parameters including the device geometry, material dielectric constants, applied gate potentials, the work function difference between the nanowire and superconductor, and even on the charge density in the wire. Also, the actual nanowires have finite thickness, so the effective electrostatic potential vary across the transverse profile of the wire. This variation is critical in determining key properties, such as the strength of the Rashba spin-orbit coupling and the proximity effect induced in the wire.
Moreover, as the system parameters are, in general, position-dependent, the effective potential vary along the wire, particularly in the uncovered regions. Therefore, to gain further insight into the low-energy physics of the hybrid structure, it is critical to calculate explicitly the effective electrostatic potential in the wire. We emphasize that this is a highly nontrivial task, as is involves solving  a three-dimensional Schr\"{o}dinger-Poisson problem. Following Ref. \onlinecite{Woods2018}, we address this problem using an effective theory approach that allows us to construct a multi-orbital low-energy 1D model that incorporates the information about the dependence across the transverse profile of the wire into a position-dependent orbital basis. The details of this construction can be found in Ref. \onlinecite{Woods2018}; below, we summarize the key results relevant for this work. 

\begin{figure}[t]
\begin{center}
\includegraphics[width=0.4\textwidth]{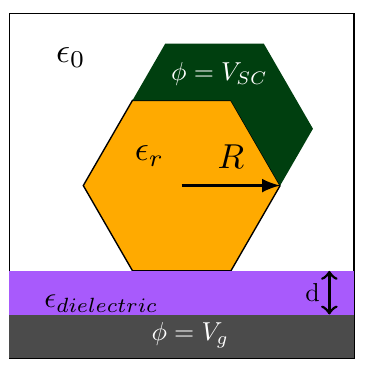}

\end{center}
\vspace{-1.5mm}

\caption{Schematic representation of the cross section of the nanowire device in a proximitized region.  The semiconductor (orange hexagon) is proximity-coupled to a superconductor (black).  There is a dielectric (purple) separating the nanowire from a back gate (gray).  The parameters $\epsilon_r$ and  $\epsilon_{dielectric}$ are the dielectric constants of the SM and the dielectric, respectively, $V_g$ is the voltage applied to the (local) back gate, and $V_{SC}$ is the work function difference between the SM wire and the SC.  In an uncovered region the cross section is similar,  except that there is no superconductor.}
\label{Fig_02}
\end{figure}

Consider a finite nanowire with a given transverse profile. We divide the wire into $N_{x}$ layers (or slices), each containing $N_{\bot}$ sites. A cross section of the system is depicted in Fig.~\ref{Fig_02}.  The system is described by the tight-binding Hamiltonian
\begin{equation}
\begin{aligned}
H_{3D}  &= \sum\limits_{i,j,m,\sigma} t^{\bot}_{ij} c_{im\sigma}^{\dagger}c_{jm\sigma}
 + \sum\limits_{i,m,n,\sigma} t^{\parallel}_{mn} c_{im\sigma}^{\dagger}c_{in\sigma} \\
 & + \sum\limits_{i,m,\sigma,\sigma^\prime} V_{im} n_{im\sigma}\delta_{\sigma\sigma^\prime} + E_Z~ c_{im\sigma}^{\dagger} \left(\sigma_{x}\right)_{\sigma\sigma^{\prime}} c_{im\sigma^{\prime}} \\
  & + \sum\limits_{i,m,\sigma,\sigma^{\prime}} i\alpha_{R}  \left[c_{i(m+1)\sigma}^{\dagger} \left(\sigma_{y}\right)_{\sigma\sigma^{\prime}} c_{im\sigma^{\prime}}  + h.c. \right] \label{H3D} 
\end{aligned}
\end{equation}
where $c_{im\sigma}^{\dagger}$ creates an electron with spin $\sigma$ localized near the site $i$ of layer $m$,  $n_{im\sigma} = c_{im\sigma}^{\dagger}c_{im\sigma}$ is the number operator, $t^{\bot}_{ij}$ and $t^{\parallel}_{mn}$ are intra- and inter-layer nearest neighbor hopping matrix elements, respectively, $E_Z$ is the (half) Zeeman splitting, and  $\alpha_{R}$ is the Rashba spin-orbit coefficient.  The electrostatic effects are described by the external potential $V_{im}$. The potential matrix elements are  $V_{im} = -e\left<i,m|V\left(\mathbf{r}\right)|i,m\right>$, where $\left|i,m\right>$ is the state centered on  site $i$ of layer $m$, and $V\left(\mathbf{r}\right)$ is the solution of the Laplace equation $\nabla^2 V\left(\mathbf{r}\right) = 0$. Note that the boundary conditions for the Laplace equation are set by the external gates $\left(V_0, V_{1}, V_{2}\dots, V_{1}^\prime, V_{2}^\prime\dots\right)$, as well as the work function difference between the superconductor and nanowire, $V_{SC}$. 

The Hilbert space of Eq. (\ref{H3D}) is quite large, as each layer contains many degrees of freedom (typically of the order $10^3$). However, we identify a layer-dependent low-energy subspace defined by the eigenstates of the auxiliary Hamiltonian 
\begin{equation}
\begin{gathered}
H^{(m)}_{aux} = \sum\limits_{i,j,k,\sigma} \!\!\left[ t^{\bot}_{ij} + \left({\hbar^{2}k^{2} \over 2m^{*}} 
\!+\!V_{i}^{(m)} \right) \delta_{ij} \right] c_{ik\sigma}^{\dagger} c_{jk\sigma} \\
 +\sum\limits_{ik\sigma\sigma^{\prime}} \alpha_{R}k ~c_{ik\sigma}^{\dagger} \left(\sigma_{y}\right)_{\sigma\sigma^{\prime}} c_{ik\sigma^{\prime}}, \label {Haux}
 \end{gathered}
\end{equation} 
where $V_{i}^{(m)} = V_{im}$. The auxiliary model, which describes an infinite wire, is defined on a lattice with a transverse profile that matches the lattice of layer $m$, i.e. the local transverse profile of the original 3D system. Note that, Hamiltonian (\ref{Haux}) represents a specific case of an infinite wire problem corresponding to an external potential  $V_{i}^{(m)} = V_{im}$ and no Zeeman field, i.e. $E_Z=0$. In other words, the auxiliary Hamiltonian $H^{(m)}_{aux}$ describes an infinite system in the presence of a translation-invariant external potential that matches the local external potential of the actual 3D wire on layer $m$.  The low-energy effective 1D Hamiltonian is constructed by projecting the Hamiltonian given by Eq. \ref{H3D} onto the subspace defined by the lowest $n_o$ eigenstates of the auxiliary Hamiltonian,
 \begin{equation}
\begin{gathered}
H_{\rm eff} = \sum_{m,n,\sigma}\sum_{\alpha,\beta} ^\bullet \tilde{t}^{\parallel}_{m\alpha,n\beta} ~ {c}_{m\alpha\sigma}^{\dagger}{c}_{n\beta\sigma} +\sum_{m,\sigma}\sum_{\alpha} ^\bullet \epsilon_{\alpha}^m ~{n}_{m\alpha\sigma} \\
+\sum_{m,\sigma\sigma^\prime}\sum_{\alpha,\beta} ^\bullet  {\Gamma}\left(\sigma_{x}\right)_{\sigma\sigma^{\prime}}\delta_{\alpha\beta}c_{m\alpha\sigma}^{\dagger}  c_{m \beta \sigma^{\prime}} \\
+\sum_{m,n,\sigma\sigma^\prime}\sum_{\alpha,\beta} ^\bullet i \alpha_{\alpha\beta}^{mn} (\sigma_y)_{\sigma\sigma^\prime} ~c_{m\alpha\sigma}^{\dagger}  c_{n \beta \sigma^{\prime}},     \label{Heff}
\end{gathered}
\end{equation}
where $m$ and $n$ label the sites of the (finite) 1D lattice, $\alpha$ and $\beta$ designate the molecular orbitals corresponding to the transverse bands of $H^{(m)}_{aux}$, $\epsilon_{\alpha}^m$ are the eigenvalues of the auxiliary Hamiltonian on layer $m$, and the summations marked by a $\bullet$ symbol are restricted to the lowest energy orbitals, i.e, $\alpha,\beta \leq n_o$.  
The hopping matrix elements $\tilde{t}^{\parallel}_{m\alpha,n\beta}$ can be written in terms of the hopping matrix $[T^\parallel]_{im,jn} = {t}^{\parallel}_{m n}~\delta_{ij}$ between layers $m$ and $n$ as
\begin{equation}
\tilde{t}^{\parallel}_{m\alpha,n\beta} = \langle\varphi_\alpha^m|T^\parallel|\varphi_\beta^n\rangle,
\end{equation}
where $|\varphi_\alpha^m\rangle$ is eigenstate $\alpha$ of Hamiltonian (\ref{Haux}). Note that, in general, the hopping matrix elements are position-dependent. The effective spin-orbit coupling matrix elements $\alpha_{\alpha\beta}^{mn}$ are calculated in a similar manner. Induced superconductivity is added to the effective model through a term
\begin{equation}
\sum_{m} \sum_{\alpha,\beta} ^\bullet \Delta_{\alpha\beta}^m c_{m\alpha\uparrow}^{\dagger}  c_{m \beta \downarrow}^{\dagger} + h.c.,
\end{equation}
where the induced pairing potential is proportional to the effective coupling $\gamma_{\alpha\beta}^{mn}$ between the semiconductor and superconductor. This effective coupling is calculated by
\begin{equation}
\gamma_{\alpha\beta}^{mn} = \langle\varphi_\alpha^m|\widetilde{\gamma}|\varphi_\beta^n\rangle,
\end{equation}
with $\widetilde{\gamma}_{ij}^{mn} = \widetilde{\gamma}_{i}^{m} \delta_{m,n}\delta_{i,j}$, where $\widetilde{\gamma}_{i}^{m}$ is zero everywhere except at the semiconductor-superconductor boundary.

The key observation behind this construction is that the transverse profiles within the finite wire are very similar to those of an infinite system with a potential matching the local potential of the 3D system. Consequently, the low-energy eigenstates defined by the auxiliary problem provide an excellent basis for the finite 3D system, as demonstrated explicitly in Ref.  \onlinecite{Woods2018}.  We note that in Eq. \ref{Heff} $\epsilon_{\alpha}^m$ plays the role of an effective potential that is position and band dependent. Also note that the effective potential $\epsilon_{\alpha}^m$ incorporates all of the information about the ``actual'' electrostatic potential $V_{im}$ and the intra-layer hopping $t_{ij}^{\bot}$. 
In the calculations, we solve Eq. (\ref{Haux}) for each ``slice'' $m$ and construct the effective Hamiltonian $H_{\rm eff}$, which is then solved numerically under the assumption that the charge density in the wire is low. We note that, in general, $H_{\rm eff}$ includes a mean-field contribution due to the Coulomb interaction of the charge inside the wire and has to be solved self-consistently.\cite{Woods2018} While the effects of interactions are not expected to change qualitatively our conclusions, it may be quantitatively significant and should be included in realistic calculations of specific hybrid devices. 

\begin{figure}[t]
\begin{center}
\includegraphics[width=0.45\textwidth]{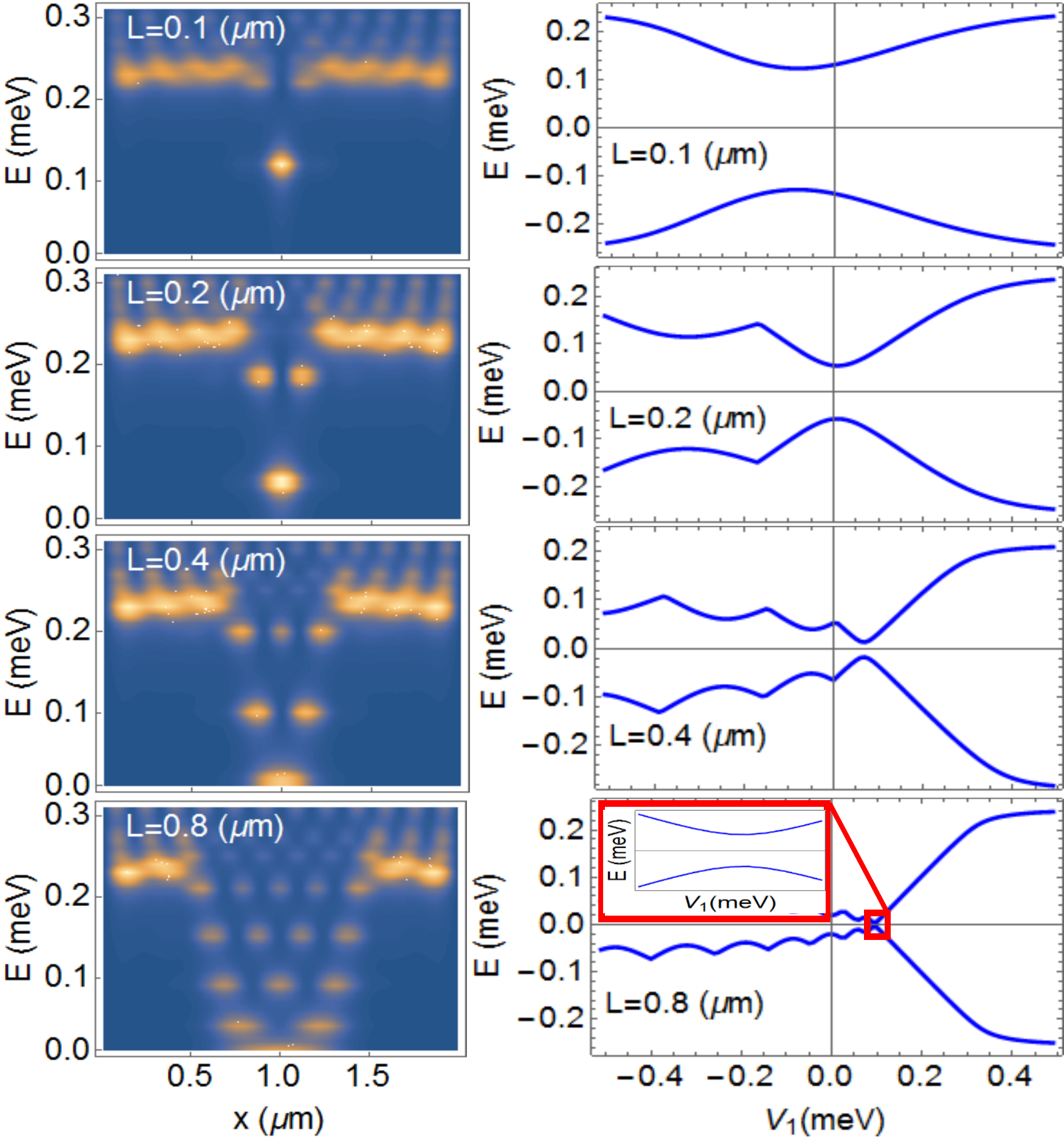}
\vspace{-1.5mm}
\end{center}
\vspace{-3.5mm}

\caption{{\em Left}: Color map of the local density of states at zero magnetic field as a function of energy and position along the wire for a system with $N_{SC}=2$ and different values $L$ of the uncovered (barrier) region length.  The potential barrier height is chosen to minimize the energy of the lowest-energy state, which is localized in the barrier region.   {\em Right}: Energy of the lowest energy states as a function of barrier height for various barrier region lengths.  Notice that the induced gap corresponding to the uncovered (barrier) region decreases strongly with $L$, becoming effectively zero in the limit of long uncovered regions $(L\rightarrow\infty)$.}
\label{Fig_1}
\end{figure}

\section{Results} \label{results} 

Before presenting our main results, it is useful to discuss some generic low-energy features of the heterostructure by focusing on a simple two-island system. Using the simplified model described in Sec. \ref{TbHam}, we determine the low-energy local density of states (LDOS) for different values of the system parameters, including the length of the uncovered region and the value of the gate potential $V_1$ in the uncovered region.
In Fig. \ref{Fig_1} (left panels) we show the LDOS at zero magnetic field as a function of energy and position along the wire for different lengths $L_u$ of the uncovered (barrier) region. The dependence of the corresponding lowest state energy on the applied gate potential $V_1$ is shown in the right panels. Notice the presence of low-energy sub-gap states localized in the uncovered region. The number of these states increases with $L_u$, while the lowest state energy decreases. However, these localized modes are always gapped, since their wave functions are not entirely confined inside the uncovered region, but leak into the nearby proximitized regions. Note that the covered regions have an induced superconducting gap of about $\Delta=0.25~$meV, while the ``effective induced gap'' for the states localized in the uncovered region is significantly lower.  The spectra shown in the right panels  demonstrate that this gap is strongly dependent on the gate potential, $V_1$, and reaches its minimum at a value of the potential  that depends on the length of the uncovered region. We emphasize that the quasi-particle gap is finite regardless of the length of the uncovered region. The minimum of the quasi-particle gap corresponds to a state localized near the center of the uncovered region and having small weight inside the proximitized segments.  Hence, when the uncovered region is long enough (i.e. longer than about $0.4~\mu$m) it can be considered to be non-superconducting for practical purposes.  In the following we will focus on systems with $L_u$ less than $0.4~\mu$m, which means that the system will always have a non-negligible gap.  For all the figures, the nearest neighbor hopping is $t_{SM}=10~meV$ and the spin orbit coupling is $\alpha=2~meV$.

\begin{figure}[t]
\begin{center}
\includegraphics[width=0.45\textwidth]{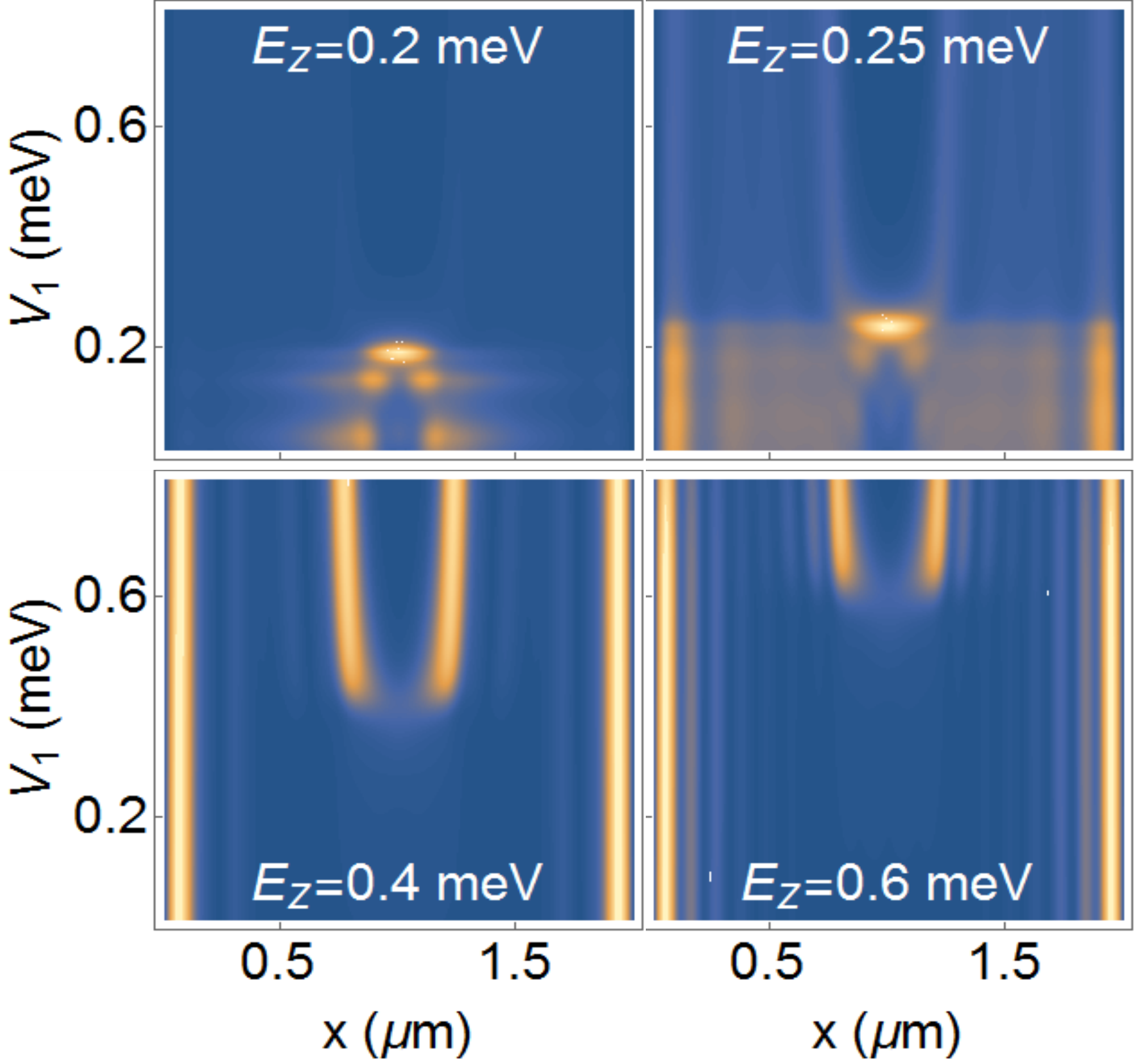}
\vspace{-3.5mm}
\end{center}
\vspace{-3.5mm}
\caption{Local density of states at zero energy as a function of position and potential barrier strength for different magnetic field values.  For $E_Z=0.2~$meV the gap of the covered region has not yet closed (since $\Delta\approx 0.25~$meV), but gapless states emerge in the barrier region for small values of the barrier potential.  The Zeeman field $E_Z=\Delta_c=0.25~$meV corresponds to a topological quantum phase transition. One or two pairs of MBSs emerge at higher field values.}
\label{Fig_2}
\end{figure}
	
The dependence of the zero-energy  LDOS on the potential barrier strength for different different values of the magnetic field is shown in Fig. \ref{Fig_2}.  For the covered regions, the induced gap is $\Delta=0.25~$meV and the chemical potential is tuned to the bottom of the band, so that a topological quantum phase transition is expected at a critical value of the Zeeman field $E_Z^c=0.25~$meV. The topological phase is signaled by the emergence of zero-energy MBSs localized at the ends of the covered regions. Note that in the uncovered region the ``effective induced gap'' has a significantly lower value, generating a finite size precursor of a topological quantum phase transition (TQPT) at Zeeman fields lower than $E_Z^c$. For example, in the upper left panel ($E_Z=0.2~$meV), for $V_1<0.2~$meV, one can clearly distinguish a nearly zero-energy mode having maxima at the ends of the uncovered region. This mode can be viewed as an Andreev bound state (ABS) consisting of two partially overlapping MBSs localized at the ends of the uncovered segment. We note that the mechanism responsible for the emergence of this mode is similar to that acting in Majorana nanowires coupled to quantum dots, which was shown to generate (trivial) ABSs that mimic the behavior of well separated MZMs when detected using local probes.
 Increasing $V_1$ is equivalent to a local increase of the chemical potential, which drives the system into a trivial phase for $V_1>0.2~$meV. The upper right panel corresponds to the critical field $E_Z^c=0.25~$meV. The closing of the bulk gap involves a delocalized mode, which is signaled by the faint LDOS present throughout the system at low values of the gate potential. For $V_1>0.2~$meV the system becomes gapped. We can understand this behavior as a manifestation of the finite size effect. Specifically, large $V_1$ implies disconnected proximitized regions, i.e. shorter ``active'' nanowires. In turn, this pushes the critical field for the topological ``transition'' (which, strictly speaking, is a finite size crossover) to higher values,  $E_Z^* > E_Z^c$. Consequently, the system with   $E_Z=0.25~$meV and $V_1>0.2~$meV is still in the trivial regime. The lower panels in Fig. \ref{Fig_2} correspond to the topologically nontrivial regime characterized by the emergence of zero-energy MBSs. The important feature is the distinction between a low-barrier (strongly coupled) regime and high-barrier (effectively decoupled) regime. The two regimes are characterized by the presence a single pair of MZMs localized at the ends of the system and two pairs of MBSs localized at the ends of the proximitized segments, respectively. The value of the crossover potential separating these regimes depends on the Zeeman field. 



\begin{figure}[t]
\begin{center}
\includegraphics[width=0.45\textwidth]{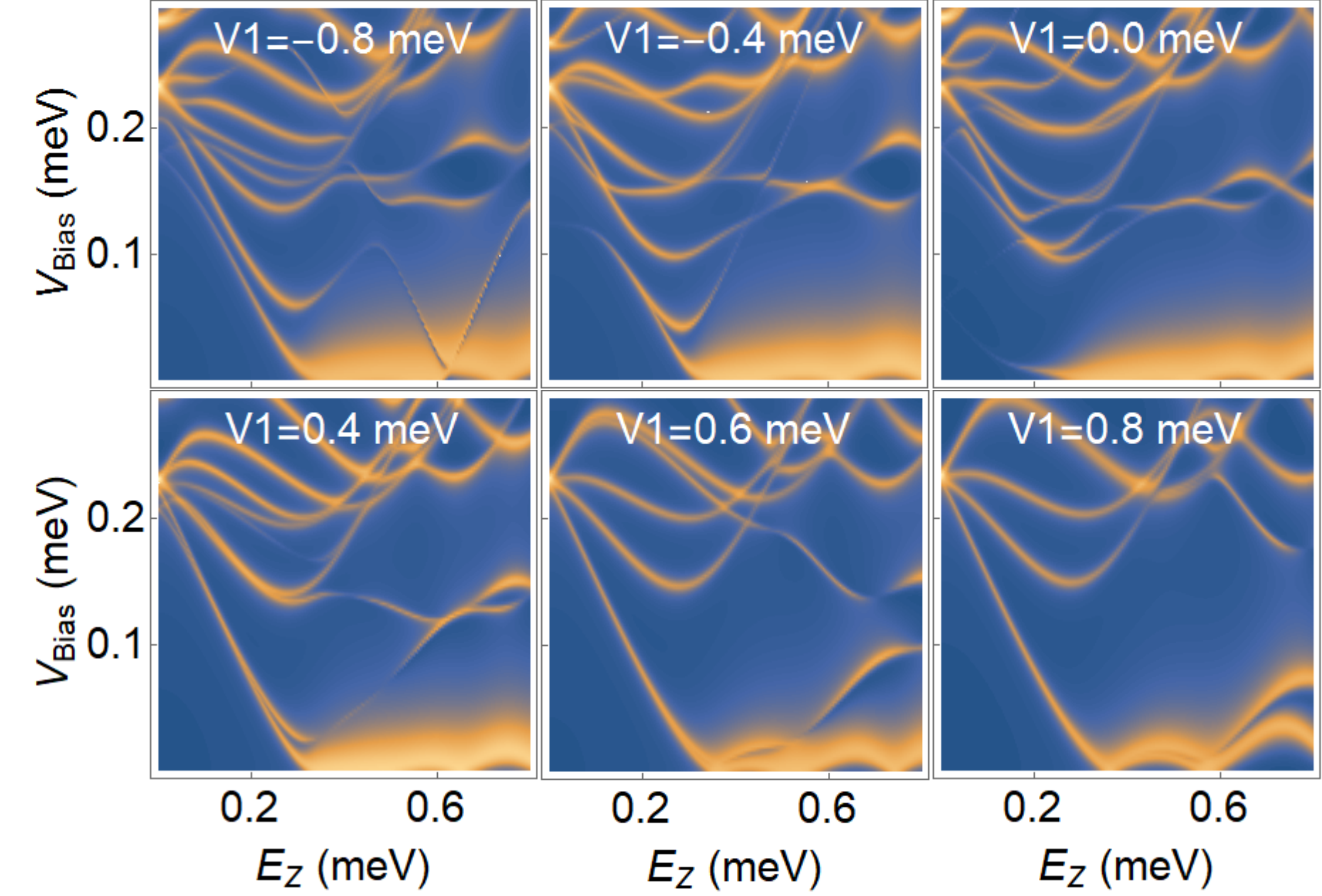}
\vspace{-0.5mm}
\end{center}
\vspace{-3.5mm}
\caption{Differential conductance as a function of magnetic field and bias potential for various back gate potential values.  Above $V_1=0.8~$meV the two covered regions are completely separated and the pairs of MBSs localized at the ends of each region overlap strongly.  The energy splitting oscillations decrease as the potential decreases to zero and do not increase dramatically as the potential becomes negative.  Notice an additional ABS crossing in the $V_1=-0.8~$meV panel (at $E_Z\approx0.6~$meV).}
\label{Fig_4}
\end{figure}

\subsection{Differential conductance calculations} \label{barrierPR}

The key question that we want to address next is how would a charge tunneling measurement reveal the basic low-energy physics discussed above. The relevant setup is shown schematically in Fig. \ref{Fig_01}, with $V_0$ acting as a tunneling barrier (see also Sec. \ref{TbHam} and Sec. \ref{Conductance}). We start with a two-island device, similar to that discussed above. 
Fig. \ref{Fig_4} shows the differential conductance as a function of magnetic field and bias potential  for six different values of the barrier potential $V_1$.  For $V_1=0$ (which corresponds to a uniform effective potential throughout the structure), the system undergoes a TQPT at $E_Z\approx 0.25~$meV, as revealed by the emergence of a robust zero-bias peak (ZBP) in the tunneling conductance (see Fig. \ref{Fig_4}). In general, for small values of $V_1$ the system behaves like a (nearly) homogeneous wire consisting of two strongly coupled segments. We will dub it as the {\em strongly coupled} regime.
Increasing $V_1$ results in a partial decoupling of the two proximitized regions and the emergence of a low-energy state in the uncovered region. This state overlaps with the MBSs localized at the ends of the system, which acquire an energy gap that oscillates with the Zeeman field, as revealed by the oscillations of the ZBP. The amplitude of the oscillations increases with $V_1$,  until the two covered regions become effectively decoupled  (for gate potentials larger than $V_1\approx 0.8~$meV).  We will refer to this high potential regime as the  {\em decoupled} (or {\em uncoupled}) regime.  For all practical purposes, a hybrid system in the uncoupled regime behaves as two separate short wires. Note that the ``critical'' Zeeman field associated with the emergence of the ZBP increases with $V_1$, which is a finite size effect already meanioned in the previous section. 
For values of $V_1$  of the order of the induced gap, i.e. above the strongly coupled regime, but below the decoupled regime, there is an {\em intermediate} regime characterized by a partial protection of the  ZBP.  Finally, for negative values of  $V_1$, i.e. when $V_1(i)$ corresponds to a potential well,  the ``critical'' Zeeman energy starts to increase again, signaling a reduction of the coupling between the proximitized segments. However,  in this {\em potential well} regime, the oscillations of the ZBP do not increase, suggesting that there is no additional low-energy state that could hybridize with the MZMs localized at the ends of the chain. Nonetheless, for particular values of the gate potential and Zeeman field, Andreev bound states (ABSs) trapped  in the uncovered region can cross zero energy, as shown  in the top right panel of Fig. \ref{Fig_4}.  These crossing can destroy the topological protection of the Majorana mode, but they can be easily avoided by tuning $V_1$.  Below, we will discuss in some detail  the challenges associated with effectively turning off the potential barrier (i.e. realizing the strongly coupled regime), as well as the low-energy physics in the potential well regime;  for now, let us focus on the intermediate and uncoupled regimes.

\begin{figure}[t]
\begin{center}
\includegraphics[width=0.45\textwidth]{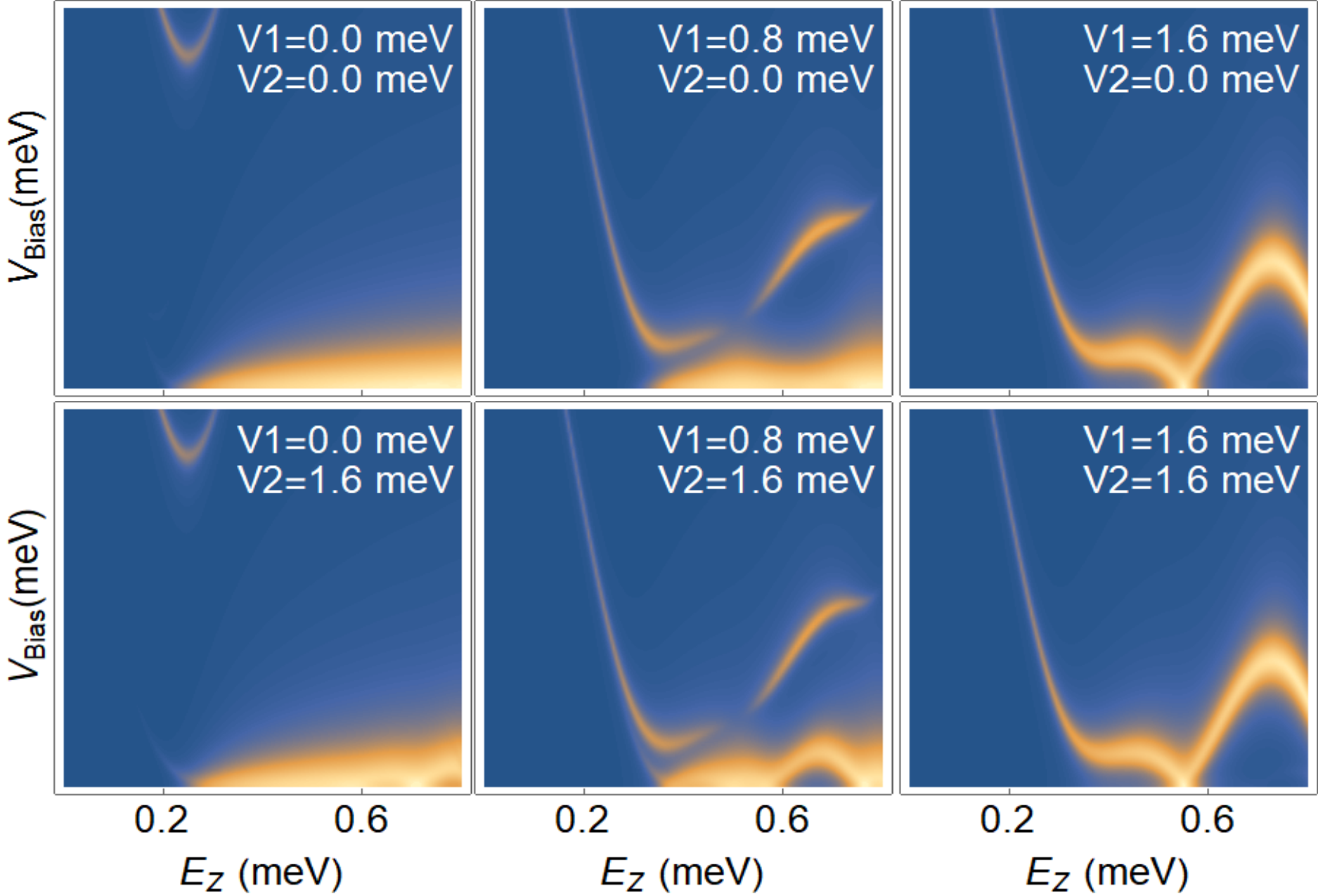}
\vspace{-0.5mm}
\end{center}
\vspace{-3.5mm}
\caption{Differential conductance as a function of magnetic field and bias potential for a three-island device. The panels correspond to different values of the back gate potentials in the uncovered regions, $V_1$ (which is closer to the tunneling barrier) and $V_2$ (the ``far'' uncovered region).  When $V_1$ is turned off,  $V_2$ does not have a significant effect on the ZBP profile.  Interestingly, the effect of $V_2$ on the ZBP increases at intermediate values of $V_1$.  As expected, large values of $V_1$ result in effectively disconnecting the first covered region (hence, $V_2$ once again has a negligible effect on the tunneling conductance).}
\label{Fig_5}
\end{figure}

To better understand the intermediate regime, let us explore the behavior of the tunnel conductance for a system with two barriers (i.e. a three-island chain). Fig. \ref{Fig_5} shows the differential conductance for a wire with two barrier regions as a function of magnetic field and bias potential for several values of potential barrier strengths.  When both barriers are turned off (i.e., when both are in the strongly coupled regime), the ZBP is extremely robust as there is a single pair of MZMs separated by the entire length of the wire. In this case, the phase transition occurs at $E_Z^c=\Delta=0.25~$meV and there are no noticeable splitting osculations. 
Increasing $V_2$ (i.e. the ``far gate'' potential) has little effect on the ZBP, as the first two covered segments are long enough to ensure its protection. 
On the other hand, when the first barrier is in the decoupled regime ($V_1=1.6~$meV), there are two low-energy fermionic states, although only one is visible in the differential conductance.  Again, the visible peak is unaffected by $V_2$, as segments two and three are effectively decoupled from segment one, which is the only one coupled to the normal lead.    By contrast, an interesting behavior can be observed in the intermediate regime, $V_1=0.8~$meV. In this case $V_2$ has a sizable effect on the ZBP suggesting the presence of  a low-energy state that couples to the end-of-wire MBSs.  

\begin{figure}[t]
\begin{center}
\includegraphics[width=0.5\textwidth]{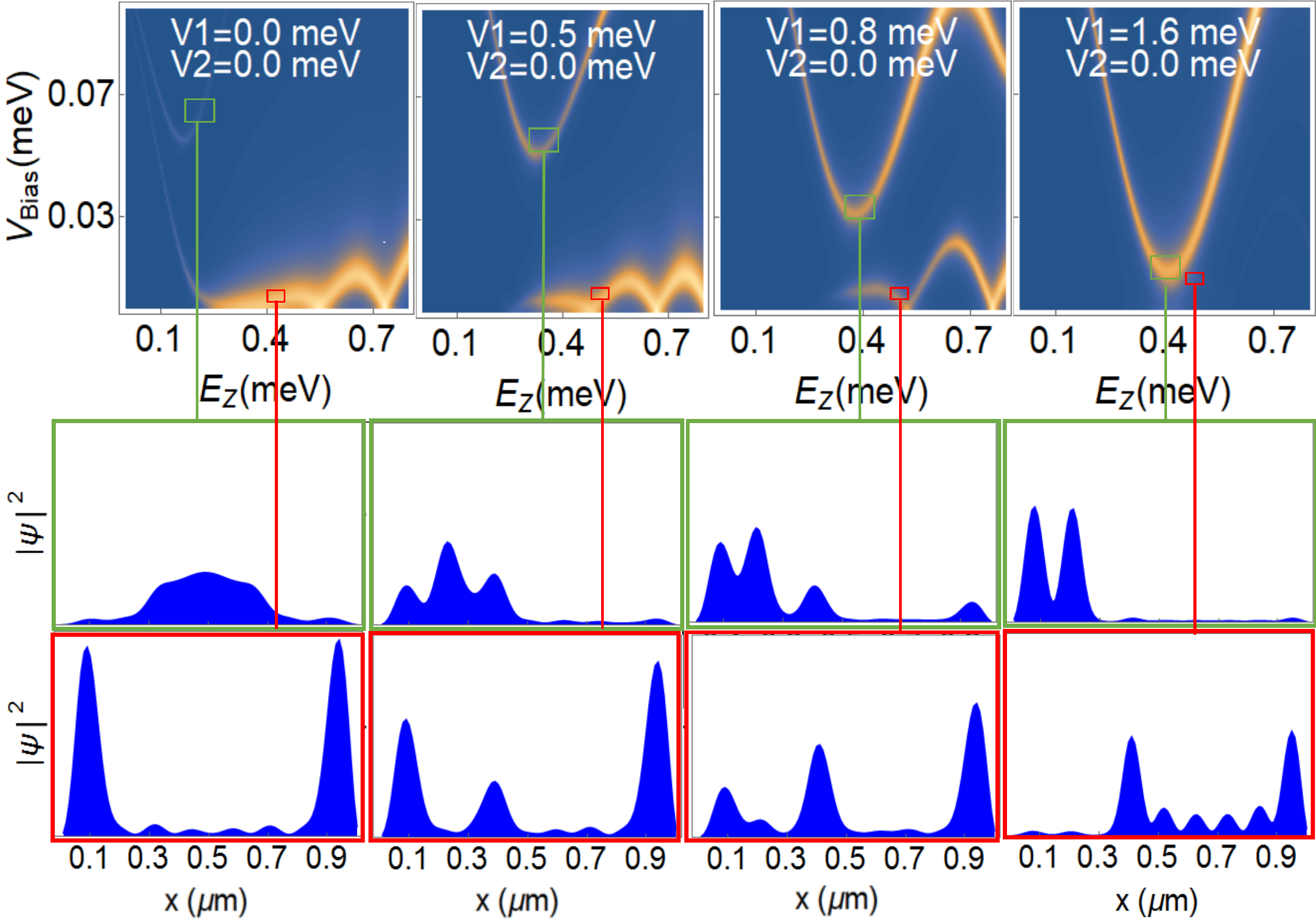}
\vspace{-0.5mm}
\end{center}
\vspace{-1.5mm}
\caption{{\em Top}: Differential conductance as a function of magnetic field and bias potential for various strengths of the first barrier potential in a three-island device.  {\em Bottom}: The lowest-energy visible state (red border/lowest panels) and the second-lowest visible state (green border/middle panels).  When $V_1$ is turned off the lowest state is composed of two MBSs localized at either end of the wire, while the second lowest state is a bulk state that is barely visible.  As $V_1$ increases, the first state shifts its weight from the ends of the wire to the edges of the combined second and third covered regions (eventually becoming invisible), while the second state lowers its minimum energy and shifts its weight to the edges of the first covered region (becoming more visible).}
\label{Fig_6}
\end{figure}

To support the picture described above, it is instructive to calculte the probability distribution of the low energy states and track its change as the first gate potential  increases from $V_1=0$ (strongly coupled regime) to the intermediate regime and then into the uncoupled regime. The second barrier is turned off, $V_2=0$. The results are shown in Fig. \ref{Fig_6}.   The top panels represent the differential conductance as a function of magnetic field and bias voltage.  The next row (green border) depicts the probability distribution along the wire for the second lowest energy state, while the third row (red border) shows the probability distribution for the lowest state.  
When $V_1$ is off, the lowest energy state is composed of two well separated Majorana  modes that live at either end of the wire.  Since the wire is fairly short,  there are some visible oscillations in the ZBP even in this regime.  The second state is a bulk state having most of its weight near the middle of the wire, with very little weight at the ends, which makes it nearly invisible in the tunneling conductance.  
Going into the intermediate regime couples the lowest energy states,  both having finite weight at all four edges of the active regions, i.e. region one and the strongly coupled segments two and three. Note that even though the lowest energy mode starts to lose protection, it still generates a visible ZBP (although having significant splitting oscillations).  Furthermore, both states are clearly visible in the differential conductance because both have weight at the end of the wire.  Finally, in the decoupled regime, the lowest energy state is practically contained inside regions two and three and has negligible weight at the left end of the wire, therefore losing most of its visibility in the conductance.  On the other hand, the second state, which is clearly visible, has shifted all its weight to the edges of the first covered region.  As the first region is short, the Majorana modes that it hosts are highly overlapping, which results in a (finite energy) ABS that does not generate a ZBP, but disperses with the Zeeman field.  

\begin{figure}[t]
\begin{center}
\includegraphics[width=0.4\textwidth]{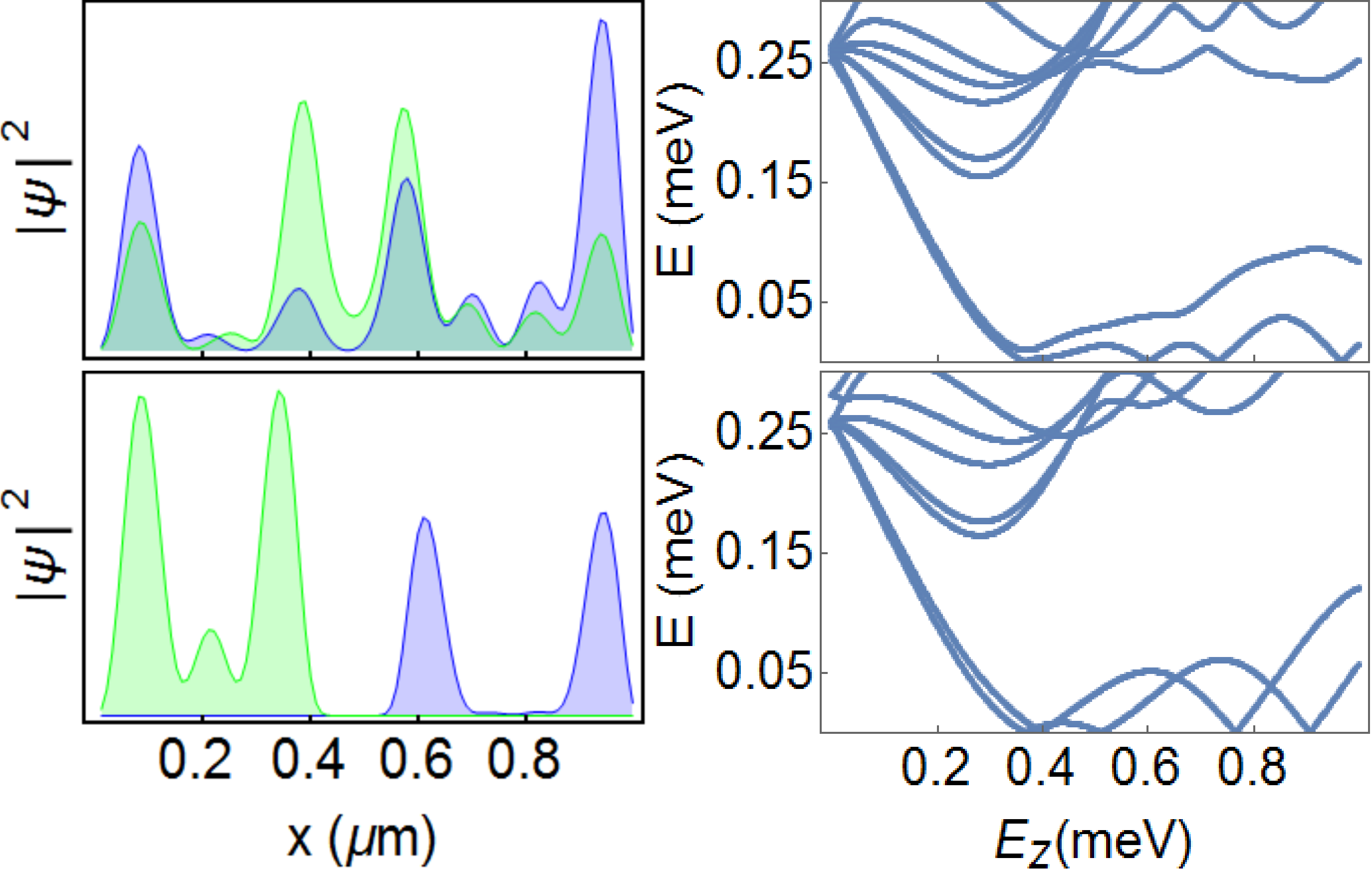}
\vspace{-1.5mm}
\end{center}
\vspace{-1.5mm}
\caption{{\em Right}: Probability distribution as a function of position for the lowest two states of a two-island system at intermediate barrier strength (upper panel) and high barrier strength (lower panel).  The Zeeman field is $E_Z=0.5~$meV). {\em  Left}: Energy as a function of magnetic field for the same barrier strengths. When the potential barrier in the uncovered region is strong (bottom), the proximitized regions are effectively disconnected, each supporting a pair of (partially overlapping) MBSs. At intermediate values of the barrier potential (top), the two covered regions are strongly coupled. The four coupled MBSs generate a low-energy fermionic mode that tends to stick at zero energy and a mode that acquires a finite gap.}
\label{Fig_7}
\end{figure}

The results shown in Fig. \ref{Fig_6} suggest that the pinning of the ZBP increases as the lowest two energy states become more coupled. To better understand this behavior, we consider once again a two-island system having a single barrier region that cuts the wire in half and we calculate the spectrum as a function of the Zeeman field, as well as the probability distribution of the lowest lying states. The results are shown in Fig. \ref{Fig_7}.  The left panels show the probability distribution for the lowest and second lowest energy modes, while the right panels show the dependence of the energy levels on the magnetic field.  When the barrier is in the uncoupled regime (bottom) the energy levels are uncorrelated.  The lowest state consists of highly overlapping pairs of Majorana modes that live at the edges of the two covered regions.  On the other hand, when the barrier is in the intermediate regime the two lowest energy levels anti-cross and split.  The splitting of these levels pushes one of them down toward zero energy, while the other acquires a finite gap.  It is this ``splitting'' effect that is responsible for the apparent increase in protection of the ZBP in the conductance calculations.  
However, examining the probability distribution reveals that the lowest energy mode, although relatively well pinned to zero energy, is an overlap of four MBSs localized at the ends of the covered regions. The major practical problem is that, in certain parameter regimes,  the presence of the MBSs localized near the middle of the wire cannot be infered based on charge tunneling measurements at one end of the system. In other words, in a chain of coupled superconducting islands one could easily obtain an apparent exponential protection of the ZBP by turning off the potential barriers, but the observation of a robust ZBP  does not guarantee the presence of a single pair of MZMs localized at the ends of the chain. 

\begin{figure}[t]
\begin{center}
\includegraphics[width=0.5\textwidth]{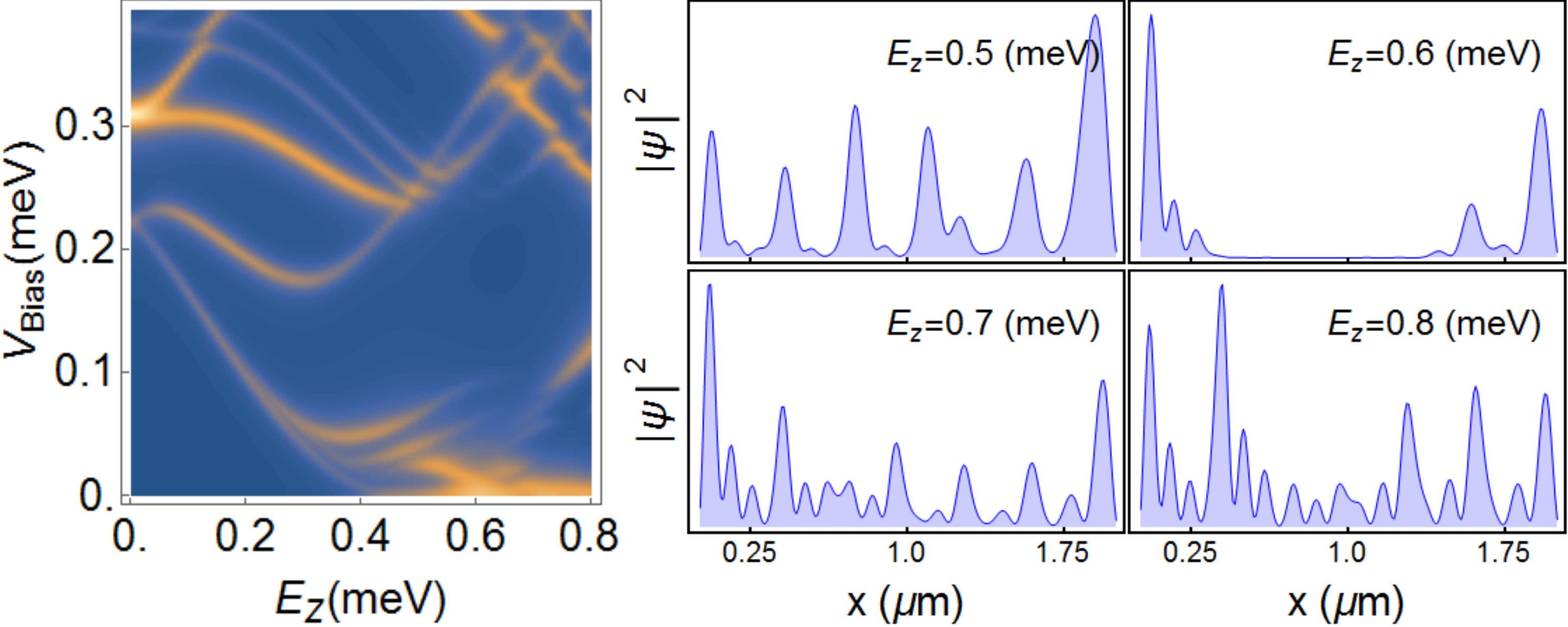}
\vspace{-1.5mm}
\end{center}
\vspace{-1.5mm}
\caption{Low-energy states in a six-island device. {\em  Left}: Differential conductance as a function of magnetic field and bias potential.  {\em Right}: Probability distribution  as a function of position for the lowest energy (fermionic) mode.  All five barriers are tuned to the intermediate regime. Notice that the extended ZBP is not associated with the presence of well-separated MZMs.  Instead,  the wave function of the lowest-energy mode has significant weight at the edges of each of the six covered regions.}
\label{Fig_9}
\end{figure}

The fact that a well pinned ZBP does not guarantee well separated edge modes is dramatically demonstrated in wires with many uncovered regions. In Fig. \ref{Fig_9} we consider six covered segments separated by five barrier regions.  In the left panel we show the tunnel conductance and in the right panels we plot the spatial profile of the lowest energy mode (i.e. the mode responsible for the ZBP visible in the left panel) at four different magnetic fields.  Notice that the conductance shows a very well pinned ZBP even though the wave function does not represent two Majorana modes separated by the length of the wire. It is important to note that the lowest energy mode has the strongest conductance response of the lowest energy states, hence it would be  easy to mistake it for well separated end modes.  Quantitatively, we note  that the lowest state pins to zero around $0.4~$meV, which is a Zeeman field larger than the predicted critical field for the topological phase transition.  However, the actual $g$ factor of the wire  is unknown and it is extracted from the slope of the lowest energy mode as a function of the external field.  In other words, the Zeeman energy is related to the external magnetic field by {\em assuming} that the ZBP emerges at a Zeeman splitting equal to the induced gap.  In addition, the chemical potential could be shifted away from the bottom of the band. This makes it extremely difficult to conclude, based on experimental data,  that the ZBP does not emerge at the predicted critical field.

It is clear that a well pinned ZBP does not guarantee well separated MZMs.  In order for the hybrid structure to be useful as a  platform for topological quantum computation one has to be sure that the lowest energy fermionic mode consists of a single pair of MZMs localized at the ends of the system.  This could be realized in the strongly coupled regime, when the effective potential is (approximately) uniform throughout the wire. However, practically turning off the potential barriers can be challenging, as we show explicitly in the next section. The previous discussion, based on a simplified tight-binding model that {\em assumes} certain profiles for the gate potentials, raises two important questions: i) How can one discriminate between a low-energy mode that is robustly pinned to zero, but consists of several overlapping MBSs, and a pair of well separated MZMs? ii) How should one tune the gate potentials in order to eliminate the unwanted additional low-energy states that can hybridize with the MZM pair? The first question has been addressed in several previous works.\cite{Moore2018,Hansen2018,Liu2017a,Ptok2017,Rosdah2018} Here, we focus on the second question. To properly answer it, we use a more detail model of the hybrid device, capable of capturing the position dependence of the effective potential both along the wire, as well as across its transverse profile. 

\subsection{Effective potential calculations}\label{effectivePMR}

To gain a better understanding of how much control over the effective potential one can actually have in an experimental setting, we investigate the chain of coupled superconducting islands using the effective model described in Sec. \ref{EffModel}.  A schematic representation of the cross section of the device is shown in Fig. \ref{Fig_02} and we focus on a two-island system (see Fig. \ref{Fig_01}).  The system parameters are chosen to be $R = 50$ nm, $d = 30$ nm, $\epsilon_{SM} = 17.7$, $\epsilon_{diel} = 8.0$, $m^* = 0.014$, $\alpha = 250$ meV$\cdot$\AA, $a_\parallel = 10$ nm, and $a_\bot = 5$ nm, where $a_\parallel$ and $a_\bot$ are the lattice constants parallel and perpendicular to the length of the wire. Note that the potential on the two gates beneath the covered regions are always set to the same value. Laplace's equation was solved using the Fenics software package.\cite{Alnaes2015}

\begin{figure}[t]
\begin{center}
\vspace{-5mm}
\includegraphics[width=0.5\textwidth]{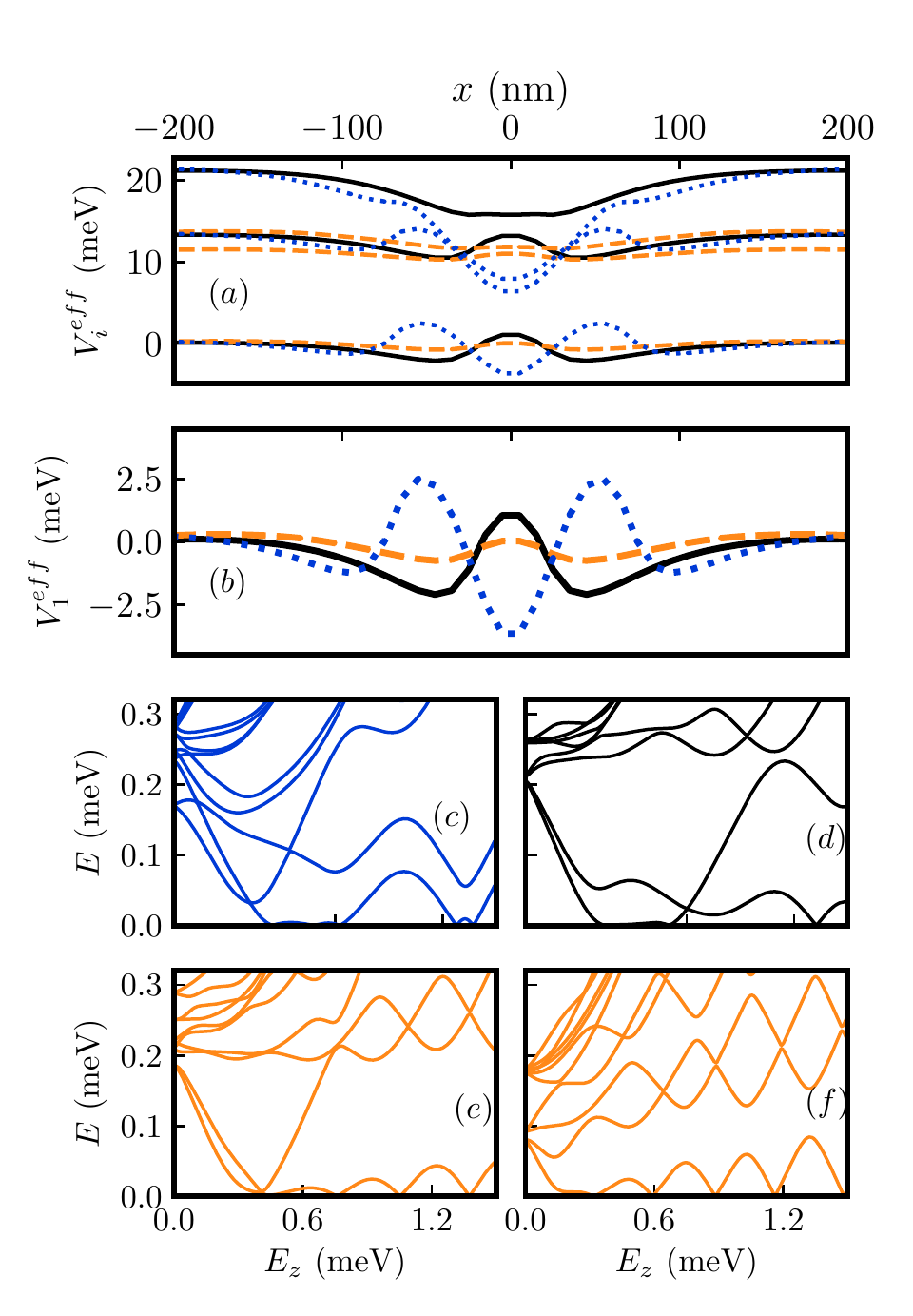}
\vspace{-4.5mm}
\end{center}
\vspace{-3.5mm}
\caption{(a) The effective potential profiles of the first three bands  for a two-island chain with different sets of system parameters. The length of the wire is $L = 1.2~\mu$m and $V_{SC} = 150$ mV, while the other parameters are are set to:  (blue dotted lines) $V_{1}^\prime=V_2^\prime = 0$, $V_{1} = 255$ mV, $L_{u} = 0.15$ $\mu$m, (black solid lines) $V_{1}^\prime=V_2^\prime = 0$, $V_{1} = 450$ mV, $L_{u} = 0.05$ $\mu$m, (orange dashed lines) $V_{1}^\prime=V_2^\prime = 100$ mV, $V_{1} = 240$ mV, $L_{u} = 0.05$ $\mu$m.  
    (b) A zoom in on the lowest energy band in (a).  (c-f) Energy spectrum as a function of Zeeman energy for a system with the chemical potential tuned to the effective potential of first band in the bulk of the covered region. Colors of lines correspond to those in (a) and (b). The (effective) induced gap is assumed to be $0.2$ meV \textit{after} gating in panels (c-e), whereas in (f) the induced gap is assumed to be $0.2$ meV \textit{before} gating.}  
\label{Fig_B1}
\end{figure}

In Fig. \ref{Fig_B1}a we show the spatial profile near the uncovered (barrier) region of the effective potential of the three lowest energy bands  for various device parameters. The effective potential of the  lowest band is set to zero in the bulk of the covered region. Note that the effective potential deep inside the covered region is assumed to be constant and equal to the potential calculated using an infinite wire. One can easily see that the effective potential is band dependent. Fig. \ref{Fig_B1}b provides a zoom-in view of the lowest energy band. The barrier gate voltage $V_{1}$ was chosen such that the effective potential is as flat as possible for the given values of the work function difference $V_{SC}$ and gate potentials $V_{j}^\prime$. Note that a positive gate potential is attractive. Also note that the flatness of the effective potential depends strongly on the specific device geometry and on the gate potential/work function values. This is reflected in the energy spectrum as a function of Zeeman energy show in Figs. \ref{Fig_B1}(c-f). For these spectra, the chemical potential is assumed to be equal to the effective potential of the lowest-energy band in the bulk of the covered region and only the first band is incorporated into the low-energy effective Hamiltonian (\ref{Heff}) since it is well separated ($\approx 10$ meV) from the other bands.  In panel (c) (i.e. for a system with a $150~$nm uncovered region),  one can see that there are two low-energy states throughout the relevant magnetic field range. This is due to the large potential inhomogeneity of the barrier region [dotted blue line in Fig. \ref{Fig_B1}(b)], which leads to the emergence of an ``unwanted' ' low-energy states. The situation is slightly improved in panel (d), where the barrier length is reduced to $0.05~\mu$m. However, even in this case a second low-energy state is present for $E_z<0.7~$meV. The effective potential can be made sufficiently flat to completely remove the second low-energy state by applying an attractive voltage on the gates below the covered regions of the wire, as shown in Fig. \ref{Fig_B1}(e). However, there are a few undesirable consequences of increasing the bottom gate voltage. First of all, it reduces the proximity induced gap due to the reduced weight of the states near the SM-SC interface, as shown in panels (d) and (e). In Fig. \ref{Fig_B1}(e) we have increased the SM-SC coupling to maintain a value of the induced gap similar to that in panels (c) and (d). On the other hand, if the coupling is maintained constant, we obtain the spectrum shown in panel (f) when the bottom gate voltage is applied. Here the induced gap has been reduced to about half of its initial value, resulting in less protection for the Majorana mode. A second problem arising from the application of the bottom gate potential is a reduction in the energy separation between certain bands [see orange dashed lines in panel (a)]. These nearly degenerate bands are due to the six-fold symmetry of the hexagon nanowires, which is lifted in the presence of a large work function difference, $V_{SC}$, and no counterbalancing gate potential. The symmetry is partially restored when the applied gate potential $V_{j}^\prime$ approaches $V_{SC}$. If the chemical potential happens to be close to these nearly degenerate bands it may be more difficult to obtain an odd number of occupied bands, i.e. to stabilize the topological phase.   

\begin{figure}[t]
\begin{center}
\includegraphics[width=0.4\textwidth]{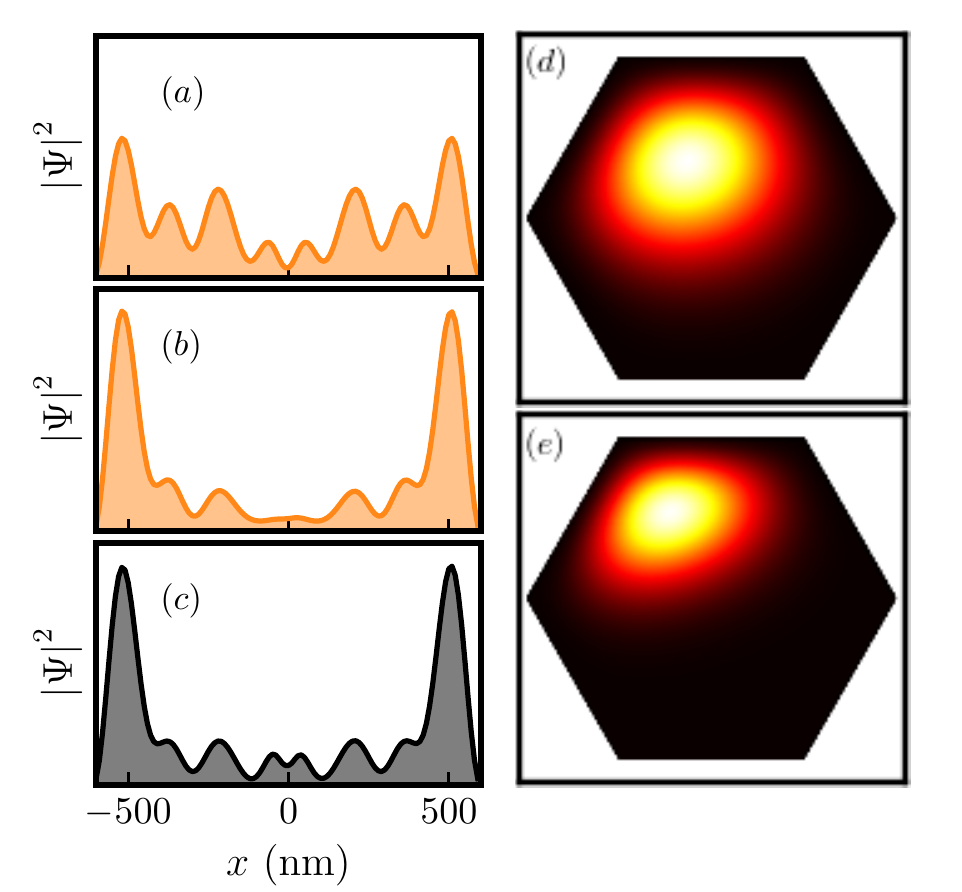}
\vspace{-1.5mm}
\end{center}
\vspace{-3.5mm}
\caption{(a-c) Probability distributions along the wire for the lowest-energy mode. The states correspond to the parameters in Fig. \ref{Fig_B1} with $E_z = 0.6$ meV and panels (a), (b), and (c) matching (f), (e), and (d), respectively. Note that in (a) the reduced gap in Fig. \ref{Fig_B1}(f) results in a more extended Majorana mode,  as compared to (b). (d,e) Transverse probability distributions of the lowest energy states. The profile in panel (d) corresponds to the modes shown in (a) and (b), while the profile in (e) corresponds to panel (c).}
\label{Fig_B2}
\end{figure}

The probability distributions for the lowest energy mode in the presence gating, as well as for $V_j^\prime=0$, are shown in 
Fig. \ref{Fig_B2}, panels (a-c). Note that these profiles correspond to the  spectra in Fig. \ref{Fig_B1}. One can clearly see that the reduced induced gap in Fig. \ref{Fig_B1}(f) results in a highly overlapping pair of Majorana modes [\ref{Fig_B2}(a)]. Also note that the MBSs in Fig. \ref{Fig_B2}(c) are slightly less well separated than in those in \ref{Fig_B2}(b) as a result of the  effective potential being less flat.  However, the effect is quite small at the chosen value of the magnetic field ($E_z=0.6~$meV). Nonetheless, as shown in Fig. \ref{Fig_B1}(d), the potential inhomogeneity in the uncovered region induces  a second low-energy state that hybridizes with the MBS pair localized at the ends of the system destroying its protection. 

The key message here is that one must be extremely careful when dealing with the electrostatic potentials created in the uncovered regions of a semiconductor-superconductor structure and, more generally, when characterizing and engineering regions with significant variations of the system parameters. Typically, such regions support (trivial) low-energy states that can compromise the topological protection of the Majorana modes, as exemplified by the calculations discussed above. If one is interested in creating a flat effective potential in the barrier region (to prevent the emergence of such unwanted low-energy states), the uncovered segment should be made as short as possible. In addition, one may have to apply gate potentials in the proximitized regions, which could generate secondary adverse effects. Hence, the natural question: is it really necessary to have a flat effective potential (i.e. to be in the strongly coupled regime) in order to realize well-protected MZMs in a quantum dot array? 

\begin{figure}[t]
\begin{center}
\vspace{-7mm}
\includegraphics[width=0.48\textwidth]{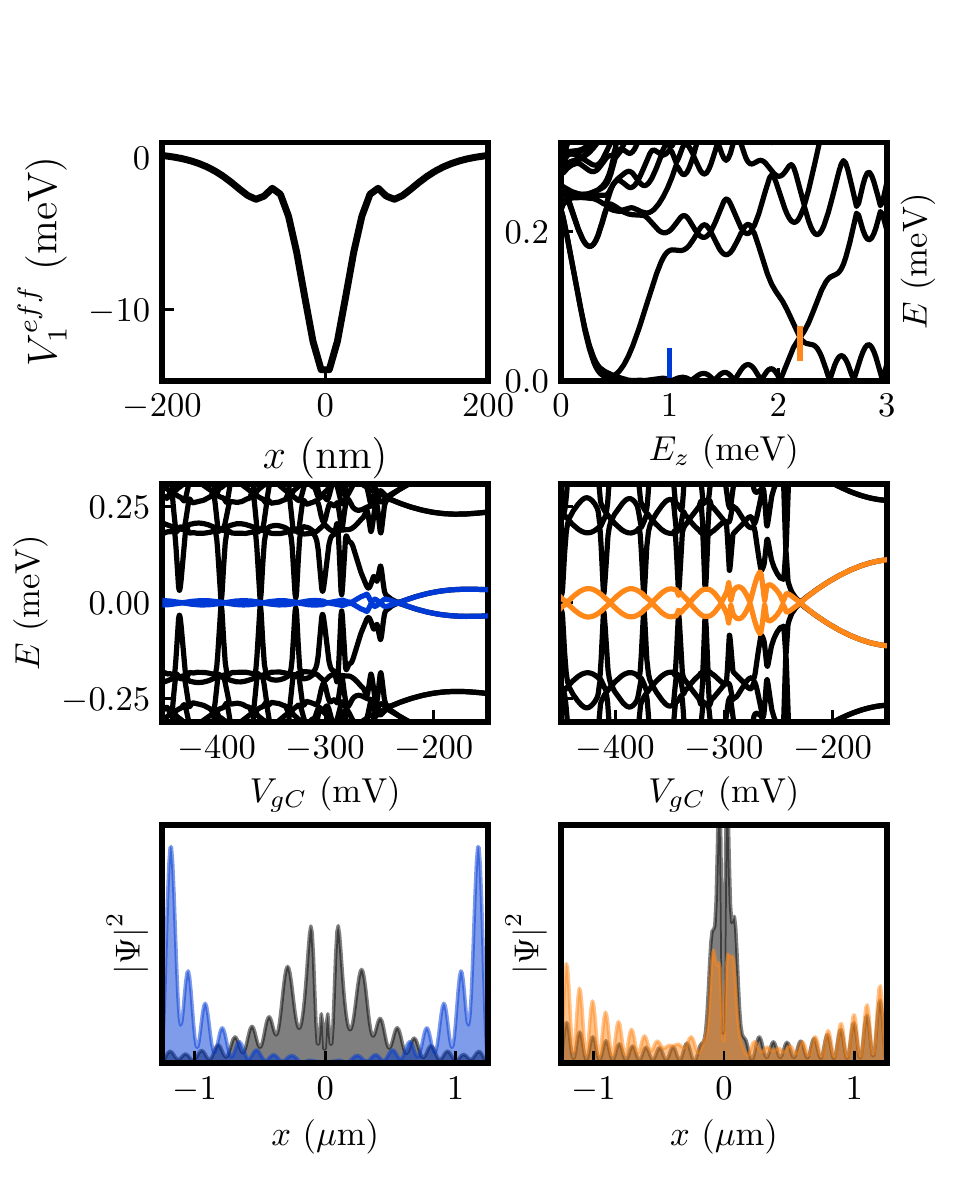}
\vspace{-3.5mm}
\end{center}
\vspace{-3.5mm}
\caption{{\em Top left}: Effective potential profile near the uncovered region for a two-island system with $V_{SC} =150$ mV, $V_{j}^\prime = 0$, $V_{1} = 295$ mV, $L_u = 0.15$ $\mu$m, and $L=2.5$ $\mu$m.  {\em Top right}: Energy spectrum as a function of Zeeman energy with the chemical potential tuned to the effective potential of first band in the bulk of the covered region.  The second lowest energy level joins the continuum except at ABS crossings.  {\em Middle left}: Energy spectrum as a function of barrier gate voltage at $E_z= 1.0$ meV.  {\em Middle right}: Energy spectrum  as a function of gate voltage at $E_z=2.2 meV$. {\em Bottom left}: Probability distributions for the two lowest energy states in the top right panel at $E_z=1.0$ meV.  {\em Bottom right}: Probability distributions for the two lowest energy states at $E_z=2.2$ meV, where there is an ABS crossing.}
\label{Fig_B3}
\end{figure}

The answer to this question is provided by the analysis of the potential well regime. Indeed,  although the potential profile cannot always be flattened so that there is a single pair of well separated MZMs, it turns out that decreasing the potential into the well regime can produce the desired outcome within a significant range of control parameters.
Fig. \ref{Fig_B3} captures the basic features of the potential well regime. The top left panel shows the profile of the effective potential for a well depth of about $10~$meV, which is rather large compared to the induced gap. Although this leads to the formation of a sub-gap ABS, the component MBSs  are trapped in the potential well, which results in a strong overlap and in the ABS acquiring a finite energy gap. Of course, these trivial ABS modes are characterized by discrete zero-energy crossings. However,  when the uncovered region is short and the potential  well is deep, the separation (in energy) of these ABS crossings is large enough so that the gate potential can be tuned to values for which there is a significant magnetic field range characterized by the presence of well separated MZMs localized at the ends of the system and separated from all other states by finite energy gap. 
Note that the characteristic energy scale of the bound states hosted by the uncovered segment can be estimated by assuming a harmonic potential and vanishing induced pairing. For the parameters used in the calculation we have $\hbar \omega = \sqrt{ {2 \hbar^2 V\left(x=0\right)/ m L_u^2}} \approx 0.6$ meV. We conclude that the bound states will move away from zero energy, above the induced gap, for a significant range of Zeeman fields. 
An example is provided in the top right panel of Fig. \ref{Fig_B3}, which covers a large Zeeman energy range. Note that for $E_Z<2.2~$meV the spectrum remains gapped above a well-protected Majorana mode. Moreover, this behavior  does not rely on fine tuning of the gate potential, as one can see in the middle left  panel, which shows the dependence of the low-energy spectrum on the applied gate voltage. There are regions of $V_{1}$ of width of the order $40~$mV for which the spectrum remains gapped separated by narrower regions where zero-energy crossings occur. 
We note that the depth of the well should not exceed certain values, as other bands may become close to zero energy and lead to the formation of additional bound states that could  ruin the Majorana protection. In the calculations shown in Fig. \ref{Fig_B3} we include only a single band to illustrate the oscillatory behavior of the low-energy states. The inclusion of multiple  bands results in the observation of only a few oscillations with respect to $V_{1}$, while above a critical value of $V_{1}$ the system has gapped low-energy bound states for parameters that are not fine tuned. Nonetheless, in the weak potential well regime the low-energy physics is well described by the single-band approximation used in the calculation. 
Finally, to clearly demonstrate the nature of the low-energy states, we calculate their spatial profile for representative values of the Zeeman field. 
The bottom left panel of Fig. \ref{Fig_B3} shows a typical probability distribution for the two lowest energy states corresponding to $E_z=1.0~$meV, i.e.when the second-lowest state is strongly gapped and the Majorana mode is well protected.  Indeed, the lowest energy fermionic state consists of a pair of well separated MZMs localized at the ends of the system, while the second-lowest mode is a bulk state. By contrast, for $E_z=2.2~$meV (which corresponds to an ABS crossing) we notice that the ABS localized in the uncovered region hybridizes strongly with the MBSs localized at the ends of the system, destroying their protection.  

\begin{figure}[t]
\begin{center}
\includegraphics[width=0.48\textwidth]{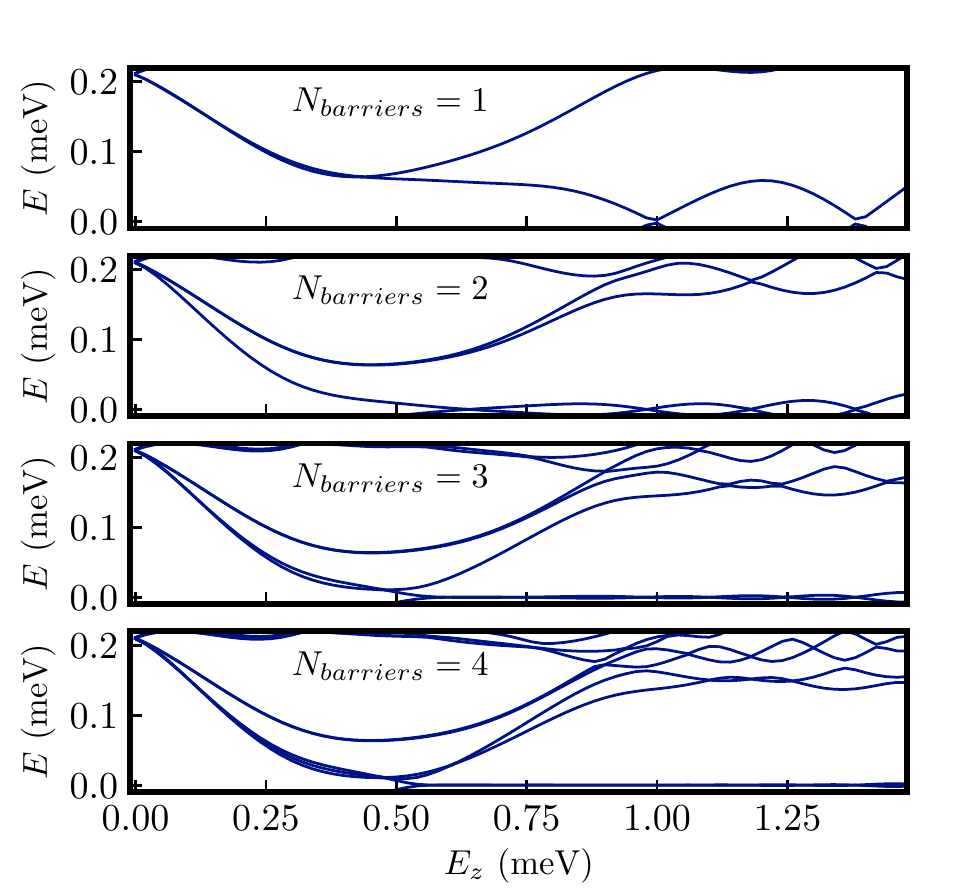}
\vspace{-0.5mm}
\end{center}
\vspace{-3.5mm}
\caption{Energy spectrum as a function of Zeeman energy for a system with increasing number of ``active'' superconducting islands. The parameters are the same as in Fig. \ref{Fig_B3}(a), with each superconducting island having a length $L_c=1$ $\mu$m. As the number of potential  barriers that are ``turned off" increases, one obtains both exponential pinning of the lowest energy mode and the opening of a significant gap between the lowest and second lowest state.}
\label{Fig_B4}
\end{figure}

The final step of this analysis involves considering an increasing number of (short) superconducting islands separated by uncovered regions with applied gate potentials in the potential well regime. Our goal is to show that one can create an effectively long wire that hosts a single pair of MZMs (at its ends) by using multiple small superconducting islands with potential wells between them. We test this idea by repeating the single barrier structure multiple times for islands with length $L_c = 1~\mu$m. The corresponding spectra are shown in Fig. \ref{Fig_B4} as a function of Zeeman splitting. Notice that in all of these spectra the second lowest energy level is strongly gapped. In addition, there is a clear increase of the zero energy pinning of the Majorana mode as the number of barriers is increased, i.e.  as the total length of the wire increases. Hence, one can create an effectively long wire by coupling multiple short 
superconducting islands through (relatively shallow) potential wells. This construction circumvents the issue of creating a flat effective potential and eliminates the difficulties associated with fine tunning the gate potential in the uncovered regions. Furthermore, since the potential wells act as traps for the the nearby low-energy states, this construction could be a useful solution for improving the protection of the Majorana mode in hybrid structures that contain (possibly unknown) sources of trivial sub-gap states. 

\section{Experimental Design}

The findings of this theoretical study are relevant for recent experiments on hybrid semiconductor-superconductor devices with several (1-4) gate electrodes underneath a proximitized semiconductor wire.\cite{Mourik2012,Chen2017} In such devices, generic settings of the gates can lead to the localization of MBSs underneath just a single gate, though, as was shown previously, the MBSs will typically leak out into other regions of the device. \cite{Chen2017} This wavefunction leakage reduces the MBS overlap, so that the corresponding low-energy mode appears pinned to zero energy in tunneling spectroscopy.

There are also ongoing experiments in which chains of superconducting quantum dots are defined along the nanowire. So far these chain devices were limited to two superconductors and two quantum dots between them, \cite{Su2017} but the fabrication of multi-dot devices should not pose fundamental challenges. These devices can be used for the quantum simulation of the Kitaev chain model.\cite{Sau2012,Fulga2013,Zhang2016} We point out that if quantum dots have dimensions similar to the islands studied here, spin-orbit interaction can lead to the formation of partially separated MBSs within each dot. A chain then would consist of pre-formed MBS pairs, and the goal would be to hybridize MBSs located left and right of each interdot barrier, while leaving one outermost pair at the chain ends uncoupled.

The primary challenge with multi-dot systems is finding the correct gate settings that would lead to well-separated MZMs. It has already been stressed that one has to carefully tune the gates under the quantum dots.\cite{Fulga2013}  Here we demonstrate that it is additionally necessary to carefully tune the barrier regions between the superconducting dots.  Not only does one have to be careful 
to tune away from potential values that support Andreev states bound to the barrier regions, but also to avoid barrier heights that allow for  multiple low energy modes spanning across multiple dots.  We propose shallow potential wells in the barrier region as the optimal regime to be realized experimentally.  

\begin{figure}[t]
\begin{center}
\includegraphics[width=0.49\textwidth]{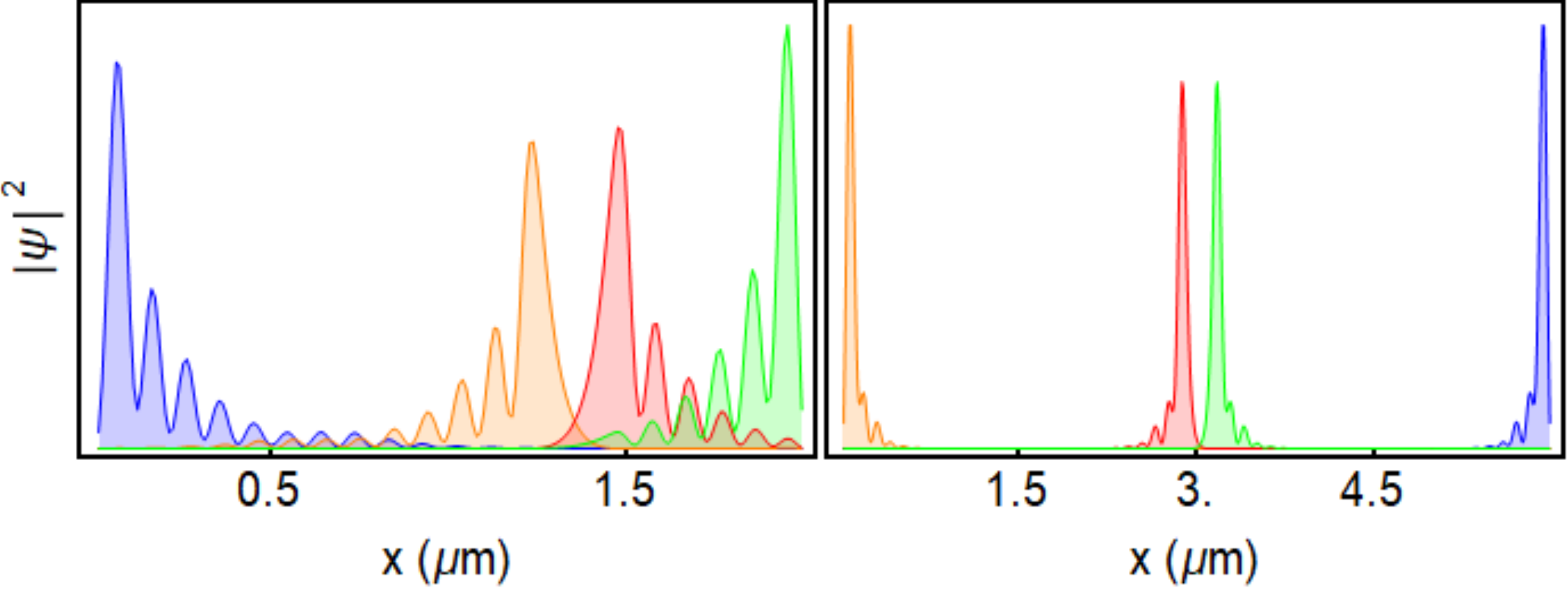}
\vspace{-0.5mm}
\end{center}
\vspace{-3.5mm}
\caption{Probability distribution as a function of position for the MBSs corresponding to the two lowest energy states in a $L=2$ $\mu$m, three-island device with $V_1=0.0$ and $V_2=0.8$ meV  (left) and a $L=6$ $\mu$m, six-island device  with all gates off except $V_3=0.8$ meV (right).  In both cases the Zeeman energy is greater than the critical field ($E_Z>\Delta$).}
\label{Fig_R1}
\end{figure}

A crucial experimental task is to test whether the system does indeed host well-separated MZMs localized at the ends (and no other low-energy state, including MBSs localized aways from the ends of the chain). Accomplishing such a task clearly requires non-local probes. Recent proposals for measuring Majorana non-locality include the use a quantum dot as a spectroscopic tool in a local transport measurement,\cite{Prada2017} as well as two-terminal setups that measure the current noise correlations\cite{Lu2012} or the spin blocking effect of MZMs.\cite{Ren2017} While this type of schemes can certainly distinguish well-separated MBSs from trivial low-energy modes localized at the ends of the system,  it is not clear that they can guarantee the absence of `false negatives' in certain (rather generic) conditions. As a specific example, let us consider a system supporting two (nearly degenerate) low-energy states which are superpositions of four MBSs, two localized at the ends of the chain and two deep inside the system (see Fig. \ref{Fig_R1}). 
The annihilation operators corresponding to the low-energy states can be expressed in terms of the Majorana operators 
\begin{eqnarray}
&\gamma_{w}&=\cos(\phi)\left(\psi_{\epsilon_1}+\psi_{-\epsilon_1}\right)-\sin(\phi)\left(\psi_{\epsilon_2}+\psi_{-\epsilon_2}\right),
\nonumber
\\
&\gamma_{x}&=\sin(\phi)\left(\psi_{\epsilon_1}+\psi_{-\epsilon_1}\right)+\cos(\phi)\left(\psi_{\epsilon_2}+\psi_{-\epsilon_2}\right),
\nonumber
\\
i&\gamma_{y}&=\cos(\theta)\left(\psi_{\epsilon_1}-\psi_{-\epsilon_1}\right)-\sin(\theta)\left(\psi_{\epsilon_2}-\psi_{-\epsilon_2}\right),
\nonumber
\\
i&\gamma_{z}&=\sin(\theta)\left(\psi_{\epsilon_1}-\psi_{-\epsilon_1}\right)+\cos(\theta)\left(\psi_{\epsilon_2}-\psi_{-\epsilon_2}\right),
\end{eqnarray}
where $\psi_{\pm\epsilon_i}$ are the $i^{th}$ positive and negative energy eigenvalues, and $\phi$ and $\theta$ are some parameter-specific rotations that decouple the MBSs.  The left panel of Fig. \ref{Fig_R1} shows the probability distribution for the two lowest energy states in a three dot device when the second barrier is tuned to the intermediate regime ($V_2=0.8$ meV), while the first remains off ($V_1=0.0$).    Although there are four MBSs, three of which are highly coupled, the only state which has significant weight at the left end of the wire is a single, well-separated MBS that has negligible overlap with all other low energy states.  A tunneling probe involving a quantum dot at the right end of the device\cite{Prada2017} will likely signal the presence of a ps-ABS\cite{Prada2017} (since there will be finite coupling to both the `green' and the `red' MBSs). By contrast, the same type of probe will signal the presence of a well-separated MBS at the left end of the system (as it only couples significantly to the `blue' Majorana). Consequently, this type of local probe can detect the presence of certain ps-ABSs, but a signal indicating their absence (i.e. well-separated MBSs) could very well be a false negative. More generally, no local probe at the left end of the wire can distinguishing the scenario illustrated in the left panel of Fig. \ref{Fig_R1} from a single low-energy state composed of a pair of MZMs localized at the two ends of the wire. Turning now to non-local probes,\cite{Lu2012,Ren2017} they can certainly distinguish between a system having a pair of MZMs and one having two (trivial) ABSs at the ends. In fact, this can also be done by performing two local measurements (one at each end of the chain). However, the key question is whether or not such a measurement is sensitive to the presence of a low-energy modes hidden deep inside the wire (in addition to the well-separated MZMs localized at the ends), a situation illustrated in the right panel of Fig. \ref{Fig_R1}. In principle, this type of `hidden' low-energy modes could be coupled to the end modes through charging effects. However, in a system characterized by a very small charging energy this coupling may not be measurable. In general, any detection scheme that aims to demonstrate the realization of MZMs localized at the ends of the system (which are separated by a finite energy gap from any other low-energy state) must be able to detect the presence of MBSs localized away from the ends of the wire (including deep inside the system). This would ensure the absence of false negatives. For the structure studied in this work -- a chain of coupled superconducting islands -- we propose that tunnel probes be attached to each island, more specifically to the uncovered (barrier) regions separating the islands. This would provide position-dependent spectroscopy, which clearly satisfies the requirement discussed above, provided the islands are short-enough (as compared to the MBS characteristic length-scale) so that no spurious low-energy state can `hide' inside an island (i.e. have negligible coupling to probes attached to the ends of the island).

\section{Conclusion} \label{conclusion}

We have studied the low-energy states that emerge in a one dimensional array of proximitized, gate controlled quantum dots. By performing extensive numerical calculations, we have shown that the realization of well-separated MZMs localized at the ends of the system by tunning the back-gate potentials is possible, but is not necessarily straightforward. Generically, a chain of coupled superconducting islands supports multiple low-energy states, which can be viewed as pairs of partially overlapping MBSs. However, the existence
of multiple MBSs throughout the system is not inconsistent with the observation of a robust zero-bias conductance peak (ZBCP) in a charge tunneling measurement at the end of the system.  Therefore, a single local measurement at the end of the system (where a MZM is expected to emerge) is insufficient for establishing the presence of well-separated MZMs. This conclusion has important practical implications regarding the interpretation of charge tunneling experiments on semiconductor-superconductor hybrid structures, because the presence of defects, impurities, or other non-homogeneities could effectively make the system a chain of proximitized quantum dots. Since such a system is generically expected to support multiple MBSs (some of which can be well-localized) and because the presence of this type of low-energy states is not inconsistent with the observation of a robust ZBCP, such observations cannot be interpreted as demonstrating the presence of well separated MZMs. Hence, using nonlocal probes represents a critical requirement for further progress in this field.

We have found that, in order to realize well-separated, topologically protected MZMs localized at the ends of a quantum dot chain, it is optimal to create potential wells inside the uncovered regions between the superconducting dots. These potential wells trap the nearby MBSs, which overlap strongly and acquire a finite gap. We have found that a sizable gap protecting the MZMs persists over a significant range of Zeeman fields and gate potentials. To optimize the stability of the MZMs, the regions separating the superconducting islands should be narrow (tens of nanometers). Realizing a flat effective potential is rather difficult, requires fine tunning, and may even be impossible if the uncovered region is too long. This observation has direct consequences for understanding the effective potential inside tunnel barrier regions, where (unintentional) quantum dots can form within a significant range of applied gate potentials. A quantitative theoretical description of the effective electrostatic potential requires solving a 3D Schr\"{o}dinger-Poisson problem that incorporates the details of the device, including its geometry. On the other hand, from the experimental point of view, engineering arrays of proximitized quantum dots represents a promising possible solution for realizing more controllable Majorana devices, provided one ensures that the gate regions, which act as traps for the nearby low-energy states, are narrow-enough and the detection scheme, which should be able to demonstrate the presence of well-separated MZMs, is based on non-local probes.   

\begin{acknowledgments}
This  work  is  supported  by  NSF DMR-1414683.  Additionally, J.P.T.S. and S.M.F are supported by NSF PIRE-1743717. S.M.F. is supported by NSF DMR-1743972, ONR and ARO.
\end{acknowledgments}

\bibliography{REFERENCES}

\begin{thebibliography}{80}%
\makeatletter
\providecommand \@ifxundefined [1]{%
 \@ifx{#1\undefined}
}%
\providecommand \@ifnum [1]{%
 \ifnum #1\expandafter \@firstoftwo
 \else \expandafter \@secondoftwo
 \fi
}%
\providecommand \@ifx [1]{%
 \ifx #1\expandafter \@firstoftwo
 \else \expandafter \@secondoftwo
 \fi
}%
\providecommand \natexlab [1]{#1}%
\providecommand \enquote  [1]{``#1''}%
\providecommand \bibnamefont  [1]{#1}%
\providecommand \bibfnamefont [1]{#1}%
\providecommand \citenamefont [1]{#1}%
\providecommand \href@noop [0]{\@secondoftwo}%
\providecommand \href [0]{\begingroup \@sanitize@url \@href}%
\providecommand \@href[1]{\@@startlink{#1}\@@href}%
\providecommand \@@href[1]{\endgroup#1\@@endlink}%
\providecommand \@sanitize@url [0]{\catcode `\\12\catcode `\$12\catcode
  `\&12\catcode `\#12\catcode `\^12\catcode `\_12\catcode `\%12\relax}%
\providecommand \@@startlink[1]{}%
\providecommand \@@endlink[0]{}%
\providecommand \url  [0]{\begingroup\@sanitize@url \@url }%
\providecommand \@url [1]{\endgroup\@href {#1}{\urlprefix }}%
\providecommand \urlprefix  [0]{URL }%
\providecommand \Eprint [0]{\href }%
\providecommand \doibase [0]{http://dx.doi.org/}%
\providecommand \selectlanguage [0]{\@gobble}%
\providecommand \bibinfo  [0]{\@secondoftwo}%
\providecommand \bibfield  [0]{\@secondoftwo}%
\providecommand \translation [1]{[#1]}%
\providecommand \BibitemOpen [0]{}%
\providecommand \bibitemStop [0]{}%
\providecommand \bibitemNoStop [0]{.\EOS\space}%
\providecommand \EOS [0]{\spacefactor3000\relax}%
\providecommand \BibitemShut  [1]{\csname bibitem#1\endcsname}%
\let\auto@bib@innerbib\@empty
\bibitem [{\citenamefont {Majorana}(1937)}]{Majorana1937}%
  \BibitemOpen
  \bibfield  {author} {\bibinfo {author} {\bibfnamefont {E.}~\bibnamefont
  {Majorana}},\ }\href@noop {} {\bibfield  {journal} {\bibinfo  {journal}
  {Nuovo Cimento}\ }\textbf {\bibinfo {volume} {5}},\ \bibinfo {pages} {171}
  (\bibinfo {year} {1937})}\BibitemShut {NoStop}%
\bibitem [{\citenamefont {Read}\ and\ \citenamefont {Green}(2000)}]{Read2000}%
  \BibitemOpen
  \bibfield  {author} {\bibinfo {author} {\bibfnamefont {N.}~\bibnamefont
  {Read}}\ and\ \bibinfo {author} {\bibfnamefont {D.}~\bibnamefont {Green}},\
  }\href {\doibase 10.1103/PhysRevB.61.10267} {\bibfield  {journal} {\bibinfo
  {journal} {Phys. Rev. B}\ }\textbf {\bibinfo {volume} {61}},\ \bibinfo
  {pages} {10267} (\bibinfo {year} {2000})}\BibitemShut {NoStop}%
\bibitem [{\citenamefont {Kitaev}(2001)}]{Kitaev2001}%
  \BibitemOpen
  \bibfield  {author} {\bibinfo {author} {\bibfnamefont {A.~Y.}\ \bibnamefont
  {Kitaev}},\ }\href {\doibase 10.1070/1063-7869/44/10S/S29} {\bibfield
  {journal} {\bibinfo  {journal} {Physics-Uspekhi}\ }\textbf {\bibinfo {volume}
  {44}},\ \bibinfo {pages} {131} (\bibinfo {year} {2001})}\BibitemShut
  {NoStop}%
\bibitem [{\citenamefont {Fu}\ and\ \citenamefont {Kane}(2008)}]{Fu2008}%
  \BibitemOpen
  \bibfield  {author} {\bibinfo {author} {\bibfnamefont {L.}~\bibnamefont
  {Fu}}\ and\ \bibinfo {author} {\bibfnamefont {C.~L.}\ \bibnamefont {Kane}},\
  }\href {\doibase 10.1103/PhysRevLett.100.096407} {\bibfield  {journal}
  {\bibinfo  {journal} {Phys. Rev. Lett.}\ }\textbf {\bibinfo {volume} {100}},\
  \bibinfo {pages} {096407} (\bibinfo {year} {2008})}\BibitemShut {NoStop}%
\bibitem [{\citenamefont {Fu}\ and\ \citenamefont {Kane}(2009)}]{Fu2009}%
  \BibitemOpen
  \bibfield  {author} {\bibinfo {author} {\bibfnamefont {L.}~\bibnamefont
  {Fu}}\ and\ \bibinfo {author} {\bibfnamefont {C.~L.}\ \bibnamefont {Kane}},\
  }\href {\doibase 10.1103/PhysRevB.79.161408} {\bibfield  {journal} {\bibinfo
  {journal} {Phys. Rev. B}\ }\textbf {\bibinfo {volume} {79}},\ \bibinfo
  {pages} {161408} (\bibinfo {year} {2009})}\BibitemShut {NoStop}%
\bibitem [{\citenamefont {Vazifeh}\ and\ \citenamefont
  {Franz}(2013)}]{Vazifeh2013}%
  \BibitemOpen
  \bibfield  {author} {\bibinfo {author} {\bibfnamefont {M.~M.}\ \bibnamefont
  {Vazifeh}}\ and\ \bibinfo {author} {\bibfnamefont {M.}~\bibnamefont
  {Franz}},\ }\href {\doibase 10.1103/PhysRevLett.111.206802} {\bibfield
  {journal} {\bibinfo  {journal} {Phys. Rev. Lett.}\ }\textbf {\bibinfo
  {volume} {111}},\ \bibinfo {pages} {206802} (\bibinfo {year}
  {2013})}\BibitemShut {NoStop}%
\bibitem [{\citenamefont {Nadj-Perge}\ \emph {et~al.}(2013)\citenamefont
  {Nadj-Perge}, \citenamefont {Drozdov}, \citenamefont {Bernevig},\ and\
  \citenamefont {Yazdani}}]{NadjPerge2013}%
  \BibitemOpen
  \bibfield  {author} {\bibinfo {author} {\bibfnamefont {S.}~\bibnamefont
  {Nadj-Perge}}, \bibinfo {author} {\bibfnamefont {I.~K.}\ \bibnamefont
  {Drozdov}}, \bibinfo {author} {\bibfnamefont {B.~A.}\ \bibnamefont
  {Bernevig}}, \ and\ \bibinfo {author} {\bibfnamefont {A.}~\bibnamefont
  {Yazdani}},\ }\href {\doibase 10.1103/PhysRevB.88.020407} {\bibfield
  {journal} {\bibinfo  {journal} {Phys. Rev. B}\ }\textbf {\bibinfo {volume}
  {88}},\ \bibinfo {pages} {020407} (\bibinfo {year} {2013})}\BibitemShut
  {NoStop}%
\bibitem [{\citenamefont {Pientka}\ \emph {et~al.}(2013)\citenamefont
  {Pientka}, \citenamefont {Glazman},\ and\ \citenamefont {von
  Oppen}}]{Pientka2013}%
  \BibitemOpen
  \bibfield  {author} {\bibinfo {author} {\bibfnamefont {F.}~\bibnamefont
  {Pientka}}, \bibinfo {author} {\bibfnamefont {L.~I.}\ \bibnamefont
  {Glazman}}, \ and\ \bibinfo {author} {\bibfnamefont {F.}~\bibnamefont {von
  Oppen}},\ }\href {\doibase 10.1103/PhysRevB.88.155420} {\bibfield  {journal}
  {\bibinfo  {journal} {Phys. Rev. B}\ }\textbf {\bibinfo {volume} {88}},\
  \bibinfo {pages} {155420} (\bibinfo {year} {2013})}\BibitemShut {NoStop}%
\bibitem [{\citenamefont {Sau}\ \emph {et~al.}(2010{\natexlab{a}})\citenamefont
  {Sau}, \citenamefont {Lutchyn}, \citenamefont {Tewari},\ and\ \citenamefont
  {Das~Sarma}}]{Sau2010a}%
  \BibitemOpen
  \bibfield  {author} {\bibinfo {author} {\bibfnamefont {J.~D.}\ \bibnamefont
  {Sau}}, \bibinfo {author} {\bibfnamefont {R.~M.}\ \bibnamefont {Lutchyn}},
  \bibinfo {author} {\bibfnamefont {S.}~\bibnamefont {Tewari}}, \ and\ \bibinfo
  {author} {\bibfnamefont {S.}~\bibnamefont {Das~Sarma}},\ }\href {\doibase
  10.1103/PhysRevLett.104.040502} {\bibfield  {journal} {\bibinfo  {journal}
  {Phys. Rev. Lett.}\ }\textbf {\bibinfo {volume} {104}},\ \bibinfo {pages}
  {040502} (\bibinfo {year} {2010}{\natexlab{a}})}\BibitemShut {NoStop}%
\bibitem [{\citenamefont {Alicea}(2010)}]{Alicea2010}%
  \BibitemOpen
  \bibfield  {author} {\bibinfo {author} {\bibfnamefont {J.}~\bibnamefont
  {Alicea}},\ }\href {\doibase 10.1103/PhysRevB.81.125318} {\bibfield
  {journal} {\bibinfo  {journal} {Phys. Rev. B}\ }\textbf {\bibinfo {volume}
  {81}},\ \bibinfo {pages} {125318} (\bibinfo {year} {2010})}\BibitemShut
  {NoStop}%
\bibitem [{\citenamefont {Oreg}\ \emph {et~al.}(2010)\citenamefont {Oreg},
  \citenamefont {Refael},\ and\ \citenamefont {von Oppen}}]{Oreg2010}%
  \BibitemOpen
  \bibfield  {author} {\bibinfo {author} {\bibfnamefont {Y.}~\bibnamefont
  {Oreg}}, \bibinfo {author} {\bibfnamefont {G.}~\bibnamefont {Refael}}, \ and\
  \bibinfo {author} {\bibfnamefont {F.}~\bibnamefont {von Oppen}},\ }\href
  {\doibase 10.1103/PhysRevLett.105.177002} {\bibfield  {journal} {\bibinfo
  {journal} {Phys. Rev. Lett.}\ }\textbf {\bibinfo {volume} {105}},\ \bibinfo
  {pages} {177002} (\bibinfo {year} {2010})}\BibitemShut {NoStop}%
\bibitem [{\citenamefont {Lutchyn}\ \emph {et~al.}(2010)\citenamefont
  {Lutchyn}, \citenamefont {Sau},\ and\ \citenamefont
  {Das~Sarma}}]{Lutchyn2010}%
  \BibitemOpen
  \bibfield  {author} {\bibinfo {author} {\bibfnamefont {R.~M.}\ \bibnamefont
  {Lutchyn}}, \bibinfo {author} {\bibfnamefont {J.~D.}\ \bibnamefont {Sau}}, \
  and\ \bibinfo {author} {\bibfnamefont {S.}~\bibnamefont {Das~Sarma}},\ }\href
  {\doibase 10.1103/PhysRevLett.105.077001} {\bibfield  {journal} {\bibinfo
  {journal} {Phys. Rev. Lett.}\ }\textbf {\bibinfo {volume} {105}},\ \bibinfo
  {pages} {077001} (\bibinfo {year} {2010})}\BibitemShut {NoStop}%
\bibitem [{\citenamefont {Sau}\ \emph {et~al.}(2010{\natexlab{b}})\citenamefont
  {Sau}, \citenamefont {Tewari}, \citenamefont {Lutchyn}, \citenamefont
  {Stanescu},\ and\ \citenamefont {Das~Sarma}}]{Sau2010}%
  \BibitemOpen
  \bibfield  {author} {\bibinfo {author} {\bibfnamefont {J.~D.}\ \bibnamefont
  {Sau}}, \bibinfo {author} {\bibfnamefont {S.}~\bibnamefont {Tewari}},
  \bibinfo {author} {\bibfnamefont {R.~M.}\ \bibnamefont {Lutchyn}}, \bibinfo
  {author} {\bibfnamefont {T.~D.}\ \bibnamefont {Stanescu}}, \ and\ \bibinfo
  {author} {\bibfnamefont {S.}~\bibnamefont {Das~Sarma}},\ }\href {\doibase
  10.1103/PhysRevB.82.214509} {\bibfield  {journal} {\bibinfo  {journal} {Phys.
  Rev. B}\ }\textbf {\bibinfo {volume} {82}},\ \bibinfo {pages} {214509}
  (\bibinfo {year} {2010}{\natexlab{b}})}\BibitemShut {NoStop}%
\bibitem [{\citenamefont {Mourik}\ \emph {et~al.}(2012)\citenamefont {Mourik},
  \citenamefont {Zuo}, \citenamefont {Frolov}, \citenamefont {Plissard},
  \citenamefont {Bakkers},\ and\ \citenamefont {Kouwenhoven}}]{Mourik2012}%
  \BibitemOpen
  \bibfield  {author} {\bibinfo {author} {\bibfnamefont {V.}~\bibnamefont
  {Mourik}}, \bibinfo {author} {\bibfnamefont {K.}~\bibnamefont {Zuo}},
  \bibinfo {author} {\bibfnamefont {S.~M.}\ \bibnamefont {Frolov}}, \bibinfo
  {author} {\bibfnamefont {S.~R.}\ \bibnamefont {Plissard}}, \bibinfo {author}
  {\bibfnamefont {E.~P. A.~M.}\ \bibnamefont {Bakkers}}, \ and\ \bibinfo
  {author} {\bibfnamefont {L.~P.}\ \bibnamefont {Kouwenhoven}},\ }\href
  {\doibase 10.1126/science.1222360} {\bibfield  {journal} {\bibinfo  {journal}
  {Science}\ }\textbf {\bibinfo {volume} {336}},\ \bibinfo {pages} {1003}
  (\bibinfo {year} {2012})}\BibitemShut {NoStop}%
\bibitem [{\citenamefont {Deng}\ \emph {et~al.}(2012)\citenamefont {Deng},
  \citenamefont {Yu}, \citenamefont {Huang}, \citenamefont {Larsson},
  \citenamefont {Caroff},\ and\ \citenamefont {Xu}}]{Deng2012}%
  \BibitemOpen
  \bibfield  {author} {\bibinfo {author} {\bibfnamefont {M.~T.}\ \bibnamefont
  {Deng}}, \bibinfo {author} {\bibfnamefont {C.~L.}\ \bibnamefont {Yu}},
  \bibinfo {author} {\bibfnamefont {G.~Y.}\ \bibnamefont {Huang}}, \bibinfo
  {author} {\bibfnamefont {M.}~\bibnamefont {Larsson}}, \bibinfo {author}
  {\bibfnamefont {P.}~\bibnamefont {Caroff}}, \ and\ \bibinfo {author}
  {\bibfnamefont {H.~Q.}\ \bibnamefont {Xu}},\ }\href {\doibase
  10.1021/nl303758w} {\bibfield  {journal} {\bibinfo  {journal} {Nano Letters}\
  }\textbf {\bibinfo {volume} {12}},\ \bibinfo {pages} {6414} (\bibinfo {year}
  {2012})}\BibitemShut {NoStop}%
\bibitem [{\citenamefont {Das}\ \emph {et~al.}(2012)\citenamefont {Das},
  \citenamefont {Ronen}, \citenamefont {Most}, \citenamefont {Oreg},
  \citenamefont {Heiblum},\ and\ \citenamefont {Shtrikman}}]{Das2012}%
  \BibitemOpen
  \bibfield  {author} {\bibinfo {author} {\bibfnamefont {A.}~\bibnamefont
  {Das}}, \bibinfo {author} {\bibfnamefont {Y.}~\bibnamefont {Ronen}}, \bibinfo
  {author} {\bibfnamefont {Y.}~\bibnamefont {Most}}, \bibinfo {author}
  {\bibfnamefont {Y.}~\bibnamefont {Oreg}}, \bibinfo {author} {\bibfnamefont
  {M.}~\bibnamefont {Heiblum}}, \ and\ \bibinfo {author} {\bibfnamefont
  {H.}~\bibnamefont {Shtrikman}},\ }\href@noop {} {\bibfield  {journal}
  {\bibinfo  {journal} {Nature Physics}\ }\textbf {\bibinfo {volume} {8}},\
  \bibinfo {pages} {887} (\bibinfo {year} {2012})}\BibitemShut {NoStop}%
\bibitem [{\citenamefont {Rokhinson}\ \emph {et~al.}(2012)\citenamefont
  {Rokhinson}, \citenamefont {Liu},\ and\ \citenamefont
  {Furdyna}}]{Rokhinson2012}%
  \BibitemOpen
  \bibfield  {author} {\bibinfo {author} {\bibfnamefont {L.~P.}\ \bibnamefont
  {Rokhinson}}, \bibinfo {author} {\bibfnamefont {X.}~\bibnamefont {Liu}}, \
  and\ \bibinfo {author} {\bibfnamefont {J.~K.}\ \bibnamefont {Furdyna}},\
  }\href {\doibase 10.1038/nphys2429} {\bibfield  {journal} {\bibinfo
  {journal} {Nature Physics}\ }\textbf {\bibinfo {volume} {8}},\ \bibinfo
  {pages} {795} (\bibinfo {year} {2012})}\BibitemShut {NoStop}%
\bibitem [{\citenamefont {Churchill}\ \emph {et~al.}(2013)\citenamefont
  {Churchill}, \citenamefont {Fatemi}, \citenamefont {Grove-Rasmussen},
  \citenamefont {Deng}, \citenamefont {Caroff}, \citenamefont {Xu},\ and\
  \citenamefont {Marcus}}]{Churchill2013}%
  \BibitemOpen
  \bibfield  {author} {\bibinfo {author} {\bibfnamefont {H.~O.~H.}\
  \bibnamefont {Churchill}}, \bibinfo {author} {\bibfnamefont {V.}~\bibnamefont
  {Fatemi}}, \bibinfo {author} {\bibfnamefont {K.}~\bibnamefont
  {Grove-Rasmussen}}, \bibinfo {author} {\bibfnamefont {M.~T.}\ \bibnamefont
  {Deng}}, \bibinfo {author} {\bibfnamefont {P.}~\bibnamefont {Caroff}},
  \bibinfo {author} {\bibfnamefont {H.~Q.}\ \bibnamefont {Xu}}, \ and\ \bibinfo
  {author} {\bibfnamefont {C.~M.}\ \bibnamefont {Marcus}},\ }\href {\doibase
  10.1103/PhysRevB.87.241401} {\bibfield  {journal} {\bibinfo  {journal} {Phys.
  Rev. B}\ }\textbf {\bibinfo {volume} {87}},\ \bibinfo {pages} {241401}
  (\bibinfo {year} {2013})}\BibitemShut {NoStop}%
\bibitem [{\citenamefont {Finck}\ \emph {et~al.}(2013)\citenamefont {Finck},
  \citenamefont {Van~Harlingen}, \citenamefont {Mohseni}, \citenamefont
  {Jung},\ and\ \citenamefont {Li}}]{Finck2013}%
  \BibitemOpen
  \bibfield  {author} {\bibinfo {author} {\bibfnamefont {A.~D.~K.}\
  \bibnamefont {Finck}}, \bibinfo {author} {\bibfnamefont {D.~J.}\ \bibnamefont
  {Van~Harlingen}}, \bibinfo {author} {\bibfnamefont {P.~K.}\ \bibnamefont
  {Mohseni}}, \bibinfo {author} {\bibfnamefont {K.}~\bibnamefont {Jung}}, \
  and\ \bibinfo {author} {\bibfnamefont {X.}~\bibnamefont {Li}},\ }\href
  {\doibase 10.1103/PhysRevLett.110.126406} {\bibfield  {journal} {\bibinfo
  {journal} {Phys. Rev. Lett.}\ }\textbf {\bibinfo {volume} {110}},\ \bibinfo
  {pages} {126406} (\bibinfo {year} {2013})}\BibitemShut {NoStop}%
\bibitem [{\citenamefont {Nadj-Perge}\ \emph {et~al.}(2014)\citenamefont
  {Nadj-Perge}, \citenamefont {Drozdov}, \citenamefont {Li}, \citenamefont
  {Chen}, \citenamefont {Jeon}, \citenamefont {Seo}, \citenamefont {MacDonald},
  \citenamefont {Bernevig},\ and\ \citenamefont {Yazdani}}]{NadjPerge2014}%
  \BibitemOpen
  \bibfield  {author} {\bibinfo {author} {\bibfnamefont {S.}~\bibnamefont
  {Nadj-Perge}}, \bibinfo {author} {\bibfnamefont {I.~K.}\ \bibnamefont
  {Drozdov}}, \bibinfo {author} {\bibfnamefont {J.}~\bibnamefont {Li}},
  \bibinfo {author} {\bibfnamefont {H.}~\bibnamefont {Chen}}, \bibinfo {author}
  {\bibfnamefont {S.}~\bibnamefont {Jeon}}, \bibinfo {author} {\bibfnamefont
  {J.}~\bibnamefont {Seo}}, \bibinfo {author} {\bibfnamefont {A.~H.}\
  \bibnamefont {MacDonald}}, \bibinfo {author} {\bibfnamefont {B.~A.}\
  \bibnamefont {Bernevig}}, \ and\ \bibinfo {author} {\bibfnamefont
  {A.}~\bibnamefont {Yazdani}},\ }\href {\doibase 10.1126/science.1259327}
  {\bibfield  {journal} {\bibinfo  {journal} {Science}\ }\textbf {\bibinfo
  {volume} {346}},\ \bibinfo {pages} {602} (\bibinfo {year}
  {2014})}\BibitemShut {NoStop}%
\bibitem [{\citenamefont {Wiedenmann}\ \emph {et~al.}(2016)\citenamefont
  {Wiedenmann}, \citenamefont {Bocquillon}, \citenamefont {Deacon},
  \citenamefont {Hartinger}, \citenamefont {Herrmann}, \citenamefont
  {Klapwijk}, \citenamefont {Maier}, \citenamefont {Ames}, \citenamefont
  {Brune}, \citenamefont {Gould}, \citenamefont {Oiwa}, \citenamefont
  {Ishibashi}, \citenamefont {Tarucha}, \citenamefont {Buhmann},\ and\
  \citenamefont {Molenkamp}}]{Wiedenmann2016}%
  \BibitemOpen
  \bibfield  {author} {\bibinfo {author} {\bibfnamefont {J.}~\bibnamefont
  {Wiedenmann}}, \bibinfo {author} {\bibfnamefont {E.}~\bibnamefont
  {Bocquillon}}, \bibinfo {author} {\bibfnamefont {R.~S.}\ \bibnamefont
  {Deacon}}, \bibinfo {author} {\bibfnamefont {S.}~\bibnamefont {Hartinger}},
  \bibinfo {author} {\bibfnamefont {O.}~\bibnamefont {Herrmann}}, \bibinfo
  {author} {\bibfnamefont {T.~M.}\ \bibnamefont {Klapwijk}}, \bibinfo {author}
  {\bibfnamefont {L.}~\bibnamefont {Maier}}, \bibinfo {author} {\bibfnamefont
  {C.}~\bibnamefont {Ames}}, \bibinfo {author} {\bibfnamefont {C.}~\bibnamefont
  {Brune}}, \bibinfo {author} {\bibfnamefont {C.}~\bibnamefont {Gould}},
  \bibinfo {author} {\bibfnamefont {A.}~\bibnamefont {Oiwa}}, \bibinfo {author}
  {\bibfnamefont {K.}~\bibnamefont {Ishibashi}}, \bibinfo {author}
  {\bibfnamefont {S.}~\bibnamefont {Tarucha}}, \bibinfo {author} {\bibfnamefont
  {H.}~\bibnamefont {Buhmann}}, \ and\ \bibinfo {author} {\bibfnamefont
  {L.~W.}\ \bibnamefont {Molenkamp}},\ }\href {\doibase 10.1038/ncomms10303}
  {\bibfield  {journal} {\bibinfo  {journal} {Nature Communications}\ }\textbf
  {\bibinfo {volume} {7}},\ \bibinfo {pages} {10303} (\bibinfo {year}
  {2016})}\BibitemShut {NoStop}%
\bibitem [{\citenamefont {Chang}\ \emph {et~al.}(2015)\citenamefont {Chang},
  \citenamefont {Albrecht}, \citenamefont {Jespersen}, \citenamefont
  {Kuemmeth}, \citenamefont {Krogstrup}, \citenamefont {Nyg{\r a}rd},\ and\
  \citenamefont {Marcus}}]{Chang2015}%
  \BibitemOpen
  \bibfield  {author} {\bibinfo {author} {\bibfnamefont {W.}~\bibnamefont
  {Chang}}, \bibinfo {author} {\bibfnamefont {S.~M.}\ \bibnamefont {Albrecht}},
  \bibinfo {author} {\bibfnamefont {T.~S.}\ \bibnamefont {Jespersen}}, \bibinfo
  {author} {\bibfnamefont {F.}~\bibnamefont {Kuemmeth}}, \bibinfo {author}
  {\bibfnamefont {P.}~\bibnamefont {Krogstrup}}, \bibinfo {author}
  {\bibfnamefont {J.}~\bibnamefont {Nyg{\r a}rd}}, \ and\ \bibinfo {author}
  {\bibfnamefont {C.~M.}\ \bibnamefont {Marcus}},\ }\href {\doibase
  10.1038/nnano.2014.306} {\bibfield  {journal} {\bibinfo  {journal} {Nat
  Nano}\ }\textbf {\bibinfo {volume} {10}},\ \bibinfo {pages} {232} (\bibinfo
  {year} {2015})}\BibitemShut {NoStop}%
\bibitem [{\citenamefont {Albrecht}\ \emph {et~al.}(2016)\citenamefont
  {Albrecht}, \citenamefont {Higginbotham}, \citenamefont {Madsen},
  \citenamefont {Kuemmeth}, \citenamefont {Jespersen}, \citenamefont {Nyg{\r
  a}rd}, \citenamefont {Krogstrup},\ and\ \citenamefont
  {Marcus}}]{Albrecht2016}%
  \BibitemOpen
  \bibfield  {author} {\bibinfo {author} {\bibfnamefont {S.~M.}\ \bibnamefont
  {Albrecht}}, \bibinfo {author} {\bibfnamefont {A.~P.}\ \bibnamefont
  {Higginbotham}}, \bibinfo {author} {\bibfnamefont {M.}~\bibnamefont
  {Madsen}}, \bibinfo {author} {\bibfnamefont {F.}~\bibnamefont {Kuemmeth}},
  \bibinfo {author} {\bibfnamefont {T.~S.}\ \bibnamefont {Jespersen}}, \bibinfo
  {author} {\bibfnamefont {J.}~\bibnamefont {Nyg{\r a}rd}}, \bibinfo {author}
  {\bibfnamefont {P.}~\bibnamefont {Krogstrup}}, \ and\ \bibinfo {author}
  {\bibfnamefont {C.~M.}\ \bibnamefont {Marcus}},\ }\href {\doibase
  10.1038/nature17162} {\bibfield  {journal} {\bibinfo  {journal} {Nature}\
  }\textbf {\bibinfo {volume} {531}},\ \bibinfo {pages} {206} (\bibinfo {year}
  {2016})}\BibitemShut {NoStop}%
\bibitem [{\citenamefont {Deng}\ \emph {et~al.}(2016)\citenamefont {Deng},
  \citenamefont {Vaitiekenas}, \citenamefont {Hansen}, \citenamefont {Danon},
  \citenamefont {Leijnse}, \citenamefont {Flensberg}, \citenamefont {Nyg{\r
  a}rd}, \citenamefont {Krogstrup},\ and\ \citenamefont {Marcus}}]{Deng2016}%
  \BibitemOpen
  \bibfield  {author} {\bibinfo {author} {\bibfnamefont {M.~T.}\ \bibnamefont
  {Deng}}, \bibinfo {author} {\bibfnamefont {S.}~\bibnamefont {Vaitiekenas}},
  \bibinfo {author} {\bibfnamefont {E.~B.}\ \bibnamefont {Hansen}}, \bibinfo
  {author} {\bibfnamefont {J.}~\bibnamefont {Danon}}, \bibinfo {author}
  {\bibfnamefont {M.}~\bibnamefont {Leijnse}}, \bibinfo {author} {\bibfnamefont
  {K.}~\bibnamefont {Flensberg}}, \bibinfo {author} {\bibfnamefont
  {J.}~\bibnamefont {Nyg{\r a}rd}}, \bibinfo {author} {\bibfnamefont
  {P.}~\bibnamefont {Krogstrup}}, \ and\ \bibinfo {author} {\bibfnamefont
  {C.~M.}\ \bibnamefont {Marcus}},\ }\href {\doibase 10.1126/science.aaf3961}
  {\bibfield  {journal} {\bibinfo  {journal} {Science}\ }\textbf {\bibinfo
  {volume} {354}},\ \bibinfo {pages} {1557} (\bibinfo {year}
  {2016})}\BibitemShut {NoStop}%
\bibitem [{\citenamefont {Chen}\ \emph {et~al.}(2017)\citenamefont {Chen},
  \citenamefont {Yu}, \citenamefont {Stenger}, \citenamefont {Hocevar},
  \citenamefont {Car}, \citenamefont {Plissard}, \citenamefont {Bakkers},
  \citenamefont {Stanescu},\ and\ \citenamefont {Frolov}}]{Chen2017}%
  \BibitemOpen
  \bibfield  {author} {\bibinfo {author} {\bibfnamefont {J.}~\bibnamefont
  {Chen}}, \bibinfo {author} {\bibfnamefont {P.}~\bibnamefont {Yu}}, \bibinfo
  {author} {\bibfnamefont {J.}~\bibnamefont {Stenger}}, \bibinfo {author}
  {\bibfnamefont {M.}~\bibnamefont {Hocevar}}, \bibinfo {author} {\bibfnamefont
  {D.}~\bibnamefont {Car}}, \bibinfo {author} {\bibfnamefont {S.~R.}\
  \bibnamefont {Plissard}}, \bibinfo {author} {\bibfnamefont {E.~P. A.~M.}\
  \bibnamefont {Bakkers}}, \bibinfo {author} {\bibfnamefont {T.~D.}\
  \bibnamefont {Stanescu}}, \ and\ \bibinfo {author} {\bibfnamefont {S.~M.}\
  \bibnamefont {Frolov}},\ }\href {\doibase 10.1126/sciadv.1701476} {\bibfield
  {journal} {\bibinfo  {journal} {Science Advances}\ }\textbf {\bibinfo
  {volume} {3}},\ \bibinfo {pages} {e1701476} (\bibinfo {year}
  {2017})}\BibitemShut {NoStop}%
\bibitem [{\citenamefont {Zhang}\ \emph
  {et~al.}(2017{\natexlab{a}})\citenamefont {Zhang}, \citenamefont {Gul},
  \citenamefont {Conesa-Boj}, \citenamefont {Nowak}, \citenamefont {Wimmer},
  \citenamefont {Zuo}, \citenamefont {Mourik}, \citenamefont {de~Vries},
  \citenamefont {van Veen}, \citenamefont {de~Moor}, \citenamefont {Bommer},
  \citenamefont {van Woerkom}, \citenamefont {Car}, \citenamefont {Plissard},
  \citenamefont {Bakkers}, \citenamefont {Quintero-Perez}, \citenamefont
  {Cassidy}, \citenamefont {Koelling}, \citenamefont {Goswami}, \citenamefont
  {Watanabe}, \citenamefont {Taniguchi},\ and\ \citenamefont
  {Kouwenhoven}}]{Zhang2017}%
  \BibitemOpen
  \bibfield  {author} {\bibinfo {author} {\bibfnamefont {H.}~\bibnamefont
  {Zhang}}, \bibinfo {author} {\bibfnamefont {O.}~\bibnamefont {Gul}}, \bibinfo
  {author} {\bibfnamefont {S.}~\bibnamefont {Conesa-Boj}}, \bibinfo {author}
  {\bibfnamefont {M.~P.}\ \bibnamefont {Nowak}}, \bibinfo {author}
  {\bibfnamefont {M.}~\bibnamefont {Wimmer}}, \bibinfo {author} {\bibfnamefont
  {K.}~\bibnamefont {Zuo}}, \bibinfo {author} {\bibfnamefont {V.}~\bibnamefont
  {Mourik}}, \bibinfo {author} {\bibfnamefont {F.~K.}\ \bibnamefont
  {de~Vries}}, \bibinfo {author} {\bibfnamefont {J.}~\bibnamefont {van Veen}},
  \bibinfo {author} {\bibfnamefont {M.~W.}\ \bibnamefont {de~Moor}}, \bibinfo
  {author} {\bibfnamefont {J.~D.}\ \bibnamefont {Bommer}}, \bibinfo {author}
  {\bibfnamefont {D.~J.}\ \bibnamefont {van Woerkom}}, \bibinfo {author}
  {\bibfnamefont {D.}~\bibnamefont {Car}}, \bibinfo {author} {\bibfnamefont
  {S.~R.}\ \bibnamefont {Plissard}}, \bibinfo {author} {\bibfnamefont {E.~P.}\
  \bibnamefont {Bakkers}}, \bibinfo {author} {\bibfnamefont {M.}~\bibnamefont
  {Quintero-Perez}}, \bibinfo {author} {\bibfnamefont {M.~C.}\ \bibnamefont
  {Cassidy}}, \bibinfo {author} {\bibfnamefont {S.}~\bibnamefont {Koelling}},
  \bibinfo {author} {\bibfnamefont {S.}~\bibnamefont {Goswami}}, \bibinfo
  {author} {\bibfnamefont {K.}~\bibnamefont {Watanabe}}, \bibinfo {author}
  {\bibfnamefont {T.}~\bibnamefont {Taniguchi}}, \ and\ \bibinfo {author}
  {\bibfnamefont {L.~P.}\ \bibnamefont {Kouwenhoven}},\ }\href {\doibase
  10.1038/ncomms16025} {\bibfield  {journal} {\bibinfo  {journal} {Nat.
  Commun.}\ }\textbf {\bibinfo {volume} {8}},\ \bibinfo {pages} {16025}
  (\bibinfo {year} {2017}{\natexlab{a}})}\BibitemShut {NoStop}%
\bibitem [{\citenamefont {Nichele}\ \emph {et~al.}(2017)\citenamefont
  {Nichele}, \citenamefont {Drachmann}, \citenamefont {Whiticar}, \citenamefont
  {O'Farrell}, \citenamefont {Suominen}, \citenamefont {Fornieri},
  \citenamefont {Wang}, \citenamefont {Gardner}, \citenamefont {Thomas},
  \citenamefont {Hatke}, \citenamefont {Krogstrup}, \citenamefont {Manfra},
  \citenamefont {Flensberg},\ and\ \citenamefont {Marcus}}]{Nichele2017}%
  \BibitemOpen
  \bibfield  {author} {\bibinfo {author} {\bibfnamefont {F.}~\bibnamefont
  {Nichele}}, \bibinfo {author} {\bibfnamefont {A.~C.~C.}\ \bibnamefont
  {Drachmann}}, \bibinfo {author} {\bibfnamefont {A.~M.}\ \bibnamefont
  {Whiticar}}, \bibinfo {author} {\bibfnamefont {E.~C.~T.}\ \bibnamefont
  {O'Farrell}}, \bibinfo {author} {\bibfnamefont {H.~J.}\ \bibnamefont
  {Suominen}}, \bibinfo {author} {\bibfnamefont {A.}~\bibnamefont {Fornieri}},
  \bibinfo {author} {\bibfnamefont {T.}~\bibnamefont {Wang}}, \bibinfo {author}
  {\bibfnamefont {G.~C.}\ \bibnamefont {Gardner}}, \bibinfo {author}
  {\bibfnamefont {C.}~\bibnamefont {Thomas}}, \bibinfo {author} {\bibfnamefont
  {A.~T.}\ \bibnamefont {Hatke}}, \bibinfo {author} {\bibfnamefont
  {P.}~\bibnamefont {Krogstrup}}, \bibinfo {author} {\bibfnamefont {M.~J.}\
  \bibnamefont {Manfra}}, \bibinfo {author} {\bibfnamefont {K.}~\bibnamefont
  {Flensberg}}, \ and\ \bibinfo {author} {\bibfnamefont {C.~M.}\ \bibnamefont
  {Marcus}},\ }\href {\doibase 10.1103/PhysRevLett.119.136803} {\bibfield
  {journal} {\bibinfo  {journal} {Phys. Rev. Lett.}\ }\textbf {\bibinfo
  {volume} {119}},\ \bibinfo {pages} {136803} (\bibinfo {year}
  {2017})}\BibitemShut {NoStop}%
\bibitem [{\citenamefont {Zhang}\ \emph
  {et~al.}(2017{\natexlab{b}})\citenamefont {Zhang}, \citenamefont {Liu},
  \citenamefont {Gazibegovic}, \citenamefont {Xu}, \citenamefont {Logan},
  \citenamefont {Wang}, \citenamefont {van Loo}, \citenamefont {Bommer},
  \citenamefont {de~Moor}, \citenamefont {Car}, \citenamefont {het Veld},
  \citenamefont {van Veldhoven}, \citenamefont {Koelling}, \citenamefont
  {Verheijen}, \citenamefont {Pendharkar}, \citenamefont {Pennachio},
  \citenamefont {Shojaei}, \citenamefont {Lee}, \citenamefont {Palmstrom},
  \citenamefont {Bakkers}, \citenamefont {Sarma},\ and\ \citenamefont
  {Kouwenhoven}}]{Zhang2018}%
  \BibitemOpen
  \bibfield  {author} {\bibinfo {author} {\bibfnamefont {H.}~\bibnamefont
  {Zhang}}, \bibinfo {author} {\bibfnamefont {C.-X.}\ \bibnamefont {Liu}},
  \bibinfo {author} {\bibfnamefont {S.}~\bibnamefont {Gazibegovic}}, \bibinfo
  {author} {\bibfnamefont {D.}~\bibnamefont {Xu}}, \bibinfo {author}
  {\bibfnamefont {J.~A.}\ \bibnamefont {Logan}}, \bibinfo {author}
  {\bibfnamefont {G.}~\bibnamefont {Wang}}, \bibinfo {author} {\bibfnamefont
  {N.}~\bibnamefont {van Loo}}, \bibinfo {author} {\bibfnamefont {J.~D.}\
  \bibnamefont {Bommer}}, \bibinfo {author} {\bibfnamefont {M.~W.}\
  \bibnamefont {de~Moor}}, \bibinfo {author} {\bibfnamefont {D.}~\bibnamefont
  {Car}}, \bibinfo {author} {\bibfnamefont {R.~L. M.~O.}\ \bibnamefont {het
  Veld}}, \bibinfo {author} {\bibfnamefont {P.~J.}\ \bibnamefont {van
  Veldhoven}}, \bibinfo {author} {\bibfnamefont {S.}~\bibnamefont {Koelling}},
  \bibinfo {author} {\bibfnamefont {M.~A.}\ \bibnamefont {Verheijen}}, \bibinfo
  {author} {\bibfnamefont {M.}~\bibnamefont {Pendharkar}}, \bibinfo {author}
  {\bibfnamefont {D.~J.}\ \bibnamefont {Pennachio}}, \bibinfo {author}
  {\bibfnamefont {B.}~\bibnamefont {Shojaei}}, \bibinfo {author} {\bibfnamefont
  {J.~S.}\ \bibnamefont {Lee}}, \bibinfo {author} {\bibfnamefont {C.~J.}\
  \bibnamefont {Palmstrom}}, \bibinfo {author} {\bibfnamefont {E.~P.}\
  \bibnamefont {Bakkers}}, \bibinfo {author} {\bibfnamefont {S.~D.}\
  \bibnamefont {Sarma}}, \ and\ \bibinfo {author} {\bibfnamefont {L.~P.}\
  \bibnamefont {Kouwenhoven}},\ }\href {https://arxiv.org/abs/1710.10701}
  {\bibfield  {journal} {\bibinfo  {journal} {e-print arXiv:1710.10701}\ }
  (\bibinfo {year} {2017}{\natexlab{b}})}\BibitemShut {NoStop}%
\bibitem [{\citenamefont {Shabani}\ \emph {et~al.}(2016)\citenamefont
  {Shabani}, \citenamefont {Kjaergaard}, \citenamefont {Suominen},
  \citenamefont {Kim}, \citenamefont {Nichele}, \citenamefont {Pakrouski},
  \citenamefont {Stankevic}, \citenamefont {Lutchyn}, \citenamefont
  {Krogstrup}, \citenamefont {Feidenhans'l}, \citenamefont {Kraemer},
  \citenamefont {Nayak}, \citenamefont {Troyer}, \citenamefont {Marcus},\ and\
  \citenamefont {Palmstr\o{}m}}]{Shabani2016}%
  \BibitemOpen
  \bibfield  {author} {\bibinfo {author} {\bibfnamefont {J.}~\bibnamefont
  {Shabani}}, \bibinfo {author} {\bibfnamefont {M.}~\bibnamefont {Kjaergaard}},
  \bibinfo {author} {\bibfnamefont {H.~J.}\ \bibnamefont {Suominen}}, \bibinfo
  {author} {\bibfnamefont {Y.}~\bibnamefont {Kim}}, \bibinfo {author}
  {\bibfnamefont {F.}~\bibnamefont {Nichele}}, \bibinfo {author} {\bibfnamefont
  {K.}~\bibnamefont {Pakrouski}}, \bibinfo {author} {\bibfnamefont
  {T.}~\bibnamefont {Stankevic}}, \bibinfo {author} {\bibfnamefont {R.~M.}\
  \bibnamefont {Lutchyn}}, \bibinfo {author} {\bibfnamefont {P.}~\bibnamefont
  {Krogstrup}}, \bibinfo {author} {\bibfnamefont {R.}~\bibnamefont
  {Feidenhans'l}}, \bibinfo {author} {\bibfnamefont {S.}~\bibnamefont
  {Kraemer}}, \bibinfo {author} {\bibfnamefont {C.}~\bibnamefont {Nayak}},
  \bibinfo {author} {\bibfnamefont {M.}~\bibnamefont {Troyer}}, \bibinfo
  {author} {\bibfnamefont {C.~M.}\ \bibnamefont {Marcus}}, \ and\ \bibinfo
  {author} {\bibfnamefont {C.~J.}\ \bibnamefont {Palmstr\o{}m}},\ }\href
  {\doibase 10.1103/PhysRevB.93.155402} {\bibfield  {journal} {\bibinfo
  {journal} {Phys. Rev. B}\ }\textbf {\bibinfo {volume} {93}},\ \bibinfo
  {pages} {155402} (\bibinfo {year} {2016})}\BibitemShut {NoStop}%
\bibitem [{\citenamefont {Kjaergaard}\ \emph {et~al.}(2016)\citenamefont
  {Kjaergaard}, \citenamefont {Nichele}, \citenamefont {Suominen},
  \citenamefont {Nowak}, \citenamefont {Wimmer}, \citenamefont {Akhmerov},
  \citenamefont {Folk}, \citenamefont {Flensberg}, \citenamefont {Shabani},
  \citenamefont {Palmstrøm},\ and\ \citenamefont {Marcus}}]{Kjaergaard2016}%
  \BibitemOpen
  \bibfield  {author} {\bibinfo {author} {\bibfnamefont {M.}~\bibnamefont
  {Kjaergaard}}, \bibinfo {author} {\bibfnamefont {F.}~\bibnamefont {Nichele}},
  \bibinfo {author} {\bibfnamefont {H.~J.}\ \bibnamefont {Suominen}}, \bibinfo
  {author} {\bibfnamefont {M.~P.}\ \bibnamefont {Nowak}}, \bibinfo {author}
  {\bibfnamefont {M.}~\bibnamefont {Wimmer}}, \bibinfo {author} {\bibfnamefont
  {A.~R.}\ \bibnamefont {Akhmerov}}, \bibinfo {author} {\bibfnamefont {J.~A.}\
  \bibnamefont {Folk}}, \bibinfo {author} {\bibfnamefont {K.}~\bibnamefont
  {Flensberg}}, \bibinfo {author} {\bibfnamefont {J.}~\bibnamefont {Shabani}},
  \bibinfo {author} {\bibfnamefont {C.~J.}\ \bibnamefont {Palmstrøm}}, \ and\
  \bibinfo {author} {\bibfnamefont {C.~M.}\ \bibnamefont {Marcus}},\ }\href
  {\doibase doi:10.1038/ncomms12841} {\bibfield  {journal} {\bibinfo  {journal}
  {Nature Communications}\ }\textbf {\bibinfo {volume} {7}},\ \bibinfo {pages}
  {12841} (\bibinfo {year} {2016})}\BibitemShut {NoStop}%
\bibitem [{\citenamefont {Suominen}\ \emph {et~al.}(2017)\citenamefont
  {Suominen}, \citenamefont {Kjaergaard}, \citenamefont {Hamilton},
  \citenamefont {Shabani}, \citenamefont {Palmstr\o{}m}, \citenamefont
  {Marcus},\ and\ \citenamefont {Nichele}}]{Suominen2017}%
  \BibitemOpen
  \bibfield  {author} {\bibinfo {author} {\bibfnamefont {H.~J.}\ \bibnamefont
  {Suominen}}, \bibinfo {author} {\bibfnamefont {M.}~\bibnamefont
  {Kjaergaard}}, \bibinfo {author} {\bibfnamefont {A.~R.}\ \bibnamefont
  {Hamilton}}, \bibinfo {author} {\bibfnamefont {J.}~\bibnamefont {Shabani}},
  \bibinfo {author} {\bibfnamefont {C.~J.}\ \bibnamefont {Palmstr\o{}m}},
  \bibinfo {author} {\bibfnamefont {C.~M.}\ \bibnamefont {Marcus}}, \ and\
  \bibinfo {author} {\bibfnamefont {F.}~\bibnamefont {Nichele}},\ }\href
  {\doibase 10.1103/PhysRevLett.119.176805} {\bibfield  {journal} {\bibinfo
  {journal} {Phys. Rev. Lett.}\ }\textbf {\bibinfo {volume} {119}},\ \bibinfo
  {pages} {176805} (\bibinfo {year} {2017})}\BibitemShut {NoStop}%
\bibitem [{\citenamefont {Alicea}\ \emph {et~al.}(2011)\citenamefont {Alicea},
  \citenamefont {Oreg}, \citenamefont {Refael}, \citenamefont {von Oppen},\
  and\ \citenamefont {Fisher}}]{Alicea2011}%
  \BibitemOpen
  \bibfield  {author} {\bibinfo {author} {\bibfnamefont {J.}~\bibnamefont
  {Alicea}}, \bibinfo {author} {\bibfnamefont {Y.}~\bibnamefont {Oreg}},
  \bibinfo {author} {\bibfnamefont {G.}~\bibnamefont {Refael}}, \bibinfo
  {author} {\bibfnamefont {F.}~\bibnamefont {von Oppen}}, \ and\ \bibinfo
  {author} {\bibfnamefont {M.~P.~A.}\ \bibnamefont {Fisher}},\ }\href {\doibase
  10.1038/nphys1915} {\bibfield  {journal} {\bibinfo  {journal} {Nature
  Physics}\ }\textbf {\bibinfo {volume} {7}},\ \bibinfo {pages} {412} (\bibinfo
  {year} {2011})}\BibitemShut {NoStop}%
\bibitem [{\citenamefont {Sau}\ \emph {et~al.}(2011)\citenamefont {Sau},
  \citenamefont {Clarke},\ and\ \citenamefont {Tewari}}]{Sau2011}%
  \BibitemOpen
  \bibfield  {author} {\bibinfo {author} {\bibfnamefont {J.~D.}\ \bibnamefont
  {Sau}}, \bibinfo {author} {\bibfnamefont {D.~J.}\ \bibnamefont {Clarke}}, \
  and\ \bibinfo {author} {\bibfnamefont {S.}~\bibnamefont {Tewari}},\ }\href
  {\doibase 10.1103/PhysRevB.84.094505} {\bibfield  {journal} {\bibinfo
  {journal} {Phys. Rev. B}\ }\textbf {\bibinfo {volume} {84}},\ \bibinfo
  {pages} {094505} (\bibinfo {year} {2011})}\BibitemShut {NoStop}%
\bibitem [{\citenamefont {Aasen}\ \emph {et~al.}(2016)\citenamefont {Aasen},
  \citenamefont {Hell}, \citenamefont {Mishmash}, \citenamefont {Higginbotham},
  \citenamefont {Danon}, \citenamefont {Leijnse}, \citenamefont {Jespersen},
  \citenamefont {Folk}, \citenamefont {Marcus}, \citenamefont {Flensberg},\
  and\ \citenamefont {Alicea}}]{Aasen2016}%
  \BibitemOpen
  \bibfield  {author} {\bibinfo {author} {\bibfnamefont {D.}~\bibnamefont
  {Aasen}}, \bibinfo {author} {\bibfnamefont {M.}~\bibnamefont {Hell}},
  \bibinfo {author} {\bibfnamefont {R.~V.}\ \bibnamefont {Mishmash}}, \bibinfo
  {author} {\bibfnamefont {A.}~\bibnamefont {Higginbotham}}, \bibinfo {author}
  {\bibfnamefont {J.}~\bibnamefont {Danon}}, \bibinfo {author} {\bibfnamefont
  {M.}~\bibnamefont {Leijnse}}, \bibinfo {author} {\bibfnamefont {T.~S.}\
  \bibnamefont {Jespersen}}, \bibinfo {author} {\bibfnamefont {J.~A.}\
  \bibnamefont {Folk}}, \bibinfo {author} {\bibfnamefont {C.~M.}\ \bibnamefont
  {Marcus}}, \bibinfo {author} {\bibfnamefont {K.}~\bibnamefont {Flensberg}}, \
  and\ \bibinfo {author} {\bibfnamefont {J.}~\bibnamefont {Alicea}},\ }\href
  {\doibase 10.1103/PhysRevX.6.031016} {\bibfield  {journal} {\bibinfo
  {journal} {Phys. Rev. X}\ }\textbf {\bibinfo {volume} {6}},\ \bibinfo {pages}
  {031016} (\bibinfo {year} {2016})}\BibitemShut {NoStop}%
\bibitem [{\citenamefont {Karzig}\ \emph {et~al.}(2017)\citenamefont {Karzig},
  \citenamefont {Knapp}, \citenamefont {Lutchyn}, \citenamefont {Bonderson},
  \citenamefont {Hastings}, \citenamefont {Nayak}, \citenamefont {Alicea},
  \citenamefont {Flensberg}, \citenamefont {Plugge}, \citenamefont {Oreg},
  \citenamefont {Marcus},\ and\ \citenamefont {Freedman}}]{Karzig2017}%
  \BibitemOpen
  \bibfield  {author} {\bibinfo {author} {\bibfnamefont {T.}~\bibnamefont
  {Karzig}}, \bibinfo {author} {\bibfnamefont {C.}~\bibnamefont {Knapp}},
  \bibinfo {author} {\bibfnamefont {R.~M.}\ \bibnamefont {Lutchyn}}, \bibinfo
  {author} {\bibfnamefont {P.}~\bibnamefont {Bonderson}}, \bibinfo {author}
  {\bibfnamefont {M.~B.}\ \bibnamefont {Hastings}}, \bibinfo {author}
  {\bibfnamefont {C.}~\bibnamefont {Nayak}}, \bibinfo {author} {\bibfnamefont
  {J.}~\bibnamefont {Alicea}}, \bibinfo {author} {\bibfnamefont
  {K.}~\bibnamefont {Flensberg}}, \bibinfo {author} {\bibfnamefont
  {S.}~\bibnamefont {Plugge}}, \bibinfo {author} {\bibfnamefont
  {Y.}~\bibnamefont {Oreg}}, \bibinfo {author} {\bibfnamefont {C.~M.}\
  \bibnamefont {Marcus}}, \ and\ \bibinfo {author} {\bibfnamefont {M.~H.}\
  \bibnamefont {Freedman}},\ }\href {\doibase 10.1103/PhysRevB.95.235305}
  {\bibfield  {journal} {\bibinfo  {journal} {Phys. Rev. B}\ }\textbf {\bibinfo
  {volume} {95}},\ \bibinfo {pages} {235305} (\bibinfo {year}
  {2017})}\BibitemShut {NoStop}%
\bibitem [{\citenamefont {Litinski}\ \emph {et~al.}(2017)\citenamefont
  {Litinski}, \citenamefont {Kesselring}, \citenamefont {Eisert},\ and\
  \citenamefont {von Oppen}}]{Litinski2017}%
  \BibitemOpen
  \bibfield  {author} {\bibinfo {author} {\bibfnamefont {D.}~\bibnamefont
  {Litinski}}, \bibinfo {author} {\bibfnamefont {M.~S.}\ \bibnamefont
  {Kesselring}}, \bibinfo {author} {\bibfnamefont {J.}~\bibnamefont {Eisert}},
  \ and\ \bibinfo {author} {\bibfnamefont {F.}~\bibnamefont {von Oppen}},\
  }\href {\doibase 10.1103/PhysRevX.7.031048} {\bibfield  {journal} {\bibinfo
  {journal} {Phys. Rev. X}\ }\textbf {\bibinfo {volume} {7}},\ \bibinfo {pages}
  {031048} (\bibinfo {year} {2017})}\BibitemShut {NoStop}%
\bibitem [{\citenamefont {Takei}\ \emph {et~al.}(2013)\citenamefont {Takei},
  \citenamefont {Fregoso}, \citenamefont {Hui}, \citenamefont {Lobos},\ and\
  \citenamefont {Das~Sarma}}]{Takei2013}%
  \BibitemOpen
  \bibfield  {author} {\bibinfo {author} {\bibfnamefont {S.}~\bibnamefont
  {Takei}}, \bibinfo {author} {\bibfnamefont {B.~M.}\ \bibnamefont {Fregoso}},
  \bibinfo {author} {\bibfnamefont {H.-Y.}\ \bibnamefont {Hui}}, \bibinfo
  {author} {\bibfnamefont {A.~M.}\ \bibnamefont {Lobos}}, \ and\ \bibinfo
  {author} {\bibfnamefont {S.}~\bibnamefont {Das~Sarma}},\ }\href {\doibase
  10.1103/PhysRevLett.110.186803} {\bibfield  {journal} {\bibinfo  {journal}
  {Phys. Rev. Lett.}\ }\textbf {\bibinfo {volume} {110}},\ \bibinfo {pages}
  {186803} (\bibinfo {year} {2013})}\BibitemShut {NoStop}%
\bibitem [{\citenamefont {Stanescu}\ \emph {et~al.}(2014)\citenamefont
  {Stanescu}, \citenamefont {Lutchyn},\ and\ \citenamefont
  {Das~Sarma}}]{Stanescu2014}%
  \BibitemOpen
  \bibfield  {author} {\bibinfo {author} {\bibfnamefont {T.~D.}\ \bibnamefont
  {Stanescu}}, \bibinfo {author} {\bibfnamefont {R.~M.}\ \bibnamefont
  {Lutchyn}}, \ and\ \bibinfo {author} {\bibfnamefont {S.}~\bibnamefont
  {Das~Sarma}},\ }\href {\doibase 10.1103/PhysRevB.90.085302} {\bibfield
  {journal} {\bibinfo  {journal} {Phys. Rev. B}\ }\textbf {\bibinfo {volume}
  {90}},\ \bibinfo {pages} {085302} (\bibinfo {year} {2014})}\BibitemShut
  {NoStop}%
\bibitem [{\citenamefont {Kells}\ \emph {et~al.}(2012)\citenamefont {Kells},
  \citenamefont {Meidan},\ and\ \citenamefont {Brouwer}}]{Kells2012}%
  \BibitemOpen
  \bibfield  {author} {\bibinfo {author} {\bibfnamefont {G.}~\bibnamefont
  {Kells}}, \bibinfo {author} {\bibfnamefont {D.}~\bibnamefont {Meidan}}, \
  and\ \bibinfo {author} {\bibfnamefont {P.~W.}\ \bibnamefont {Brouwer}},\
  }\href {\doibase 10.1103/PhysRevB.86.100503} {\bibfield  {journal} {\bibinfo
  {journal} {Phys. Rev. B}\ }\textbf {\bibinfo {volume} {86}},\ \bibinfo
  {pages} {100503} (\bibinfo {year} {2012})}\BibitemShut {NoStop}%
\bibitem [{\citenamefont {Chevallier}\ \emph {et~al.}(2012)\citenamefont
  {Chevallier}, \citenamefont {Sticlet}, \citenamefont {Simon},\ and\
  \citenamefont {Bena}}]{Chevallier2012}%
  \BibitemOpen
  \bibfield  {author} {\bibinfo {author} {\bibfnamefont {D.}~\bibnamefont
  {Chevallier}}, \bibinfo {author} {\bibfnamefont {D.}~\bibnamefont {Sticlet}},
  \bibinfo {author} {\bibfnamefont {P.}~\bibnamefont {Simon}}, \ and\ \bibinfo
  {author} {\bibfnamefont {C.}~\bibnamefont {Bena}},\ }\href {\doibase
  10.1103/PhysRevB.85.235307} {\bibfield  {journal} {\bibinfo  {journal} {Phys.
  Rev. B}\ }\textbf {\bibinfo {volume} {85}},\ \bibinfo {pages} {235307}
  (\bibinfo {year} {2012})}\BibitemShut {NoStop}%
\bibitem [{\citenamefont {Prada}\ \emph {et~al.}(2012)\citenamefont {Prada},
  \citenamefont {San-Jose},\ and\ \citenamefont {Aguado}}]{Prada2012}%
  \BibitemOpen
  \bibfield  {author} {\bibinfo {author} {\bibfnamefont {E.}~\bibnamefont
  {Prada}}, \bibinfo {author} {\bibfnamefont {P.}~\bibnamefont {San-Jose}}, \
  and\ \bibinfo {author} {\bibfnamefont {R.}~\bibnamefont {Aguado}},\ }\href
  {\doibase 10.1103/PhysRevB.86.180503} {\bibfield  {journal} {\bibinfo
  {journal} {Phys. Rev. B}\ }\textbf {\bibinfo {volume} {86}},\ \bibinfo
  {pages} {180503} (\bibinfo {year} {2012})}\BibitemShut {NoStop}%
\bibitem [{\citenamefont {Roy}\ \emph {et~al.}(2013)\citenamefont {Roy},
  \citenamefont {Bondyopadhaya},\ and\ \citenamefont {Tewari}}]{Roy2013}%
  \BibitemOpen
  \bibfield  {author} {\bibinfo {author} {\bibfnamefont {D.}~\bibnamefont
  {Roy}}, \bibinfo {author} {\bibfnamefont {N.}~\bibnamefont {Bondyopadhaya}},
  \ and\ \bibinfo {author} {\bibfnamefont {S.}~\bibnamefont {Tewari}},\ }\href
  {\doibase 10.1103/PhysRevB.88.020502} {\bibfield  {journal} {\bibinfo
  {journal} {Phys. Rev. B}\ }\textbf {\bibinfo {volume} {88}},\ \bibinfo
  {pages} {020502} (\bibinfo {year} {2013})}\BibitemShut {NoStop}%
\bibitem [{\citenamefont {San-Jose}\ \emph {et~al.}(2013)\citenamefont
  {San-Jose}, \citenamefont {Cayao}, \citenamefont {Prada},\ and\ \citenamefont
  {Aguado}}]{SanJose2013}%
  \BibitemOpen
  \bibfield  {author} {\bibinfo {author} {\bibfnamefont {P.}~\bibnamefont
  {San-Jose}}, \bibinfo {author} {\bibfnamefont {J.}~\bibnamefont {Cayao}},
  \bibinfo {author} {\bibfnamefont {E.}~\bibnamefont {Prada}}, \ and\ \bibinfo
  {author} {\bibfnamefont {R.}~\bibnamefont {Aguado}},\ }\href
  {http://stacks.iop.org/1367-2630/15/i=7/a=075019} {\bibfield  {journal}
  {\bibinfo  {journal} {New Journal of Physics}\ }\textbf {\bibinfo {volume}
  {15}},\ \bibinfo {pages} {075019} (\bibinfo {year} {2013})}\BibitemShut
  {NoStop}%
\bibitem [{\citenamefont {Ojanen}(2013)}]{Ojanen2013}%
  \BibitemOpen
  \bibfield  {author} {\bibinfo {author} {\bibfnamefont {T.}~\bibnamefont
  {Ojanen}},\ }\href {\doibase 10.1103/PhysRevB.87.100506} {\bibfield
  {journal} {\bibinfo  {journal} {Phys. Rev. B}\ }\textbf {\bibinfo {volume}
  {87}},\ \bibinfo {pages} {100506} (\bibinfo {year} {2013})}\BibitemShut
  {NoStop}%
\bibitem [{\citenamefont {Stanescu}\ and\ \citenamefont
  {Tewari}(2014)}]{Stanescu2014a}%
  \BibitemOpen
  \bibfield  {author} {\bibinfo {author} {\bibfnamefont {T.~D.}\ \bibnamefont
  {Stanescu}}\ and\ \bibinfo {author} {\bibfnamefont {S.}~\bibnamefont
  {Tewari}},\ }\href {\doibase 10.1103/PhysRevB.89.220507} {\bibfield
  {journal} {\bibinfo  {journal} {Phys. Rev. B}\ }\textbf {\bibinfo {volume}
  {89}},\ \bibinfo {pages} {220507} (\bibinfo {year} {2014})}\BibitemShut
  {NoStop}%
\bibitem [{\citenamefont {Lee}\ \emph {et~al.}(2014)\citenamefont {Lee},
  \citenamefont {Jiang}, \citenamefont {Houzet}, \citenamefont {Aguado},
  \citenamefont {Lieber},\ and\ \citenamefont {Franceschi}}]{Lee2014}%
  \BibitemOpen
  \bibfield  {author} {\bibinfo {author} {\bibfnamefont {E.~J.~H.}\
  \bibnamefont {Lee}}, \bibinfo {author} {\bibfnamefont {X.}~\bibnamefont
  {Jiang}}, \bibinfo {author} {\bibfnamefont {M.}~\bibnamefont {Houzet}},
  \bibinfo {author} {\bibfnamefont {R.}~\bibnamefont {Aguado}}, \bibinfo
  {author} {\bibfnamefont {C.~M.}\ \bibnamefont {Lieber}}, \ and\ \bibinfo
  {author} {\bibfnamefont {S.~D.}\ \bibnamefont {Franceschi}},\ }\href
  {\doibase 10.1038/nnano.2013.267} {\bibfield  {journal} {\bibinfo  {journal}
  {Nature Nanotechnology}\ }\textbf {\bibinfo {volume} {9}},\ \bibinfo {pages}
  {79} (\bibinfo {year} {2014})}\BibitemShut {NoStop}%
\bibitem [{\citenamefont {Cayao}\ \emph {et~al.}(2015)\citenamefont {Cayao},
  \citenamefont {Prada}, \citenamefont {San-Jose},\ and\ \citenamefont
  {Aguado}}]{Cayao2015}%
  \BibitemOpen
  \bibfield  {author} {\bibinfo {author} {\bibfnamefont {J.}~\bibnamefont
  {Cayao}}, \bibinfo {author} {\bibfnamefont {E.}~\bibnamefont {Prada}},
  \bibinfo {author} {\bibfnamefont {P.}~\bibnamefont {San-Jose}}, \ and\
  \bibinfo {author} {\bibfnamefont {R.}~\bibnamefont {Aguado}},\ }\href
  {\doibase 10.1103/PhysRevB.91.024514} {\bibfield  {journal} {\bibinfo
  {journal} {Phys. Rev. B}\ }\textbf {\bibinfo {volume} {91}},\ \bibinfo
  {pages} {024514} (\bibinfo {year} {2015})}\BibitemShut {NoStop}%
\bibitem [{\citenamefont {Klinovaja}\ and\ \citenamefont
  {Loss}(2015)}]{Klinovaja2015}%
  \BibitemOpen
  \bibfield  {author} {\bibinfo {author} {\bibfnamefont {J.}~\bibnamefont
  {Klinovaja}}\ and\ \bibinfo {author} {\bibfnamefont {D.}~\bibnamefont
  {Loss}},\ }\href {\doibase 10.1140/epjb/e2015-50882-2} {\bibfield  {journal}
  {\bibinfo  {journal} {The European Physical Journal B}\ }\textbf {\bibinfo
  {volume} {88}},\ \bibinfo {pages} {62} (\bibinfo {year} {2015})}\BibitemShut
  {NoStop}%
\bibitem [{\citenamefont {Fleckenstein}\ \emph {et~al.}(2017)\citenamefont
  {Fleckenstein}, \citenamefont {Dom\'inguez}, \citenamefont {Ziani},\ and\
  \citenamefont {Trauzettel}}]{Fleckenstein2017}%
  \BibitemOpen
  \bibfield  {author} {\bibinfo {author} {\bibfnamefont {C.}~\bibnamefont
  {Fleckenstein}}, \bibinfo {author} {\bibfnamefont {F.}~\bibnamefont
  {Dom\'inguez}}, \bibinfo {author} {\bibfnamefont {N.~T.}\ \bibnamefont
  {Ziani}}, \ and\ \bibinfo {author} {\bibfnamefont {B.}~\bibnamefont
  {Trauzettel}},\ }\href {https://arxiv.org/abs/1710.08866} {\bibfield
  {journal} {\bibinfo  {journal} {arXiv:1710.08866}\ } (\bibinfo {year}
  {2017})}\BibitemShut {NoStop}%
\bibitem [{\citenamefont {Liu}\ \emph {et~al.}(2017{\natexlab{a}})\citenamefont
  {Liu}, \citenamefont {Sau}, \citenamefont {Stanescu},\ and\ \citenamefont
  {Das~Sarma}}]{Liu2017a}%
  \BibitemOpen
  \bibfield  {author} {\bibinfo {author} {\bibfnamefont {C.-X.}\ \bibnamefont
  {Liu}}, \bibinfo {author} {\bibfnamefont {J.~D.}\ \bibnamefont {Sau}},
  \bibinfo {author} {\bibfnamefont {T.~D.}\ \bibnamefont {Stanescu}}, \ and\
  \bibinfo {author} {\bibfnamefont {S.}~\bibnamefont {Das~Sarma}},\ }\href
  {\doibase 10.1103/PhysRevB.96.075161} {\bibfield  {journal} {\bibinfo
  {journal} {Phys. Rev. B}\ }\textbf {\bibinfo {volume} {96}},\ \bibinfo
  {pages} {075161} (\bibinfo {year} {2017}{\natexlab{a}})}\BibitemShut
  {NoStop}%
\bibitem [{\citenamefont {Moore}\ \emph {et~al.}(2018)\citenamefont {Moore},
  \citenamefont {Stanescu},\ and\ \citenamefont {Tewari}}]{Moore2018}%
  \BibitemOpen
  \bibfield  {author} {\bibinfo {author} {\bibfnamefont {C.}~\bibnamefont
  {Moore}}, \bibinfo {author} {\bibfnamefont {T.~D.}\ \bibnamefont {Stanescu}},
  \ and\ \bibinfo {author} {\bibfnamefont {S.}~\bibnamefont {Tewari}},\ }\href
  {\doibase 10.1103/PhysRevB.97.165302} {\bibfield  {journal} {\bibinfo
  {journal} {Phys. Rev. B}\ }\textbf {\bibinfo {volume} {97}},\ \bibinfo
  {pages} {165302} (\bibinfo {year} {2018})}\BibitemShut {NoStop}%
\bibitem [{\citenamefont {Nayak}\ \emph {et~al.}(2008)\citenamefont {Nayak},
  \citenamefont {Simon}, \citenamefont {Stern}, \citenamefont {Freedman},\ and\
  \citenamefont {Das~Sarma}}]{Nayak2008}%
  \BibitemOpen
  \bibfield  {author} {\bibinfo {author} {\bibfnamefont {C.}~\bibnamefont
  {Nayak}}, \bibinfo {author} {\bibfnamefont {S.~H.}\ \bibnamefont {Simon}},
  \bibinfo {author} {\bibfnamefont {A.}~\bibnamefont {Stern}}, \bibinfo
  {author} {\bibfnamefont {M.}~\bibnamefont {Freedman}}, \ and\ \bibinfo
  {author} {\bibfnamefont {S.}~\bibnamefont {Das~Sarma}},\ }\href {\doibase
  10.1103/RevModPhys.80.1083} {\bibfield  {journal} {\bibinfo  {journal} {Rev.
  Mod. Phys.}\ }\textbf {\bibinfo {volume} {80}},\ \bibinfo {pages} {1083}
  (\bibinfo {year} {2008})}\BibitemShut {NoStop}%
\bibitem [{\citenamefont {Sarma}\ \emph {et~al.}(2015)\citenamefont {Sarma},
  \citenamefont {Freedman},\ and\ \citenamefont {Nayak}}]{DSarma2015}%
  \BibitemOpen
  \bibfield  {author} {\bibinfo {author} {\bibfnamefont {S.~D.}\ \bibnamefont
  {Sarma}}, \bibinfo {author} {\bibfnamefont {M.}~\bibnamefont {Freedman}}, \
  and\ \bibinfo {author} {\bibfnamefont {C.}~\bibnamefont {Nayak}},\ }\href
  {\doibase 10.1038/npjqi.2015.1} {\bibfield  {journal} {\bibinfo  {journal}
  {Npj Quantum Information}\ }\textbf {\bibinfo {volume} {1}},\ \bibinfo
  {pages} {15001} (\bibinfo {year} {2015})}\BibitemShut {NoStop}%
\bibitem [{\citenamefont {Stanescu}(2017)}]{Stanescu2017}%
  \BibitemOpen
  \bibfield  {author} {\bibinfo {author} {\bibfnamefont {T.~D.}\ \bibnamefont
  {Stanescu}},\ }\href@noop {} {\emph {\bibinfo {title} {Introduction to
  topological quantum matter and quantum computation}}}\ (\bibinfo  {publisher}
  {CRC Press, Taylor \& Francis Group},\ \bibinfo {year} {2017})\BibitemShut
  {NoStop}%
\bibitem [{\citenamefont {Stanescu}\ and\ \citenamefont
  {Tewari}(2016)}]{Stanescu2016}%
  \BibitemOpen
  \bibfield  {author} {\bibinfo {author} {\bibfnamefont {T.~D.}\ \bibnamefont
  {Stanescu}}\ and\ \bibinfo {author} {\bibfnamefont {S.}~\bibnamefont
  {Tewari}},\ }\href
  {http://journal.library.iisc.ernet.in/index.php/iisc/article/view/4607}
  {\bibfield  {journal} {\bibinfo  {journal} {Journal IISc}\ }\textbf {\bibinfo
  {volume} {96}},\ \bibinfo {pages} {107} (\bibinfo {year} {2016})}\BibitemShut
  {NoStop}%
\bibitem [{\citenamefont {Prada}\ \emph {et~al.}(2017)\citenamefont {Prada},
  \citenamefont {Aguado},\ and\ \citenamefont {San-Jose}}]{Prada2017}%
  \BibitemOpen
  \bibfield  {author} {\bibinfo {author} {\bibfnamefont {E.}~\bibnamefont
  {Prada}}, \bibinfo {author} {\bibfnamefont {R.}~\bibnamefont {Aguado}}, \
  and\ \bibinfo {author} {\bibfnamefont {P.}~\bibnamefont {San-Jose}},\ }\href
  {\doibase 10.1103/PhysRevB.96.085418} {\bibfield  {journal} {\bibinfo
  {journal} {Phys. Rev. B}\ }\textbf {\bibinfo {volume} {96}},\ \bibinfo
  {pages} {085418} (\bibinfo {year} {2017})}\BibitemShut {NoStop}%
\bibitem [{\citenamefont {Clarke}(2017)}]{Clarke2017}%
  \BibitemOpen
  \bibfield  {author} {\bibinfo {author} {\bibfnamefont {D.~J.}\ \bibnamefont
  {Clarke}},\ }\href {\doibase 10.1103/PhysRevB.96.201109} {\bibfield
  {journal} {\bibinfo  {journal} {Phys. Rev. B}\ }\textbf {\bibinfo {volume}
  {96}},\ \bibinfo {pages} {201109} (\bibinfo {year} {2017})}\BibitemShut
  {NoStop}%
\bibitem [{\citenamefont {Deng}\ \emph {et~al.}(2017)\citenamefont {Deng},
  \citenamefont {Vaitiekenas}, \citenamefont {Prada}, \citenamefont {San-Jose},
  \citenamefont {Nygard}, \citenamefont {Krogstrup}, \citenamefont {Aguado},\
  and\ \citenamefont {Marcus}}]{Deng2017}%
  \BibitemOpen
  \bibfield  {author} {\bibinfo {author} {\bibfnamefont {M.~T.}\ \bibnamefont
  {Deng}}, \bibinfo {author} {\bibfnamefont {S.}~\bibnamefont {Vaitiekenas}},
  \bibinfo {author} {\bibfnamefont {E.}~\bibnamefont {Prada}}, \bibinfo
  {author} {\bibfnamefont {P.}~\bibnamefont {San-Jose}}, \bibinfo {author}
  {\bibfnamefont {J.}~\bibnamefont {Nygard}}, \bibinfo {author} {\bibfnamefont
  {P.}~\bibnamefont {Krogstrup}}, \bibinfo {author} {\bibfnamefont
  {R.}~\bibnamefont {Aguado}}, \ and\ \bibinfo {author} {\bibfnamefont {C.~M.}\
  \bibnamefont {Marcus}},\ }\href {https://arxiv.org/abs/1712.03536} {\bibfield
   {journal} {\bibinfo  {journal} {arXiv:1712.03536}\ } (\bibinfo {year}
  {2017})}\BibitemShut {NoStop}%
\bibitem [{\citenamefont {Schuray}\ \emph {et~al.}(2017)\citenamefont
  {Schuray}, \citenamefont {Weithofer},\ and\ \citenamefont
  {Recher}}]{Schuray2017}%
  \BibitemOpen
  \bibfield  {author} {\bibinfo {author} {\bibfnamefont {A.}~\bibnamefont
  {Schuray}}, \bibinfo {author} {\bibfnamefont {L.}~\bibnamefont {Weithofer}},
  \ and\ \bibinfo {author} {\bibfnamefont {P.}~\bibnamefont {Recher}},\ }\href
  {\doibase 10.1103/PhysRevB.96.085417} {\bibfield  {journal} {\bibinfo
  {journal} {Phys. Rev. B}\ }\textbf {\bibinfo {volume} {96}},\ \bibinfo
  {pages} {085417} (\bibinfo {year} {2017})}\BibitemShut {NoStop}%
\bibitem [{\citenamefont {Sau}\ and\ \citenamefont {Sarma}(2012)}]{Sau2012a}%
  \BibitemOpen
  \bibfield  {author} {\bibinfo {author} {\bibfnamefont {J.~D.}\ \bibnamefont
  {Sau}}\ and\ \bibinfo {author} {\bibfnamefont {S.~D.}\ \bibnamefont
  {Sarma}},\ }\href {\doibase 10.1038/ncomms1966} {\bibfield  {journal}
  {\bibinfo  {journal} {Nature Communications}\ }\textbf {\bibinfo {volume}
  {3}},\ \bibinfo {pages} {964} (\bibinfo {year} {2012})}\BibitemShut {NoStop}%
\bibitem [{\citenamefont {Pientka}\ \emph {et~al.}(2012)\citenamefont
  {Pientka}, \citenamefont {Kells}, \citenamefont {Romito}, \citenamefont
  {Brouwer},\ and\ \citenamefont {von Oppen}}]{Pientka2012}%
  \BibitemOpen
  \bibfield  {author} {\bibinfo {author} {\bibfnamefont {F.}~\bibnamefont
  {Pientka}}, \bibinfo {author} {\bibfnamefont {G.}~\bibnamefont {Kells}},
  \bibinfo {author} {\bibfnamefont {A.}~\bibnamefont {Romito}}, \bibinfo
  {author} {\bibfnamefont {P.~W.}\ \bibnamefont {Brouwer}}, \ and\ \bibinfo
  {author} {\bibfnamefont {F.}~\bibnamefont {von Oppen}},\ }\href {\doibase
  10.1103/PhysRevLett.109.227006} {\bibfield  {journal} {\bibinfo  {journal}
  {Phys. Rev. Lett.}\ }\textbf {\bibinfo {volume} {109}},\ \bibinfo {pages}
  {227006} (\bibinfo {year} {2012})}\BibitemShut {NoStop}%
\bibitem [{\citenamefont {Lin}\ \emph {et~al.}(2012)\citenamefont {Lin},
  \citenamefont {Sau},\ and\ \citenamefont {Das~Sarma}}]{Lin2012}%
  \BibitemOpen
  \bibfield  {author} {\bibinfo {author} {\bibfnamefont {C.-H.}\ \bibnamefont
  {Lin}}, \bibinfo {author} {\bibfnamefont {J.~D.}\ \bibnamefont {Sau}}, \ and\
  \bibinfo {author} {\bibfnamefont {S.}~\bibnamefont {Das~Sarma}},\ }\href
  {\doibase 10.1103/PhysRevB.86.224511} {\bibfield  {journal} {\bibinfo
  {journal} {Phys. Rev. B}\ }\textbf {\bibinfo {volume} {86}},\ \bibinfo
  {pages} {224511} (\bibinfo {year} {2012})}\BibitemShut {NoStop}%
\bibitem [{\citenamefont {Rainis}\ \emph {et~al.}(2013)\citenamefont {Rainis},
  \citenamefont {Trifunovic}, \citenamefont {Klinovaja},\ and\ \citenamefont
  {Loss}}]{Rainis2013}%
  \BibitemOpen
  \bibfield  {author} {\bibinfo {author} {\bibfnamefont {D.}~\bibnamefont
  {Rainis}}, \bibinfo {author} {\bibfnamefont {L.}~\bibnamefont {Trifunovic}},
  \bibinfo {author} {\bibfnamefont {J.}~\bibnamefont {Klinovaja}}, \ and\
  \bibinfo {author} {\bibfnamefont {D.}~\bibnamefont {Loss}},\ }\href {\doibase
  10.1103/PhysRevB.87.024515} {\bibfield  {journal} {\bibinfo  {journal} {Phys.
  Rev. B}\ }\textbf {\bibinfo {volume} {87}},\ \bibinfo {pages} {024515}
  (\bibinfo {year} {2013})}\BibitemShut {NoStop}%
\bibitem [{\citenamefont {Yan}\ and\ \citenamefont {Wan}(2014)}]{Yan2014}%
  \BibitemOpen
  \bibfield  {author} {\bibinfo {author} {\bibfnamefont {Z.}~\bibnamefont
  {Yan}}\ and\ \bibinfo {author} {\bibfnamefont {S.}~\bibnamefont {Wan}},\
  }\href {http://stacks.iop.org/1367-2630/16/i=9/a=093004} {\bibfield
  {journal} {\bibinfo  {journal} {New Journal of Physics}\ }\textbf {\bibinfo
  {volume} {16}},\ \bibinfo {pages} {093004} (\bibinfo {year}
  {2014})}\BibitemShut {NoStop}%
\bibitem [{\citenamefont {Setiawan}\ \emph {et~al.}(2015)\citenamefont
  {Setiawan}, \citenamefont {Brydon}, \citenamefont {Sau},\ and\ \citenamefont
  {Das~Sarma}}]{Setiwan2015}%
  \BibitemOpen
  \bibfield  {author} {\bibinfo {author} {\bibfnamefont {F.}~\bibnamefont
  {Setiawan}}, \bibinfo {author} {\bibfnamefont {P.~M.~R.}\ \bibnamefont
  {Brydon}}, \bibinfo {author} {\bibfnamefont {J.~D.}\ \bibnamefont {Sau}}, \
  and\ \bibinfo {author} {\bibfnamefont {S.}~\bibnamefont {Das~Sarma}},\ }\href
  {\doibase 10.1103/PhysRevB.91.214513} {\bibfield  {journal} {\bibinfo
  {journal} {Phys. Rev. B}\ }\textbf {\bibinfo {volume} {91}},\ \bibinfo
  {pages} {214513} (\bibinfo {year} {2015})}\BibitemShut {NoStop}%
\bibitem [{\citenamefont {Liu}\ \emph {et~al.}(2017{\natexlab{b}})\citenamefont
  {Liu}, \citenamefont {Sau},\ and\ \citenamefont {Das~Sarma}}]{Liu2017}%
  \BibitemOpen
  \bibfield  {author} {\bibinfo {author} {\bibfnamefont {C.-X.}\ \bibnamefont
  {Liu}}, \bibinfo {author} {\bibfnamefont {J.~D.}\ \bibnamefont {Sau}}, \ and\
  \bibinfo {author} {\bibfnamefont {S.}~\bibnamefont {Das~Sarma}},\ }\href
  {\doibase 10.1103/PhysRevB.95.054502} {\bibfield  {journal} {\bibinfo
  {journal} {Phys. Rev. B}\ }\textbf {\bibinfo {volume} {95}},\ \bibinfo
  {pages} {054502} (\bibinfo {year} {2017}{\natexlab{b}})}\BibitemShut
  {NoStop}%
\bibitem [{\citenamefont {Stenger}\ and\ \citenamefont
  {Stanescu}(2017)}]{Stenger2017}%
  \BibitemOpen
  \bibfield  {author} {\bibinfo {author} {\bibfnamefont {J.}~\bibnamefont
  {Stenger}}\ and\ \bibinfo {author} {\bibfnamefont {T.~D.}\ \bibnamefont
  {Stanescu}},\ }\href {\doibase 10.1103/PhysRevB.96.214516} {\bibfield
  {journal} {\bibinfo  {journal} {Phys. Rev. B}\ }\textbf {\bibinfo {volume}
  {96}},\ \bibinfo {pages} {214516} (\bibinfo {year} {2017})}\BibitemShut
  {NoStop}%
\bibitem [{\citenamefont {Woods}\ \emph {et~al.}(2018)\citenamefont {Woods},
  \citenamefont {Stanescu},\ and\ \citenamefont {Das~Sarma}}]{Woods2018}%
  \BibitemOpen
  \bibfield  {author} {\bibinfo {author} {\bibfnamefont {B.~D.}\ \bibnamefont
  {Woods}}, \bibinfo {author} {\bibfnamefont {T.~D.}\ \bibnamefont {Stanescu}},
  \ and\ \bibinfo {author} {\bibfnamefont {S.}~\bibnamefont {Das~Sarma}},\
  }\href@noop {} {\bibfield  {journal} {\bibinfo  {journal} {arXiv:1801.02630}\
  } (\bibinfo {year} {2018})}\BibitemShut {NoStop}%
\bibitem [{\citenamefont {Stanescu}\ and\ \citenamefont
  {Tewari}(2013)}]{Stanescu2013}%
  \BibitemOpen
  \bibfield  {author} {\bibinfo {author} {\bibfnamefont {T.~D.}\ \bibnamefont
  {Stanescu}}\ and\ \bibinfo {author} {\bibfnamefont {S.}~\bibnamefont
  {Tewari}},\ }\href {http://stacks.iop.org/0953-8984/25/i=23/a=233201}
  {\bibfield  {journal} {\bibinfo  {journal} {J. Phys.: Condens. Matter}\
  }\textbf {\bibinfo {volume} {25}},\ \bibinfo {pages} {233201} (\bibinfo
  {year} {2013})}\BibitemShut {NoStop}%
\bibitem [{\citenamefont {Blonder}\ \emph {et~al.}(1982)\citenamefont
  {Blonder}, \citenamefont {Tinkham},\ and\ \citenamefont
  {Klapwijk}}]{Blonder1982}%
  \BibitemOpen
  \bibfield  {author} {\bibinfo {author} {\bibfnamefont {G.~E.}\ \bibnamefont
  {Blonder}}, \bibinfo {author} {\bibfnamefont {M.}~\bibnamefont {Tinkham}}, \
  and\ \bibinfo {author} {\bibfnamefont {T.~M.}\ \bibnamefont {Klapwijk}},\
  }\href {\doibase 10.1103/PhysRevB.25.4515} {\bibfield  {journal} {\bibinfo
  {journal} {Phys. Rev. B}\ }\textbf {\bibinfo {volume} {25}},\ \bibinfo
  {pages} {4515} (\bibinfo {year} {1982})}\BibitemShut {NoStop}%
\bibitem [{\citenamefont {Hansen}\ \emph {et~al.}(2018)\citenamefont {Hansen},
  \citenamefont {Danon},\ and\ \citenamefont {Flensberg}}]{Hansen2018}%
  \BibitemOpen
  \bibfield  {author} {\bibinfo {author} {\bibfnamefont {E.~B.}\ \bibnamefont
  {Hansen}}, \bibinfo {author} {\bibfnamefont {J.}~\bibnamefont {Danon}}, \
  and\ \bibinfo {author} {\bibfnamefont {K.}~\bibnamefont {Flensberg}},\ }\href
  {\doibase 10.1103/PhysRevB.97.041411} {\bibfield  {journal} {\bibinfo
  {journal} {Phys. Rev. B}\ }\textbf {\bibinfo {volume} {97}},\ \bibinfo
  {pages} {041411} (\bibinfo {year} {2018})}\BibitemShut {NoStop}%
\bibitem [{\citenamefont {Ptok}\ \emph {et~al.}(2017)\citenamefont {Ptok},
  \citenamefont {Kobia\l{}ka},\ and\ \citenamefont {Doma\ifmmode~\acute{n}\else
  \'{n}\fi{}ski}}]{Ptok2017}%
  \BibitemOpen
  \bibfield  {author} {\bibinfo {author} {\bibfnamefont {A.}~\bibnamefont
  {Ptok}}, \bibinfo {author} {\bibfnamefont {A.}~\bibnamefont {Kobia\l{}ka}}, \
  and\ \bibinfo {author} {\bibfnamefont {T.}~\bibnamefont
  {Doma\ifmmode~\acute{n}\else \'{n}\fi{}ski}},\ }\href {\doibase
  10.1103/PhysRevB.96.195430} {\bibfield  {journal} {\bibinfo  {journal} {Phys.
  Rev. B}\ }\textbf {\bibinfo {volume} {96}},\ \bibinfo {pages} {195430}
  (\bibinfo {year} {2017})}\BibitemShut {NoStop}%
\bibitem [{\citenamefont {Rosdahl}\ \emph {et~al.}(2018)\citenamefont
  {Rosdahl}, \citenamefont {Vuik}, \citenamefont {Kjaergaard},\ and\
  \citenamefont {Akhmerov}}]{Rosdah2018}%
  \BibitemOpen
  \bibfield  {author} {\bibinfo {author} {\bibfnamefont {T.~O.}\ \bibnamefont
  {Rosdahl}}, \bibinfo {author} {\bibfnamefont {A.}~\bibnamefont {Vuik}},
  \bibinfo {author} {\bibfnamefont {M.}~\bibnamefont {Kjaergaard}}, \ and\
  \bibinfo {author} {\bibfnamefont {A.~R.}\ \bibnamefont {Akhmerov}},\ }\href
  {\doibase 10.1103/PhysRevB.97.045421} {\bibfield  {journal} {\bibinfo
  {journal} {Phys. Rev. B}\ }\textbf {\bibinfo {volume} {97}},\ \bibinfo
  {pages} {045421} (\bibinfo {year} {2018})}\BibitemShut {NoStop}%
\bibitem [{\citenamefont {Alnaes}\ \emph {et~al.}(2015)\citenamefont {Alnaes},
  \citenamefont {Blechta}, \citenamefont {Hake}, \citenamefont {Johansson},
  \citenamefont {Kehlet}, \citenamefont {Logg}, \citenamefont {Richardson},
  \citenamefont {Ring}, \citenamefont {Rognes},\ and\ \citenamefont
  {Wells}}]{Alnaes2015}%
  \BibitemOpen
  \bibfield  {author} {\bibinfo {author} {\bibfnamefont {M.}~\bibnamefont
  {Alnaes}}, \bibinfo {author} {\bibfnamefont {J.}~\bibnamefont {Blechta}},
  \bibinfo {author} {\bibfnamefont {J.}~\bibnamefont {Hake}}, \bibinfo {author}
  {\bibfnamefont {A.}~\bibnamefont {Johansson}}, \bibinfo {author}
  {\bibfnamefont {B.}~\bibnamefont {Kehlet}}, \bibinfo {author} {\bibfnamefont
  {A.}~\bibnamefont {Logg}}, \bibinfo {author} {\bibfnamefont {C.}~\bibnamefont
  {Richardson}}, \bibinfo {author} {\bibfnamefont {J.}~\bibnamefont {Ring}},
  \bibinfo {author} {\bibfnamefont {M.}~\bibnamefont {Rognes}}, \ and\ \bibinfo
  {author} {\bibfnamefont {G.}~\bibnamefont {Wells}},\ }\href {\doibase
  10.11588/ans.2015.100.20553} {\bibfield  {journal} {\bibinfo  {journal}
  {Archive of Numerical Software}\ }\textbf {\bibinfo {volume} {3}},\ \bibinfo
  {pages} {100} (\bibinfo {year} {2015})}\BibitemShut {NoStop}%
\bibitem [{\citenamefont {Su}\ \emph {et~al.}(2017)\citenamefont {Su},
  \citenamefont {Tacla}, \citenamefont {Hocevar}, \citenamefont {Car},
  \citenamefont {Plissard}, \citenamefont {Bakkers}, \citenamefont {Daley},
  \citenamefont {Pekker},\ and\ \citenamefont {Frolov}}]{Su2017}%
  \BibitemOpen
  \bibfield  {author} {\bibinfo {author} {\bibfnamefont {Z.}~\bibnamefont
  {Su}}, \bibinfo {author} {\bibfnamefont {A.~B.}\ \bibnamefont {Tacla}},
  \bibinfo {author} {\bibfnamefont {M.}~\bibnamefont {Hocevar}}, \bibinfo
  {author} {\bibfnamefont {D.}~\bibnamefont {Car}}, \bibinfo {author}
  {\bibfnamefont {S.~R.}\ \bibnamefont {Plissard}}, \bibinfo {author}
  {\bibfnamefont {E.~P. A.~M.}\ \bibnamefont {Bakkers}}, \bibinfo {author}
  {\bibfnamefont {A.~J.}\ \bibnamefont {Daley}}, \bibinfo {author}
  {\bibfnamefont {D.}~\bibnamefont {Pekker}}, \ and\ \bibinfo {author}
  {\bibfnamefont {S.~M.}\ \bibnamefont {Frolov}},\ }\href {\doibase
  10.1038/s41467-017-00665-7} {\bibfield  {journal} {\bibinfo  {journal}
  {Nature Communications}\ }\textbf {\bibinfo {volume} {8}},\ \bibinfo {pages}
  {585} (\bibinfo {year} {2017})}\BibitemShut {NoStop}%
\bibitem [{\citenamefont {Sau}\ \emph {et~al.}(2012)\citenamefont {Sau},
  \citenamefont {Tewari},\ and\ \citenamefont {Das~Sarma}}]{Sau2012}%
  \BibitemOpen
  \bibfield  {author} {\bibinfo {author} {\bibfnamefont {J.~D.}\ \bibnamefont
  {Sau}}, \bibinfo {author} {\bibfnamefont {S.}~\bibnamefont {Tewari}}, \ and\
  \bibinfo {author} {\bibfnamefont {S.}~\bibnamefont {Das~Sarma}},\ }\href
  {\doibase 10.1103/PhysRevB.85.064512} {\bibfield  {journal} {\bibinfo
  {journal} {Phys. Rev. B}\ }\textbf {\bibinfo {volume} {85}},\ \bibinfo
  {pages} {064512} (\bibinfo {year} {2012})}\BibitemShut {NoStop}%
\bibitem [{\citenamefont {Fulga}\ \emph {et~al.}(2013)\citenamefont {Fulga},
  \citenamefont {Haim}, \citenamefont {Akhmerov},\ and\ \citenamefont
  {Oreg}}]{Fulga2013}%
  \BibitemOpen
  \bibfield  {author} {\bibinfo {author} {\bibfnamefont {I.~C.}\ \bibnamefont
  {Fulga}}, \bibinfo {author} {\bibfnamefont {A.}~\bibnamefont {Haim}},
  \bibinfo {author} {\bibfnamefont {A.~R.}\ \bibnamefont {Akhmerov}}, \ and\
  \bibinfo {author} {\bibfnamefont {Y.}~\bibnamefont {Oreg}},\ }\href {\doibase
  10.1088/1367-2630/15/4/045020} {\bibfield  {journal} {\bibinfo  {journal}
  {New Journal of Physics}\ }\textbf {\bibinfo {volume} {15}},\ \bibinfo
  {pages} {045020} (\bibinfo {year} {2013})}\BibitemShut {NoStop}%
\bibitem [{\citenamefont {Zhang}\ and\ \citenamefont {Nori}(2016)}]{Zhang2016}%
  \BibitemOpen
  \bibfield  {author} {\bibinfo {author} {\bibfnamefont {P.}~\bibnamefont
  {Zhang}}\ and\ \bibinfo {author} {\bibfnamefont {F.}~\bibnamefont {Nori}},\
  }\href {http://stacks.iop.org/1367-2630/18/i=4/a=043033} {\bibfield
  {journal} {\bibinfo  {journal} {New Journal of Physics}\ }\textbf {\bibinfo
  {volume} {18}},\ \bibinfo {pages} {043033} (\bibinfo {year}
  {2016})}\BibitemShut {NoStop}%
\bibitem [{\citenamefont {L\"u}\ \emph {et~al.}(2012)\citenamefont {L\"u},
  \citenamefont {Lu},\ and\ \citenamefont {Shen}}]{Lu2012}%
  \BibitemOpen
  \bibfield  {author} {\bibinfo {author} {\bibfnamefont {H.-F.}\ \bibnamefont
  {L\"u}}, \bibinfo {author} {\bibfnamefont {H.-Z.}\ \bibnamefont {Lu}}, \ and\
  \bibinfo {author} {\bibfnamefont {S.-Q.}\ \bibnamefont {Shen}},\ }\href
  {\doibase 10.1103/PhysRevB.86.075318} {\bibfield  {journal} {\bibinfo
  {journal} {Phys. Rev. B}\ }\textbf {\bibinfo {volume} {86}},\ \bibinfo
  {pages} {075318} (\bibinfo {year} {2012})}\BibitemShut {NoStop}%
\bibitem [{\citenamefont {Ren}\ \emph {et~al.}(2017)\citenamefont {Ren},
  \citenamefont {Yang}, \citenamefont {Xiang}, \citenamefont {Wang},\ and\
  \citenamefont {Tian}}]{Ren2017}%
  \BibitemOpen
  \bibfield  {author} {\bibinfo {author} {\bibfnamefont {C.}~\bibnamefont
  {Ren}}, \bibinfo {author} {\bibfnamefont {J.}~\bibnamefont {Yang}}, \bibinfo
  {author} {\bibfnamefont {J.}~\bibnamefont {Xiang}}, \bibinfo {author}
  {\bibfnamefont {S.}~\bibnamefont {Wang}}, \ and\ \bibinfo {author}
  {\bibfnamefont {H.}~\bibnamefont {Tian}},\ }\href {\doibase
  10.7566/JPSJ.86.124715} {\bibfield  {journal} {\bibinfo  {journal} {Journal
  of the Physical Society of Japan}\ }\textbf {\bibinfo {volume} {86}},\
  \bibinfo {pages} {124715} (\bibinfo {year} {2017})}\BibitemShut {NoStop}%
\end{thebibliography}%

\end{document}